\documentclass[14pt]{extarticle}

\usepackage{amsmath}
\usepackage{amssymb}
\usepackage{appendix}
\usepackage{amsthm}
\usepackage{enumitem}
\usepackage{hyperref}

\usepackage[round]{natbib}

\usepackage[margin=0.8in]{geometry}
\usepackage{mathtools}
\mathtoolsset{showonlyrefs=true}
\usepackage{bm}
\usepackage{setspace}

\numberwithin{equation}{section}

\theoremstyle{plain}
\newtheorem{theorem}{Theorem}[section]
\newtheorem{corollary}[theorem]{Corollary}

\newtheorem{Aassume}[theorem]{Assumption} 
\newtheorem{proposition}[theorem]{Proposition}
\newtheorem{lemma}[theorem]{Lemma}

\newtheorem{remark}[theorem]{Remark}

\theoremstyle{remark}

\begin{document}
	
	\title{Sensitivity analysis of long-term cash flows} 

	\author{Hyungbin Park\thanks{hyungbin@snu.ac.kr, hyungbin2015@gmail.com } \\ \\
		\normalsize{Department of Mathematical Sciences} \\ 
		\normalsize{Seoul National University}\\
		\normalsize{1, Gwanak-ro, Gwanak-gu, Seoul, Republic of Korea}  }
	
	\date{}

	\maketitle

\begin{abstract}
	This paper conducts a sensitivity analysis of long-term cash flows.
The price of the cash flow at time zero is given by
the pricing operator of a
Markov diffusion acting on the cash flow function.
We study the extent to which the price of the cash flow is affected by small perturbations of the underlying Markov diffusion.
The main tool is the Hansen--Scheinkman decomposition,
which is a method to express the cash flow in terms of eigenvalues and eigenfunctions
of the pricing operator. By incorporating techniques developed by \cite{fournie1999applications},  the sensitivities of long-term cash flows can be
represented via simple expressions in terms of the eigenvalue and the eigenfunction.
\end{abstract}

\section{Introduction}

\subsection{Long-term sensitivity}

In quantitative finance, we often evaluate expectations in the form 
\begin{equation}\label{eqn:p_T}
p_T:=\mathbb{E}_\xi^{\mathbb{P}}[e^{-\int_0^T r(X_s)\,ds} f(X_T)]  
\end{equation}
where $\mathbb{E}_\xi^\mathbb{P}$ is an expectation, $r$ and $f$ are measurable functions, and $X=(X_t)_{t\geq0}$ is an underlying stochastic process with $X_0=\xi.$  
Many financial quantities, such as option price and expected utility, are expressed in the above form.
This article examines the sensitivities of the expectation $p_T$ for large $T$ with respect to perturbations of the underlying process $X.$ 
The process $X$ with killing rate $r$ generates a pricing operator. 
We demonstrate that the long-term sensitivities can be represented by simple expressions in terms of eigenvalues and eigenfunctions of the pricing operator.
Our analysis reveals implications of models for the long-term risk and performance of cash flows.

We begin with a quadruple of functions $(b,\sigma,r,f)$ and a vector $\xi\in\mathbb{R}^d,$ 
which satisfy Assumptions \ref{assume:SDE} - \ref{assume:f} stated below.
(In this article, the terminology ``Assumption" is used when stating the premises for the whole paper. The terminology ``condition" is used for the hypotheses of propositions, theorems, and corollaries.)
Let $(\Omega,\mathcal{F},(\mathcal{F}_t)_{t\ge0},\mathbb{P})$ be a filtered probability space that has a $d$-dimensional Brownian motion $B=(B_t)_{t\ge0}=(B_{1,t},\cdots,B_{d,t})_{t\ge0}^\top.$ The filtration $(\mathcal{F}_t)_{t\geq 0}$ is the usual completed filtration generated by the Brownian motion $B.$

\begin{Aassume} \label{assume:SDE}
	Let $b:\mathbb{R}^d\to\mathbb{R}^d$ and $\sigma:\mathbb{R}^d\to\mathbb{R}^{d\times d}$ be continuous functions and $\xi\in\mathbb{R}^d.$  The matrix $\sigma$ is invertible. 
	Assume that the stochastic differential equation (SDE)
	\begin{equation}\label{eqn:SDE_X}
	dX_t=b(X_t)\,dt+\sigma(X_t)\,dB_t\,,\;X_0=\xi
	\end{equation}
	has a unique strong solution $X$ and that the solution is non-explosive.
\end{Aassume}
\noindent In this paper, the function $b=(b_1,\cdots,b_d)^\top$ is expressed as a $d$-dimensional column vector.
As is well known, the process $X$ is a $d$-dimensional conservative Markov diffusion process.

\begin{Aassume}\label{assume:r}
	The function $r:\mathbb{R}^d\to\mathbb{R}$ is a continuous function.
\end{Aassume}

\begin{Aassume}\label{assume:f}
	The function $f:\mathbb{R}^d\to\mathbb{R}$ is
	non-negative, non-zero and measurable.
\end{Aassume}

This paper primarily conducts two types of sensitivity analyses. The first is the initial value sensitivity, known as the {\em delta}, defined as $\nabla_\xi p_T.$ The long-term behavior of delta is of interest to us. When the 
quadruple $(b,\sigma,r,f)$ and $\xi$
satisfy Assumptions \ref{assume:SDE} - \ref{assume:f} and 
admit the Hansen--Scheinkman decomposition (Assumptions \ref{assume:HS} - \ref{assume:ergodic}), we will observe that the expectation $p_T$ is asymptotically
$$p_T\simeq e^{-\lambda T}\phi(\xi)$$
with a constant $\lambda$ and a positive function $\phi.$
Here, 
for two positive functions $p_T$ and $q_T$ of $T,$ the notation
$p_{T}\simeq q_{T}$ means that the limit
$\lim_{T\rightarrow\infty}\frac{p_{T}}{q_{T}}$ converges to a positive constant.
Therefore, one can predict that the long-term asymptotic behavior is
\begin{equation}\label{eqn:delta_asym}
\nabla_\xi\ln p_T=\frac{\nabla_\xi\, p_T}{p_{T}}\,\simeq\,\frac{\nabla_\xi\,\phi(\xi)}{\phi(\xi)}\,. 
\end{equation}

The second type of sensitivity analysis consists of the drift and diffusion sensitivities, which are known as the {\em rho} and the {\em vega}, respectively. The precise definitions of rho and vega are given in Section \ref{sec:sen_drift_diffu}. Consider a perturbed quadruple $(b_\epsilon,\sigma_\epsilon,r_\epsilon,f_\epsilon)$ and an initial value $\xi,$ which satisfy Assumptions \ref{assume:pert_defi} - \ref{assume:perturbed} in Section \ref{sec:sen_drift_diffu}. Here, $\epsilon$ is the perturbation parameter.
The perturbed underlying process and the perturbed expectation are denoted by $X^\epsilon=(X_t^\epsilon)_{t\geq0}$ and
\begin{equation}\label{eqn:perturbed_quant_intro}
p_T^\epsilon:=\mathbb{E}_{\xi}^{\mathbb{P}}
[e^{-\int_{0}^{T}r_\epsilon(X_{s}^{\epsilon})ds}
f_\epsilon(X_{T}^{\epsilon})]\,,
\end{equation} respectively. 
We want to determine the long-term asymptotic behavior of 
$$\frac{\partial }{\partial\epsilon}\Big|_{\epsilon=0}
p_{T}^{\epsilon}\,.$$
Through the Hansen--Scheinkman decomposition, 
it can be observed that 
$$p_{T}^\epsilon\simeq e^{-\lambda_\epsilon T}\phi_\epsilon(\xi)$$
with a constant $\lambda_\epsilon$  and a positive function $\phi_\epsilon.$ 
When $T$ is large,
because $e^{-\lambda_\epsilon T}$ dominates the perturbed quantity
$p_{T}^\epsilon,$
we can expect
$$\frac{\partial }{\partial\epsilon}\Big|_{\epsilon=0}p_{T}^{\epsilon}\,\simeq\, -T e^{-\lambda T}\phi(\xi) \frac{\partial }{\partial\epsilon}\Big|_{\epsilon=0}\lambda_\epsilon+e^{-\lambda T} \frac{\partial }{\partial\epsilon}\Big|_{\epsilon=0}\phi_{\epsilon}(\xi)\,.$$
Therefore,   the   simple relationship  
\begin{equation}\label{eqn:pert_asym}
\frac{1}{T}\frac{\partial }{\partial\epsilon}\Big|_{\epsilon=0}\ln p_T^\epsilon=\frac{\left.\frac{\partial }{\partial\epsilon}\right|_{\epsilon=0}p_{T}^{\epsilon}}{T p_{T}}
\,\simeq\, -\frac{\partial }{\partial\epsilon}\Big|_{\epsilon=0}\lambda_\epsilon 
\end{equation}
is obtained.

The main purpose of this paper is to justify these long-term relationships in Eq.\eqref{eqn:delta_asym} and Eq.\eqref{eqn:pert_asym} in a mathematically rigorous manner.
We employ the method
of \cite{fournie1999applications}, who use Malliavin calculus for their sensitivity analysis.
(Refer to \cite{nualart2006malliavin} and \cite{di2009malliavin} for topics related to Malliavin calculus.) 
Unfortunately, the method developed by \cite{fournie1999applications} cannot be applied to functionals of the form 
$$\mathbb{E}_\xi^{\mathbb{P}}[e^{-\int_{0}^{T}r(X_{s})\,ds}f(X_{T})]\,,$$
which is the form that interests us.
Their method (for calculating the delta and the vega) is valid only for {\em discretely monitored 
	functionals} of the form 
$$\mathbb{E}_\xi^{\mathbb{P}}[f(X_{t_{1}},X_{t_{2}},\cdots,X_{t_{m}})]\,,$$
in which the process $X$ is evaluated a finite number of times up to terminal time $T.$
In our case, however, the expectation $p_T$ in Eq.\eqref{eqn:p_T} involves the term
$e^{-\int_{0}^{T}r(X_{s})\,ds}$, 
which depends on the entire path of $(X_{s})_{0\leq s\leq T}.$ The Hansen--Scheinkman decomposition is useful in overcoming this problem because 
the decomposition 
transforms path-dependent functionals  
into discretely monitored 
functionals.
Thus, by utilizing the Hansen--Scheinkman decomposition, the method developed by \cite{fournie1999applications} can successfully be applied to our cases.

The rest of this paper is structured as follows.  
The related literature is reviewed in Section \ref{sec:related}.
We explain the Hansen--Scheinkman decomposition in Section \ref{sec:HS}.  
In Section \ref{sec:delta},
the long-term sensitivity of the initial value is 
investigated, and  
Eq.\eqref{eqn:delta_asym} is justified. 
In Section \ref{sec:sen_drift_diffu},   the
long-term sensitivity of the drift and diffusion terms are demonstrated, and Eq.\eqref{eqn:pert_asym} is derived. 
Sections \ref{sec:ex_option_prices} and \ref{sec:ex_utility} present examples, and the last section summarizes the paper.
The proofs of the main results and the details of the examples are presented in the appendices.

\subsection{Related literature}
\label{sec:related}

Sensitivity analysis of the expectation $p_T$ has been studied by many authors.  \cite{fournie1999applications} investigate option price sensitivities for hedging purposes. 
They present an original probabilistic method for the numerical computations of the sensitivities by employing Malliavin calculus.
\cite{benhamou2003optimal}
utilizes the Malliavin weighting functions introduced by \cite{fournie1999applications} to investigate an efficient computational method for the Greeks, and also  derives the weighting function
with the smallest total variance.
Employing  Malliavin calculus, 
\cite{el2004computations}   compute Greeks whose underlying process has Poisson jump times and random
jump sizes.  
\cite{gobet2005sensitivity} analyze the sensitivity of  expected costs and derive an expectation form of the sensitivity that can be evaluated using Monte Carlo simulations.
For this purpose, they employ three methods: the Malliavin calculus approach, the adjoint approach, and the martingale approach. 
\cite{davis2006malliavin}
investigate Malliavin calculus for Levy processes and derive the Malliavin weight for a certain class of
jump diffusion processes.
\cite{chen2007malliavin}
derive Monte Carlo estimators of option Greeks in diffusion models illuminating the connection
between Malliavin estimators and the likelihood
ratio method.
\cite{hansen2012dynamic}, \cite{hansen2009long} and \cite{hansen2012pricing}  
investigate the long-term behavior of risk by 
developing the Hansen--Scheinkman decomposition to reveal
the long-term risk-return trade-off.

We now briefly compare the  relevant results of \cite{borovivcka2011risk} and \cite{hansen2012pricing}, who employ sensitivity analysis 
to characterize and reveal risk price dynamics encoded in the stochastic discount factor.
By exploring shock-exposure elasticities 
and shock-price elasticities, they measure the impact of
shock on the expectation of the stochastic discount factor or cash flows.
They also demonstrate the state dependence of these elasticities.

The present paper 
differs from \cite{borovivcka2011risk} and \cite{hansen2012pricing} in two fundamental ways:  the multiplicative functional form and the perturbation form.
In their research, a more general form of the stochastic discount factor or cash flow, known as the multiplicative functional, is studied. One of the most commonly used multiplicative functionals is 
\begin{equation}\label{eqn:most_common_M}
M_t=e^{\int_0^tk(X_s)\,ds+\int_0^tv(X_s)\,dB_s}\,,\;t\geq0
\end{equation} 
for two functions $k$ and $v.$
In the current paper, however, we restrict the multiplicative functional to the form of
$e^{-\int_0^T r(X_s)\,ds}$ given in Eq.\eqref{eqn:p_T}.
In fact, the most commonly used multiplicative functional above can be reduced to this simpler form, as will be shown later.
The other distinction is that their perturbation form is somewhat different from the perturbation form in this paper. Let $$H_T^\epsilon:=e^{\int_0^T\kappa_\epsilon(X_s)\,ds+\epsilon\int_0^T\alpha(X_s)\,dB_s}\,,\;t\geq0\,.$$
Here, $\kappa_\epsilon(\cdot)$ and $\alpha(\cdot)$ are given functions that define the direction of perturbations.
Define the perturbed cash flows
and the perturbed expected return
by
$$q_T^\epsilon:=\mathbb{E}^\mathbb{P}[M_TH_T^\epsilon]\quad\textnormal{and} \quad \rho_T^\epsilon:=\frac{\mathbb{E}^\mathbb{P}[M_TH_T^\epsilon]}{\mathbb{E}^\mathbb{P}[G_TM_TH_T^\epsilon]}\,,$$ 
respectively.
Here, the random variable $G_T$ is a discount factor.  

One of the main purposes of their research is to 
study the sensitivities  
$\left.\frac{\partial }{\partial\epsilon}\right|_{\epsilon=0}
\ln q_{T}^{\epsilon}$
and 
$\left.\frac{\partial }{\partial\epsilon}\right|_{\epsilon=0}
\ln\rho_{T}^{\epsilon}$ for economic interpretation. 
{\em Shock elasticity}  $\epsilon(x,T)$ (see equation (12) in \cite{borovivcka2011risk}) is a component of  
$\left.\frac{\partial }{\partial\epsilon}\right|_{\epsilon=0}
\ln q_{T}^{\epsilon}$
and reflects  sensitivity with respect to a perturbation over an instant $dt.$
Its long-term behavior $\lim_{T\rightarrow\infty}\epsilon(x,T)$ 
is the counterpart to the computation of delta in this paper. From the discussion following Result 2.2 on page 13 in \cite{borovivcka2011risk}, it is apparent that the long-term shock elasticity is consistent with Theorem \ref{thm:delta} of the present paper.
\cite{borovivcka2014shock} present a more direct method of computing the shock elasticity.
They do not provide a long-term analysis for $\left.\frac{\partial }{\partial\epsilon}\right|_{\epsilon=0}
\ln\rho_{T}^{\epsilon}.$

The quantities $q_T^\epsilon$ and $\rho_T^\epsilon$ can be reduced to the form in Eq.\eqref{eqn:perturbed_quant_intro}   with $f_\epsilon\equiv1$  
if the process $X$ is a diffusion given by Eq.\eqref{eqn:SDE_X} and if the multiplicative functional is the most common form $M$ as defined in Eq.\eqref{eqn:most_common_M}. 
Define a measure $\mathbb{Q}^\epsilon$ with the Girsanov kernel $(v+\epsilon\alpha)(X_T).$ Then 
\begin{equation}\label{eqn:Girsanov_moving}
q_T^\epsilon=\mathbb{E}^\mathbb{P}[M_TH_T^\epsilon]=
\mathbb{E}^\mathbb{P}[e^{\int_0^T(k+\kappa_\epsilon)(X_s)\,ds+\int_0^T(v+\epsilon\alpha)(X_s)\,dB_s}]
=\mathbb{E}^{\mathbb{Q}^\epsilon}[e^{\int_0^T(k +\kappa_\epsilon +\frac{1}{2}|v +\epsilon\alpha|^2)(X_s)\,ds}]\,.
\end{equation} 
This equation 
has the interpretation of moving 
from the physical measure $\mathbb{P}$ to the risk-neutral measure $\mathbb{Q}^\epsilon.$ 
The $\mathbb{Q}^\epsilon$-dynamics of $X$ is
$$dX_t=(b+\sigma(v+\epsilon\alpha))(X_t)\,dt+\sigma(X_t)\,dW_t^\epsilon$$
with a $\mathbb{Q}^\epsilon$-Brownian motion  $W^\epsilon.$
For convenience, let $\mathbb{Q}:=\mathbb{Q}^{0}$ and $W:=W^0$.
Define a process $\hat{X}^\epsilon$ as the solution of 
$$d\hat{X}_t^\epsilon=(b+\sigma(v+\epsilon\alpha))(\hat{X}_t^\epsilon)\,dt+\sigma(\hat{X}_t^\epsilon)\,dW_t\,.$$
Then the $\mathbb{Q}$-distribution of $(\hat{X}_t^\epsilon)_{t\ge0}$ is the same as the $\mathbb{Q}^\epsilon$-distribution of $(X_t)_{t\ge0}.$ 
It follows that
$$q_T^\epsilon
=\mathbb{E}^{\mathbb{Q}^\epsilon}[e^{\int_0^T(k +\kappa_\epsilon +\frac{1}{2}|v +\epsilon\alpha|^2)(X_s)\,ds}]
=\mathbb{E}^{\mathbb{Q}}[e^{\int_0^T(k+\kappa_\epsilon+\frac{1}{2}|v+\epsilon\alpha|^2)(\hat{X}_s^\epsilon)\,ds}]\,.$$
Here, $|\cdot|$ is the usual multi-dimensional Euclidean norm.
By defining $r_\epsilon:=-(k+\kappa_\epsilon +\frac{1}{2}|v+\epsilon\alpha|^2),$  we obtain  
$q_T^\epsilon=\mathbb{E}^\mathbb{Q}[e^{-\int_0^Tr_\epsilon(\hat{X}_s^\epsilon)\,ds}],$
which is the  $p_T^\epsilon$ form with $f_\epsilon\equiv1$ in Eq.\eqref{eqn:perturbed_quant_intro}. 
Similarly, the quantity $\rho_T^\epsilon$ can also be expressed as a ratio of two expectations of the form   
$p_T^\epsilon.$

\section{Hansen--Scheinkman decomposition}
\label{sec:HS}

We begin with a quadruple of functions $(b,\sigma,r,f)$ and  $\xi\in\mathbb{R}^d$ satisfying  Assumptions \ref{assume:SDE} - \ref{assume:f}. In this section,
Assumptions \ref{assume:HS} - \ref{assume:ergodic} are additionally considered in order to enable the use of the
Hansen--Scheinkman decomposition. Recall that $(\Omega,\mathcal{F},(\mathcal{F}_t)_{t\ge0},\mathbb{P})$ is a filtered probability space that has a $d$-dimensional Brownian motion $B.$ The filtration $(\mathcal{F}_t)_{t\geq 0}$ is generated by the Brownian motion $B.$ 
We define a pricing operator $\mathcal{P}$ by $$\mathcal{P}_Tf(x)=\mathbb{E}_x^{\mathbb{P}}[e^{-\int_0^T r(X_s)\,ds} f(X_T)]\,.$$
The expectation in Eq.\eqref{eqn:p_T} is expressed as $p_T=\mathcal{P}_Tf(\xi).$
For a positive measurable function $\phi$ and a real number $\lambda$ such that
\begin{equation}\label{eqn:eigen_semigp}
\mathcal{P}_T\phi(x)=e^{-\lambda T}\phi(x)\quad\textnormal{for}\; T>0\,,\,x\in\mathbb{R}^d\,, 
\end{equation}
the process
\begin{equation}\label{eqn:M}
M_t^\phi:=e^{\lambda t-\int_{0}^{t} r(X_{s})\,ds}\,\frac{\phi(X_{t})}{\phi(\xi)}\,,\;0\leq t\leq T 
\end{equation}
is a positive martingale.
A measure $\mathbb{Q}^\phi$ on
each $\mathcal{F}_T$
defined by
$$\mathbb{Q}^\phi[A]=\mathbb{E}_\xi^\mathbb{P}[\mathbb{I}_AM_T^\phi]$$
for $A\in\mathcal{F}_T$
is called the {\em eigen-measure} with respect to $\phi.$
This definition is consistent for all $T\ge0$ because
$\mathbb{E}_\xi^{\mathbb{P}}[\mathbb{I}_{A}
M_{t}^\phi]=
\mathbb{E}_\xi^{\mathbb{P}}[\mathbb{I}_{A}
M_T^\phi]$
for any $A\in\mathcal{F}_{t}$ and $0\leq t\le T.$

\begin{Aassume}\label{assume:HS}
	For a real number $\lambda$ and a positive measurable function $\phi$,
	there exists a pair $(\lambda,\phi)$ satisfying Eq.\eqref{eqn:eigen_semigp}
	such that 
	the process $X$ is recurrent under the eigen-measure $\mathbb{Q}^\phi.$ 
\end{Aassume}	
\noindent 	In this case, the discount factor $e^{-\int_0^T r(X_t)\,dt}$ can be written as
$$e^{-\int_0^T r(X_s)\,ds}=M_T^\phi e^{-\lambda T}\frac{\phi(\xi)}{\phi(X_T)}\;.$$
This expression is referred to as the {\em Hansen--Scheinkman decomposition}.  
We say that $(\lambda,\phi),$ $\lambda,$ $\phi,$ and $\mathbb{Q}^\phi$ are the {\em recurrent eigenpair}, {\em recurrent eigenvalue}, {\em recurrent eigenfunction}, and {\em recurrent eigen-measure}, respectively.
In general, a recurrent eigenpair may not exist.
However, from Assumption \ref{assume:HS}, this paper assumes that  
the quadruple of functions $(b,\sigma,r,f)$ and $\xi\in\mathbb{R}^d$ have a
recurrent eigenpair.    Several studies address the existence of the recurrent eigenpair, for example, Section 9 in \cite{hansen2009long} and
Section 5 in \cite{qin2016positive}.

It is known that the recurrent eigenpair $(\lambda,\phi)$ is unique if it exists (Proposition 7.2 in \cite{hansen2009long}
and Theorem 3.1 in \cite{qin2016positive}).
Consequently, we use notations $M$ and $\mathbb{Q}$ instead of 
$M^\phi$ and $\mathbb{Q}^\phi,$ respectively.
These notations are not confusing because 
this paper always works with the recurrent eigenpair, and
the recurrent eigenpair $(\lambda,\phi)$ is unique.

We approach the Hansen--Scheinkman decomposition with
an eigenpair of the pricing operator $\mathcal{P}.$
The decomposition can also be approached in terms of an eigenpair of the infinitesimal generator of the process. For the generator approach, see Proposition 6.1 in \cite{hansen2009long}.
Furthermore, many authors have utilized the Hansen--Scheinkman decomposition for a general class of processes (Section 2 in \cite{qin2016positive}).
However, we apply the Hansen--Scheinkman decomposition only for Markov diffusion cases.

The next assumption concerns a regularity condition of the recurrent eigenfunction.
There are several conditions in Theorem 3.1 on page 145 and Theorem 3.3 on page 148 in \cite{pinsky1995positive} that guarantee this regularity condition.

\begin{Aassume} \label{assume:diff_phi}
	The recurrent eigenfunction $\phi$ is twice continuously differentiable.
\end{Aassume}
\noindent  This assumption has two implications. First, the recurrent eigenpair $(\lambda,\phi)$ is an eigenpair of a second-order partial differential operator.
Define  operator $\mathcal{L}$ by
$$\mathcal{L}=\frac{1}{2}\sum_{i,j=1}^da_{ij}\frac{\partial^2}{\partial x_i\partial x_j}+\sum_{i=1}^db_i\frac{\partial}{\partial x_i}-r,$$ 
where $a=\sigma\sigma^\top.$
This operator $\mathcal{L}$ can be understood as the generator of $X$ with the killing rate $r.$  
Applying the Ito formula to Eq.\eqref{eqn:M}, we have
\begin{equation}\label{eqn:dM}
\frac{dM_t}{M_t}=\frac{(\mathcal{L}\phi+\lambda)(X_t)}{\phi(X_t)}\,dt+\frac{(\nabla\phi)^\top(X_t)}{\phi(X_t)}\sigma(X_t)\, dB_t,
\end{equation} 
where $\nabla\phi$ is the $d\times 1$ gradient vector of $\phi.$ Because $M$ is a martingale, the differential $dM_t$ has a $dt$-term equal to zero. Thus, the pair $(\lambda,\phi)$ satisfies
$\mathcal{L}\phi=-\lambda\phi$
on the
support of the distribution of the process $X,$
which means that $(\lambda,\phi)$ is an eigenpair of the second-order partial differential operator $-\mathcal{L}.$

Second, the above assumption implies that the process $X$ is still a Markov diffusion process under the recurrent measure $\mathbb{Q}.$ 
Define a vector-valued function 
$\varphi:={\sigma^\top\nabla \phi}/{\phi}\,;$
then, according to Eq.\eqref{eqn:dM},
$$M_t=e^{\int_0^t\varphi^\top(X_s)\, dB_s-\frac{1}{2}\int_0^t|\varphi|^2(X_s)\,ds}\,,\;0\leq t\leq T\;.$$
According to the Girsanov theorem,  
the process $W$ defined by
\begin{equation}\label{eqn:BM_Girsa}
W_{t}:=B_{t}-\int_{0}^{t}\varphi(X_{s})\,ds\,,\;0\leq t\leq T
\end{equation}
is a $\mathbb{Q}$-Brownian motion.  
The process $(\varphi(X_t))_{t\geq0}$ is the {\em Girsanov kernel} of the change in measure.
Therefore, we obtain 
\begin{equation}\label{eqn:Q-dynamics}
\begin{aligned}
dX_{t}
&=(b(X_{t})+\sigma(X_{t})\varphi(X_{t}))\, dt+\sigma(X_{t})\, dW_{t}\,,\;0\leq t\leq T \,,
\end{aligned}
\end{equation}
which 
represents the dynamics of the Markov diffusion process $X$
under the recurrent eigen-measure $\mathbb{Q}.$

We assume a stronger condition on the $\mathbb{Q}$-distribution of $X.$ 
\begin{Aassume}\label{assume:inv}
	The process $X$ has an invariant distribution $\nu$ under the recurrent eigen-measure $\mathbb{Q}.$
\end{Aassume}
\noindent This eigen-measure is referred to as a {\em stochastically stable} measure in \cite{hansen2009long}.  It is noteworthy that the invariant distribution is independent from the initial value $X_0=\xi.$ 
Regarding the existence of the invariant distribution, see Theorem 9.5 on page 185 in \cite{pinsky1995positive} and 
Lemma 2.1 and Theorem 3.1 in \cite{zhang2014quasi}.

Assumption \ref{assume:ergodic} is about the $\nu$-ergodic property of the function $f.$ 
There are several conditions  that guarantee the following $\nu$-ergodicity condition. 
An $L^2$-approach can be found in Section 3.2 in \cite{cattiaux2014long}. The
Lyapunov criteria can be found in
\cite{meyn1993stability} and in
Theorem 8.7 on page 33 of \cite{bellet2006ergodic}.

\begin{Aassume}\label{assume:ergodic}
	The function $f$ is $\nu$-ergodic, that is, $f$ satisfies 
	$$\mathbb{E}_\xi^{\mathbb{Q}}[(f/\phi)(X_T)]\to \int (f/\phi) \,d\nu\;\;\textnormal{as } T\to \infty\,,$$
	and the limit is a finite positive number.
\end{Aassume}

Assumptions \ref{assume:HS} - \ref{assume:ergodic}  have an important implication.
Using the recurrent eigen-measure $\mathbb{Q},$
it follows that  
\begin{equation}  \label{eqn:HS_tranform}
\begin{aligned}
p_T=\mathbb{E}_\xi^{\mathbb{P}}[e^{-\int_0^T r(X_s)\,ds} f(X_T)]
&=\phi(\xi)\,e^{-\lambda T}\,\mathbb{E}_\xi^{\mathbb{P}}
[M_T\,(f/\phi)(X_T)]\\
&=\phi(\xi)\,e^{-\lambda T}\,
\mathbb{E}_\xi^{\mathbb{Q}}
[(f/\phi)(X_T)]\;.
\end{aligned}
\end{equation}
The expectation $p_T$ can be written as 
\begin{equation} \label{eqn:decomposition}
\begin{aligned}
p_T=\phi(\xi)\,e^{-\lambda T} \,
\mathbb{E}_\xi^{\mathbb{Q}}[(f/\phi)(X_T)]\,.
\end{aligned}
\end{equation}
Because $\mathbb{E}_\xi^{\mathbb{Q}}
[(f/\phi)(X_T)]$ converges to a finite positive constant as $T\to\infty,$ we obtain the  equality 
$$\lim_{T\to\infty}\frac{1}{T}\ln p_T=-\lambda\,,$$
which implies that the long-term behavior of $p_T$ is determined by the recurrent eigenvalue.

Eq.\eqref{eqn:decomposition} is very useful for sensitivity analysis.
The expectation $p_T$ 
is expressed in a relatively more manageable manner.
The term $\mathbb{E}_\xi^{\mathbb{Q}}[(f/\phi)(X_T)]$ depends on the final value $X_T,$
whereas $\mathbb{E}_\xi^{\mathbb{P}}[e^{-\int_{0}^{T} r(X_{s})ds}f(X_{T})]$ depends on the entire path of $(X_s)_{0\leq s\leq T}.$ 
If the $\mathbb{Q}$-distribution of $X_T$ is known, then one can directly analyze 
the term $\mathbb{E}_\xi^{\mathbb{Q}}[(f/\phi)(X_T)].$
This advantage makes it easier to study the long-term sensitivity of cash flows.

In summary, for any given quadruple of functions $(b,\sigma,r,f)$ and initial value $\xi\in\mathbb{R}^d$ satisfying Assumptions \ref{assume:SDE} - \ref{assume:f} and \ref{assume:HS} - \ref{assume:ergodic},
we have constructed the process $X,$ the pricing operator $\mathcal{P},$ the generator $\mathcal{L},$ the martingale $M,$ the recurrent  eigen-measure $\mathbb{Q},$ 
the recurrent eigenpair $(\lambda,\phi),$ 
the Girsanov kernel $\varphi$ and the invariant distribution $\nu.$  
The notations for these objects,  
$$X,\,\mathcal{P},\,\mathcal{L},\,M,\,\mathbb{Q},\,(\lambda,\phi),\,\varphi,\,\nu\,,$$
appear frequently in the rest of this paper.
For convenience, we simply state that the quadruple $(b,\sigma,r,f)$ and the initial value $\xi$ determine these objects.
However, some factors  are superfluous in the sense that the process $X$ is not affected by $r$, 
and none of these objects is affected by $f.$

\section{Sensitivity of the initial value}
\label{sec:delta}

This section discusses the initial value sensitivity, known as the {\em delta}, defined as $\nabla_\xi p_T.$ The long-term behavior of delta is of interest to us.
For any given quadruple of functions $(b,\sigma,r,f)$ and  $\xi$ satisfying Assumptions \ref{assume:SDE} - \ref{assume:f} and \ref{assume:HS} - \ref{assume:ergodic},
the notations
$$X,\,\mathcal{P},\,\mathcal{L},\,M,\,\mathbb{Q},\,(\lambda,\phi),\,\varphi,\,\nu$$
are self-explanatory.
The zero notation, $0$, below is also used to represent the zero vector without ambiguity.

\begin{theorem} \label{thm:delta}
	Let $(b,\sigma,r,f)$ and $\xi$ be a quadruple of functions and an initial value, respectively, satisfying Assumptions \ref{assume:SDE} - \ref{assume:f} and \ref{assume:HS} - \ref{assume:ergodic}.
	If $\mathbb{E}_\xi^{\mathbb{Q}}[(f/\phi)(X_{T})]$ is continuously differentiable in $\xi$, and if $\nabla_\xi\,\mathbb{E}_\xi^{\mathbb{Q}}[(f/\phi)(X_{T})]\rightarrow0$ as $T\rightarrow\infty,$
	then
	$$\lim_{T\rightarrow\infty}\nabla_\xi\ln p_T=\frac{\nabla_\xi\,\phi}{\phi(\xi)}\,. $$
\end{theorem}
\begin{proof} 
	The functions $\phi(\xi)$ and $\mathbb{E}_\xi^{\mathbb{Q}}[(f/\phi)(X_{T})]$
	are continuously differentiable in $\xi.$   From Eq.\eqref{eqn:decomposition}, the chain rule says that $\ln p_T$ is differentiable in $\xi$ and that  
	\begin{equation}\label{eqn:delta_HS}
	\begin{aligned}
	\nabla_\xi\ln p_T=\frac{\,\nabla_\xi\,p_T\,}{p_T}=\frac{\nabla_\xi\,\phi}{\phi(\xi)}+\frac{\nabla_\xi\,\mathbb{E}_\xi^{\mathbb{Q }}[(f/\phi)(X_{T})]}{\mathbb{E}_\xi^{\mathbb{Q}}[(f/\phi)(X_{T})]}\;.
	\end{aligned}
	\end{equation}
	Because 
	$\nabla_\xi\,\mathbb{E}_\xi^{\mathbb{Q}}[(f/\phi)(X_{T})]\rightarrow0$ and $\mathbb{E}_\xi^{\mathbb{Q}}[(f/\phi)(X_{T})]$ converges to a finite positive number as $T\to\infty,$ we obtain the desired result.
\end{proof}

Next, we find a sufficient condition such that
$\mathbb{E}_\xi^{\mathbb{Q}}[(f/\phi)(X_{T})]$ is continuously differentiable in $\xi$ and $\nabla_\xi\,\mathbb{E}_\xi^{\mathbb{Q}}[(f/\phi)(X_{T})]\rightarrow0$ as $T\rightarrow\infty.$
To that end, we use the result from Section 3.2 of \cite{fournie1999applications}.
Assume that the functions $b+\sigma\varphi$ and $\sigma$ are continuously differentiable with bounded derivatives.
Let $Y=(Y_t)_{t\geq0}$ be the first variation process defined by  
\begin{equation*}
\begin{aligned}
dY_t=(b+\sigma\varphi)'(X_t)Y_t\,dt+\sum_{i=1}^{d}\sigma_i'(X_t)Y_t\,dW_{i,t}\;,\;Y_0=I_d,
\end{aligned}
\end{equation*}
where $\sigma_i$ is the $i$-th column vector of $\sigma$, and $I_d$ is the $d\times d$ identity matrix.
For a multi-valued function $g=(g_1,g_2,\cdots,g_d)^\top$ in the column vector form, the derivative $g'$ is defined as a $d\times d$ matrix 
such that the $i$-th row is $\nabla g_i$ for $i=1,2,\cdots,d.$ 
We also assume that the diffusion matrix $b+\sigma\varphi$ satisfies the uniform ellipticity condition, that is, there exists $\epsilon>0$ such that 
$y^\top(b+\sigma\varphi)^\top(x)(b+\sigma\varphi)(x)y \geq \epsilon|y|^2$
for any $x,y\in\mathbb{R}^d.$
Denote the matrix 2-norm by $|\!|\cdot|\!|$.

\begin{proposition}\label{prop:delta} 	Let $(b,\sigma,r,f)$ and $\xi$ be a quadruple of functions and an initial value, respectively, satisfying Assumptions \ref{assume:SDE} - \ref{assume:f} and \ref{assume:HS} - \ref{assume:ergodic}.
	Assume that the functions $b+\sigma\varphi$ and $\sigma$ are continuously differentiable with bounded derivatives and that $b+\sigma\varphi$ satisfies the uniform ellipticity condition.
	If there exist positive constants $p\geq 2$ and $q$ with $1/p+1/q=1$ such that
	$\mathbb{E}_\xi^\mathbb{Q}[|\!|\sigma^{-1}(X_T)Y_T|\!|^p]$ and 
	$\mathbb{E}_\xi^\mathbb{Q}[(f/\phi)^q(X_T)]$   
	are uniformly bounded in $T$ on $[0,\infty),$ then
	$\mathbb{E}_\xi^{\mathbb{Q}}[(f/\phi)(X_{T})]$ is continuously differentiable in $\xi$ and 
	$\nabla_\xi\,\mathbb{E}_\xi^{\mathbb{Q}}[(f/\phi)(X_{T})]\rightarrow0$ as $T\rightarrow\infty.$
\end{proposition}

\begin{proof}
	From Proposition 3.2 in \cite{fournie1999applications}, it follows that
	$\mathbb{E}_\xi^{\mathbb{Q}}[(f/\phi)(X_{T})]$ is continuously differentiable in $\xi$ and 
	\begin{equation}
	\label{eqn:delta_int}
	\nabla_\xi\,\mathbb{E}_\xi^{\mathbb{Q}}[(f/\phi)(X_{T})]=\frac{1}{T}\,\mathbb{E}_\xi^\mathbb{Q}\Big[(f/\phi)(X_T)\int_0^T(\sigma^{-1}(X_s)Y_s)^\top dW_s\Big]\,.
	\end{equation}	 
	This equality is proven in their paper for $p=q=2$. However, their proof is also valid for any positive $p\geq2$ and $q$ with $1/p+1/q=1$ if we replace three conditions. Replace $\phi\in L^2$ from page 400, line 15, with $f/\phi\in L^q$;  and   $\epsilon_n(x)=\mathbb{E}[\cdots]^2$ from page 400, line 19, with $\epsilon_n(x)=\mathbb{E}[\cdots]^q$; and $\psi=\mathbb{E}[\cdots]^2$ from page 400, line 25,
	with $\psi=\mathbb{E}[\cdots]^p.$
	The partial derivative  	$\nabla_\xi\,\mathbb{E}_\xi^{\mathbb{Q}}[(f/\phi)(X_{T})]$  is continuous in $\xi$ because the convergence is uniform on compact sets in the first line on page 401.
	
	Because $\mathbb{E}_\xi^\mathbb{Q}[|\!|\sigma^{-1}(X_T)Y_T|\!|^p]$ is uniformly
	bounded in $T$ on $[0,\infty),$ we can write 
	$\mathbb{E}_\xi^\mathbb{Q}[|\!|\sigma^{-1}(X_T)Y_T|\!|^p]\leq c$ for a positive constant $c.$ 
	By the Holder inequality, the Burkholder-Davis-Gundy inequality and the Jensen inequality, it follows that
	\begin{equation}
	\begin{aligned}
	\Big|\nabla_\xi\,\mathbb{E}_\xi^{\mathbb{Q}}[(f/\phi)(X_{T})]\Big|
	&\leq\frac{1}{T}\,\Big(\mathbb{E}_\xi^\mathbb{Q}[(f/\phi)^q(X_T)]\Big)^{\frac{1}{q}}\,\Big(\mathbb{E}_\xi^\mathbb{Q}\Big[\Big|\int_0^T(\sigma^{-1}(X_s)Y_s)^\top dW_s\Big|^p\Big]\Big)^{\frac{1}{p}}\\
	&\leq\frac{c_p}{T}\,\Big(\mathbb{E}_\xi^\mathbb{Q}[(f/\phi)^q(X_T)]\Big)^{\frac{1}{q}}\,\Big(\mathbb{E}_\xi^\mathbb{Q}\Big[\Big(\int_0^T|\!|\sigma^{-1}(X_s)Y_s|\!|^2ds\Big)^{\frac{p}{2}}\Big]\Big)^{\frac{1}{p}}\\
	&\leq\frac{c_p}{T^{\frac{1}{2}+\frac{1}{p}}}\,\Big(\mathbb{E}_\xi^\mathbb{Q}[(f/\phi)^q(X_T)]\Big)^{\frac{1}{q}}\,\Big(\mathbb{E}_\xi^\mathbb{Q} \Big[\int_0^T|\!|\sigma^{-1}(X_s)Y_s |\!|^{p}ds\Big]\Big)^{\frac{1}{p}}\\
	&=\frac{c_p}{T^{\frac{1}{2}+\frac{1}{p}}}\,\Big(\mathbb{E}_\xi^\mathbb{Q}[(f/\phi)^q(X_T)]\Big)^{\frac{1}{q}}\,\Big( 
	\int_0^T\mathbb{E}_\xi^\mathbb{Q} [|\!|\sigma^{-1}(X_s)Y_s |\!|^{p}]\,ds \Big)^{\frac{1}{p}}\\
	&\leq\frac{c_pc^{\frac{1}{p}}}{T^{\frac{1}{2}}}\,\Big(\mathbb{E}_\xi^\mathbb{Q}[(f/\phi)^q(X_T)]\Big)^{\frac{1}{q}} 
	\end{aligned}
	\end{equation}
	for the positive constant $c_p$ in the Burkholder-Davis-Gundy inequality.
	Because
	$\mathbb{E}_\xi^\mathbb{Q}[(f/\phi)^q(X_T)]$   
	is uniformly
	bounded in $T$ on $[0,\infty),$
	we obtain the desired result.
\end{proof}

\begin{remark}
	In \cite{fournie1999applications},
	Eq.\eqref{eqn:delta_int} holds under 
	stronger conditions of the coefficients in the dynamics of $X.$ The functions $b+\sigma\varphi$ and $\sigma$ are continuously differentiable with bounded derivatives, and $b+\sigma\varphi$ satisfies the uniform ellipticity condition.
	Recently, \cite{banos2017computing} 
	have relaxed these conditions. 
	The drift coefficient $b+\sigma\varphi$ can be in the form of
	$$b+\sigma\varphi=\tilde{B}(x)+\hat{B}(x)$$
	for  $\tilde{B}$ bounded and measurable, and $\hat{B}$ Lipschitz continuous and
	at most of linear growth in $x.$ However, their results require that the diffusion term is $\sigma(x)\equiv1.$	 
\end{remark}

\section{Sensitivities of the drift and diffusion terms}
\label{sec:sen_drift_diffu}

We now investigate the sensitivities of the drift and diffusion terms of the underlying process.
Consider the following perturbed functions.
The variable $\epsilon$ can be understood as a perturbation parameter.  
\begin{Aassume}\label{assume:pert_defi}
	Let $b_\epsilon(x),$ $\sigma_\epsilon(x),$ $r_\epsilon(x),$ $f_\epsilon(x)$ be functions 
	of two variables $(\epsilon,x)$ on $I\times\mathbb{R}^d$ for a neighborhood $I$ of $0$ such that for each $x$ they are continuously differentiable in $\epsilon$ on $I$ and 
	$b_0(x)=b(x),$ $\sigma_0(x)=\sigma(x),$
	$r_0(x)=r(x),$ $f_0(x)=f(x).$ Let $\xi_\epsilon$ be a continuously differentiable function of variable $\epsilon$ on $I$ and $\xi_0=\xi.$
\end{Aassume}

\begin{Aassume}\label{assume:perturbed}
	For each $\epsilon\in I,$ the quadruple of functions $(b_\epsilon,\sigma_\epsilon,r_\epsilon,f_\epsilon)$ and a real number $\xi_\epsilon$ satisfy Assumptions \ref{assume:SDE} - \ref{assume:f} and \ref{assume:HS} - \ref{assume:ergodic}.
\end{Aassume}
\noindent From these assumptions,
the notations
$$X^\epsilon,\,\mathcal{P}^\epsilon,\,\mathcal{L}^\epsilon,\,M^\epsilon,\,\mathbb{Q}^\epsilon,\,(\lambda_\epsilon,\phi_\epsilon),\,\varphi_\epsilon,\,\nu_\epsilon$$
are self-explanatory.
For example, the SDE 
\begin{equation}\label{eqn:SDE_perturbed}
\begin{aligned}
&dX_{t}^{\epsilon}=b_\epsilon(X_{t}^{\epsilon})\,dt+\sigma_\epsilon (X_{t}^{\epsilon})\,dB_{t}\;,\,X_0^\epsilon=\xi_\epsilon
\end{aligned}
\end{equation}
has a unique strong solution $X^\epsilon$, and the solution is non-explosive. 
It is noteworthy that Assumption \ref{assume:perturbed} guarantees that the quadruple $(b,\sigma,r,f)$ and $\xi$ also satisfy Assumptions \ref{assume:SDE} - \ref{assume:f} and \ref{assume:HS} - \ref{assume:ergodic} because $0$ is in $I.$ 

We are interested in the perturbed quantity 
\begin{equation} \label{eqn:p_T_eps}
p_{T}^{\epsilon}:=\mathbb{E}_{\xi_\epsilon}^{\mathbb{P}}
[e^{-\int_{0}^{T}r_\epsilon(X_{s}^{\epsilon})ds}
f_\epsilon(X_{T}^{\epsilon})]
\end{equation}
and the long-term behavior of its sensitivity 
$\left.\frac{\partial }{\partial \epsilon}\right|_{\epsilon=0}p_{T}^{\epsilon}.$ 
The sensitivities with respect to the drift term and the diffusion term are called {\em rho} and {\em vega}, respectively. More precisely, 
the rho value (respectively, the vega value) is defined as 
$\left.\frac{\partial }{\partial \epsilon}\right|_{\epsilon=0}p_{T}^{\epsilon}$   
when the quadruple of perturbed functions is $(b_\epsilon,\sigma,r,f)$ (respectively, $(b,\sigma_\epsilon,r,f)$) and when the initial value $\xi$
is not perturbed.

Using Eq.\eqref{eqn:decomposition}, the expectation can be expressed in a relatively more manageable manner as 
\begin{equation}\label{eqn:perturbed_quant}
\begin{aligned}
p_{T}^{\epsilon}
&=e^{-\lambda_\epsilon T}\phi_{\epsilon}(\xi_\epsilon)\,\mathbb{E}_{\xi_\epsilon}^{\mathbb{Q}^{\epsilon}}
[(f_\epsilon/\phi_{\epsilon})(X_{T}^{\epsilon})]\,.  
\end{aligned}
\end{equation}
We now apply the chain rule to this equality.

\begin{theorem} \label{thm:total_chain}
	Let $(b_\epsilon,\sigma_\epsilon,r_\epsilon,f_\epsilon)$ and $\xi_\epsilon$  be a quadruple of functions and an initial value, respectively, satisfying Assumptions \ref{assume:pert_defi} - \ref{assume:perturbed}. 
	Suppose that the following conditions hold.
	\begin{itemize}
		\item[(i)] $\lambda_\epsilon$ and 
		$\phi_{\epsilon}(\xi_\epsilon)$
		are continuously differentiable in $\epsilon$ on $I.$
		\item[(ii)] As a function of two variables $(\eta,\epsilon)\in I^2,$ the partial derivative
		$\frac{\partial}{\partial\eta}\mathbb{E}_{\xi_\eta}^{\mathbb{Q}^\epsilon}
		[(f_\eta/\phi_\eta)(X_T^\epsilon)]$ exists and 
		is continuous on $I^2,$ and moreover, 
		\begin{equation}
		\label{eqn:condi_2_conv_0}
		\lim_{T\to\infty}\frac{1}{T} \frac{\partial}{\partial \eta}\;\Big|_{\eta=0}\mathbb{E}_{\xi_\eta}^{\mathbb{Q}}
		[(f_\eta/\phi_\eta)(X_T)]=0\,.
		\end{equation}
		\item[(iii)] As a function of two variables $(\eta,\epsilon)\in I^2,$ the partial derivative	$\frac{\partial}{\partial\epsilon}\mathbb{E}_{\xi_\eta}^{\mathbb{Q}^\epsilon}
		[(f_\eta/\phi_\eta)(X_T^\epsilon)]$ exists and 
		is continuous on $I^2,$ and moreover,
		$$\lim_{T\to\infty}\frac{1}{T}\frac{\partial}{\partial \epsilon}\;\Big|_{\epsilon =0}\mathbb{E}_\xi
		^{\mathbb{Q_{\epsilon}}}
		[(f/\phi)(X_T^\epsilon)]=0\,.$$ 
	\end{itemize}
	\noindent Then, the perturbed quantity $\ln p_{T}^{\epsilon}$ is differentiable at $\epsilon=0$ and
	\begin{equation}\label{eqn:differ_rho}
	\begin{aligned}
	&\frac{1}{T}\frac{\partial}{\partial \epsilon}\;\Big|_{\epsilon =0}\ln p_T^\epsilon=\frac{\left.\frac{\partial}{\partial \epsilon}\;\right|_{\epsilon =0}p_{T}^{\epsilon}}
	{T p_{T}}\\
	=&-\frac{\partial }{\partial\epsilon}\Big|_{\epsilon=0}\lambda_\epsilon +\frac{\left.\frac{\partial}{\partial \epsilon}\;\right|_{\epsilon =0}\phi_{\epsilon}(\xi_\epsilon)}{T\phi(\xi)}
	+\frac{\left.\frac{\partial}{\partial \epsilon}\;\right|_{\epsilon =0}\mathbb{E}_{\xi_\epsilon}^{\mathbb{Q}}
		[(f_\epsilon/\phi_\epsilon)(X_T)]}
	{T\, \mathbb{E}_\xi^{\mathbb{Q}}
		[(f/\phi)(X_T)]}
	+\frac{\left.\frac{\partial}{\partial \epsilon}\;\right|_{\epsilon =0}\mathbb{E}_\xi
		^{\mathbb{Q_{\epsilon}}}
		[(f/\phi)(X_T^\epsilon)]}
	{T\,\mathbb{E}_\xi^{\mathbb{Q}}
		[(f/\phi)(X_T)]}\; .
	\end{aligned}
	\end{equation}
	Furthermore, 
	\begin{equation}\label{eqn:final_eqn}
	\lim_{T\rightarrow\infty}\frac{1}{T}\frac{\partial}{\partial\epsilon}\Big|_{\epsilon=0}\ln p_T^\epsilon
	=  -\frac{\partial }{\partial\epsilon}\Big|_{\epsilon=0}\lambda_\epsilon\;. 
	\end{equation}
\end{theorem}

\begin{proof}
	Regard $p_T^\epsilon$ as a function of four variables $(\epsilon_1,\epsilon_2,\epsilon_3,\epsilon_4)$ on $I^4$ defined by
	\begin{equation}\label{eqn:4_variable_p}
	P_T(\epsilon_1,\epsilon_2,\epsilon_3,\epsilon_4):=
	e^{-\lambda_{\epsilon_1}T}\phi_{\epsilon_2}(\xi_{\epsilon_2}) \,\mathbb{E}_{\xi_{\epsilon_3}}^{\mathbb{Q}^{\epsilon_4}}[(f_{\epsilon_3}/\phi_{\epsilon_3})(X_{T}^{\epsilon_4})] 
	\end{equation}
	so that $p_T^\epsilon=P_T(\epsilon,\epsilon,\epsilon,\epsilon).$
	The chain rule gives 
	the differentiability of $\ln p_T^\epsilon$ and the equality in Eq.\eqref{eqn:differ_rho}.
	Bearing in mind that $\mathbb{E}_\xi^{\mathbb{Q}}[(f/\phi)(X_T)]$ converges to a finite positive constant as $T\to\infty$ by Assumption \ref{assume:ergodic},
	conditions (i) - (iii)  and Eq.\eqref{eqn:differ_rho} induce   Eq.\eqref{eqn:final_eqn}.
\end{proof}

\begin{remark}
	In the above theorem, we observe that
	conditions (i) - (iii) control  the four terms in Eq.\eqref{eqn:differ_rho}. Condition (i) controls the first and the second terms. 
	Conditions (ii) and (iii) control  the third term and the last term, respectively.	 
\end{remark}

\begin{remark}
	Conditions (ii) and (iii) in Theorem \ref{thm:total_chain} guarantee that the expectation $\mathbb{E}_{\xi_\eta}^{\mathbb{Q}^\epsilon}
	[(f_\eta/\phi_\eta)(X_T^\epsilon)]$ is continuously differentiable in $(\eta,\epsilon)$ on $I^2.$ Thus, we can apply the chain rule. 
\end{remark}

We now shift our attention to 
conditions (i) - (ii) in Theorem \ref{thm:total_chain}.
In condition (i), two functions $\lambda_\epsilon$ and $\phi_{\epsilon}(\xi_\epsilon)$ are continuously differentiable in $\epsilon$ on $I$ for many financially meaningful cases.
In condition (ii), the partial derivative
$\frac{\partial}{\partial\eta}\mathbb{E}_{\xi_\eta}^{\mathbb{Q}^\epsilon}[(f_\eta/\phi_\eta)(X_T^\epsilon)]$ 
is not easy to evaluate in general.
However, if the initial value $\xi$ is not perturbed (as in  the cases of the rho and the vega), then  this partial derivative
$\frac{\partial}{\partial\eta}\mathbb{E}_{\xi}^{\mathbb{Q}^\epsilon}[(f_\eta/\phi_\eta)(X_T^\epsilon)]$ 
can be evaluated in the ordinary manner by interchanging the derivative and the integration (Theorem \ref{thm:payoff}).
Conditions (i) and (ii) can be checked case by case, so we do not go into further detail here. 

\begin{Aassume}\label{assume:condi1_2}
	Conditions (i) and (ii) in Theorem \ref{thm:total_chain} hold.
\end{Aassume}

One of the main contributions of this article is to study condition (iii) in Theorem \ref{thm:total_chain}, which  controls the last term in Eq.\eqref{eqn:differ_rho}. 
Only the last term is related to  the perturbation of the underlying process. 
We investigate sufficient conditions for (iii) to achieve the long-term asymptotic behavior specified by Eq.\eqref{eqn:final_eqn}.
The differentiability and convergence to zero 
with respect to perturbations of the drift $b_\epsilon(\cdot)$ and the volatility $\sigma_\epsilon(\cdot)$ 
are not trivial. Their sensitivities are discussed in Sections \ref{sec:rho} and \ref{sec:vega}, respectively.

\begin{remark}
	If the $\mathbb{Q}^\epsilon$-distribution of $X_T^\epsilon$ is known, then the explicit formula for the sensitivity $\left.\frac{\partial }{\partial \epsilon}\right|_{\epsilon=0}\ln p_{T}^{\epsilon}$
	can be derived from Eq.\eqref{eqn:differ_rho}.
	However, the expression is complicated.
\end{remark}

\subsection{Rho}
\label{sec:rho}

This section conducts a sensitivity analysis with respect to a perturbation of the drift term. Let $(b_\epsilon,\sigma,r_\epsilon,f_\epsilon)$ and $\xi$ be a quadruple  of functions and an initial value, respectively, satisfying Assumptions \ref{assume:pert_defi} - \ref{assume:perturbed}. (The diffusion matrix $\sigma$ and the initial value $\xi$ are not perturbed.) 
The evaluation of rho is covered by this   perturbation form. We emphasize that 
the notations 
$$X^\epsilon,\,\mathcal{P}^\epsilon,\,\mathcal{L}^\epsilon,\,M^\epsilon,\,\mathbb{Q}^\epsilon,\,(\lambda_\epsilon,\phi_\epsilon),\,\varphi_\epsilon,\,\nu_\epsilon$$
are straightforward.
The perturbed process $X^\epsilon$ is given by Eq.\eqref{eqn:SDE_perturbed}, 
\begin{equation}\label{eqn:rho_SDE}
dX_{t}^{\epsilon}=b_\epsilon(X_{t}^{\epsilon})\,dt+\sigma  (X_{t}^{\epsilon})\,dB_{t}\,,\;X_0^\epsilon=\xi. 
\end{equation}
Define $k_\epsilon:=\sigma^{-1}b_\epsilon+\varphi_\epsilon$ and $k:=k_0.$ Here, $\sigma^{-1}$ is the inverse matrix of $\sigma.$ 
Assume that both $\phi_\epsilon(x)$ and
$\nabla\phi_\epsilon(x)$ (thus, $k_\epsilon(x)$) are  continuously differentiable in $\epsilon$  on $I$ for each $x.$  
We define
\begin{equation}\label{eqn:bar_k}
\begin{aligned}
&\overline{k}_\epsilon(x):=\frac{\partial}{\partial\epsilon}\Big|_{\epsilon=0}k_\epsilon(x)\,,\\
&\overline{k}(x):=\overline{k}_0(x)\,.
\end{aligned}
\end{equation} 

Given the perturbation defined above, the main purpose of this section is to find a sufficient condition for (iii) in Theorem \ref{thm:total_chain}.
We will prove that
\begin{equation}\label{eqn:Fournie_rho_derivative}
\frac{\partial}{\partial \epsilon}\;\Big|_{\epsilon =0}\mathbb{E}_\xi^{\mathbb{Q_{\epsilon}}}[(f/\phi)(X_T^\epsilon)]=\mathbb{E}_\xi^{\mathbb{Q}}
\Big[(f/\phi)(X_T)\int_{0}^{T}\overline{k}(X_{s})\, dW_{s}\Big]\,,
\end{equation}
which is also stated in Proposition 3.1 \cite{fournie1999applications}.
However,  this proposition is not useful for many financial models because of the strict assumptions.
In their work,  
the perturbation is linear in the form $b_\epsilon=b+\epsilon\overline{b}$,
and the function $\overline{b}$ is bounded. In addition, the diffusion matrix $\sigma$ satisfies the uniform ellipticity condition, and the payoff function satisfies $\mathbb{E}_\xi^{\mathbb{Q}}[(f/\phi)^2(\cdot)]<\infty.$
We generalize their result in Proposition \ref{prop:rho} in Appendix \ref{app:pf_rho_expo_condi}.
Our generalization does not require the linear perturbation, the boundedness condition on $\overline{b},$ or the uniform ellipticity condition on $\sigma.$
Many financial models, including the examples in the present paper, satisfy the generalized conditions but not the original conditions assumed in \cite{fournie1999applications}.

We now rigorously state a sufficient condition for (iii) in Theorem \ref{thm:total_chain}.
See Appendix \ref{app:pf_rho_expo_condi} for the proof of the following theorem. 
\begin{theorem} \label{thm:rho_expo_condi} 
	Let $(b_\epsilon,\sigma,r_\epsilon,f_\epsilon)$ and $\xi$ be a quadruple of functions and an initial value, respectively, satisfying Assumptions \ref{assume:pert_defi} - \ref{assume:perturbed}. 
Assume that both $\phi_\epsilon(x)$ and
$\nabla\phi_\epsilon(x)$ (thus, $k_\epsilon(x)$) are  continuously differentiable in $\epsilon$  on $I$ for each $x$ and that there exist functions $g,\psi:\mathbb{R}^d\rightarrow\mathbb{R}$   such that	 
	\begin{align}
	&\left|\frac{\partial k_\epsilon(x)}{\partial\epsilon}\right|\leq g(x)\;, \label{eqn:g_bound}\\
	&\left|f_\epsilon(x)/\phi_\epsilon(x)\right|\leq  \psi(x) \label{eqn:psi_bound}
	\end{align}  
	for $(\epsilon,x)$ in $I\times\mathbb{R}^d.$ 
	Suppose that the following condition holds. 
	\begin{itemize} 
		\item[(i)] There exist positive constants $a,$ $c$ and $\epsilon_0$ such that for all $T>0$
		$$\mathbb{E}_\xi^{\mathbb{Q}}[e^{\epsilon_0\int_0^Tg^2(X_s)\,ds}]\leq c\,e^{aT}\,.$$ 	
	\end{itemize}
	In addition, suppose there exist positive constants $p\geq 2$ and $q$ with $1/p+1/q=1$ satisfying the following conditions.
	\begin{itemize} 
		\item[(ii)] For each $T>0,$ there is a positive number $\epsilon_1$ such that $\mathbb{E}_\xi^{\mathbb{Q}}[\int_0^Tg^{p+\epsilon_1}(X_t)\,dt]$ 
		is finite.
		\item[(iii)] As a function of $T,$ the expectation $\mathbb{E}_\xi^{\mathbb{Q}}[\psi^q(X_T)]$ is uniformly bounded on $[0,\infty).$
	\end{itemize}
	Then, 
	the partial derivative	$\frac{\partial}{\partial\epsilon}\mathbb{E}_{\xi}^{\mathbb{Q}^\epsilon}
	[(f_\eta/\phi_\eta)(X_T^\epsilon)]$ exists and 
	is continuous in $(\eta,\epsilon)$ on $I^2.$ Moreover, 
	\begin{equation}\label{eqn:rho_condi_3}
	\lim_{T\to\infty}\frac{1}{T}\frac{\partial}{\partial \epsilon}\,\Big|_{\epsilon =0}\mathbb{E}_\xi^{\mathbb{Q_{\epsilon}}}
	[(f/\phi)(X_{T}^{\epsilon})]= 0\,.
	\end{equation} 
\end{theorem}

The implications of these assumptions are as follows. First, (i) and (ii) concern the growth rate of the function $g(x),$ which controls the growth rate of the perturbation $\frac{\partial }{\partial\epsilon}k_\epsilon(x)$:
(i) implies that the expectation of $e^{\epsilon_0\int_0^Tg^2(X_s)\,ds}$ can increase  exponentially as $T\to\infty,$  and
(ii) means that $\int_0^Tg^{p+\epsilon_1}(X_t)\,dt$ has a finite expectation.
Second, (iii) controls the function $f$: because Eq\eqref{eqn:psi_bound} holds, the expectation of $(f_\epsilon/\phi_\epsilon)^q(X_T)$ is uniformly bounded in $(\epsilon,T)$ on $I\times [0,\infty).$ 
Third, the condition $1/p+1/q=1$ means that
if $g$ satisfies a stronger condition (if (ii) holds for larger $p$), then a weaker condition on $\psi$ is required ($q$ can be smaller in (iii)). 
Finally, this theorem encompasses Proposition 3.1 in \cite{fournie1999applications}.

\begin{remark} It is noteworthy that Assumption \ref{assume:ergodic} can be induced by 
	Assumption \ref{assume:inv} together with 
	(iii) in Theorem \ref{thm:rho_expo_condi}. The proof follows. Let $\hat{f}:=f/\phi\geq0.$ 
	If $\hat{f}$ is bounded, then Assumption \ref{assume:ergodic} is trivial because the process $X$ has an invariant distribution under $\mathbb{Q}.$ If $\hat{f}$ is unbounded, then it suffices to prove that
	$\lim_{L\to\infty} \mathbb{E}_\xi^\mathbb{Q}[\hat{f}(X_T)\mathbb{I}_{\hat{f}(X_T)\geq L}]=0$ uniformly in $T.$  
	This is obtained from  
	\begin{equation*}
	\begin{aligned}
	\mathbb{E}_\xi^\mathbb{Q}[\hat{f}(X_T)\mathbb{I}_{\hat{f}(X_T)\geq L}]\leq  \mathbb{E}_\xi^\mathbb{Q}\Big[\hat{f}(X_T)\,\frac{\hat{f}^{q-1}(X_T)}{L^{q-1}}\Big] 
	\leq \frac{\mathbb{E}_\xi^\mathbb{Q}[\hat{f}^{q}(X_T)]}{L^{q-1}} 
	\end{aligned}
	\end{equation*} 
	because the expectation $\mathbb{E}_\xi^\mathbb{Q}[\hat{f}^{q}(X_T)]$ is uniformly bounded in $T$ on $[0,\infty).$ 
\end{remark}

We now consider a variation of Theorem \ref{thm:rho_expo_condi}.
Conditions (i) and (ii) in Theorem \ref{thm:rho_expo_condi} are replaced by the exponential condition (i) in Theorem \ref{thm:expo_rho}.
See Appendix \ref{app:pf_rho_thm} for the proof.
Theorem \ref{thm:expo_rho} is used in Appendix \ref{app:sen_B_QTSM} to estimate the rho of a quadratic model.
\begin{theorem}\label{thm:expo_rho}
	Let $(b_\epsilon,\sigma,r_\epsilon,f_\epsilon)$ and $\xi$ be a quadruple of functions and an initial value, respectively, satisfying Assumptions \ref{assume:pert_defi} - \ref{assume:perturbed}. 
Assume that both $\phi_\epsilon(x)$ and
$\nabla\phi_\epsilon(x)$ (thus, $k_\epsilon(x)$) are  continuously differentiable in $\epsilon$  on $I$ for each $x$ and that there exist functions $g,\psi:\mathbb{R}^d\rightarrow\mathbb{R}$ satisfying Eq.\eqref{eqn:g_bound} and Eq.\eqref{eqn:psi_bound}.
	Suppose that the following conditions hold.
	\begin{itemize} 
		\item[(i)]  There exists a positive constant $\epsilon_0$ such that
		$\mathbb{E}_\xi^{\mathbb{Q}}[e^{\epsilon_0\,g^2(X_T)}]$ is uniformly bounded in $T$ on $[0,\infty).$
		\item[(ii)] There exists a constant $q>1$ such that 
		$\mathbb{E}_\xi^{\mathbb{Q}}[\psi^q(X_T)]$ is uniformly bounded in $T$ on $[0,\infty).$
	\end{itemize}	
	Then,
	the partial derivative	$\frac{\partial}{\partial\epsilon}\mathbb{E}_{\xi}^{\mathbb{Q}^\epsilon}
	[(f_\eta/\phi_\eta)(X_T^\epsilon)]$ exists and 
	is continuous in $(\eta,\epsilon)$ on $I^2.$ Moreover,  Eq.\eqref{eqn:rho_condi_3} holds.
\end{theorem}

Because the two theorems above guarantee condition (iii) in Theorem \ref{thm:total_chain}, we obtain the following corollary.
\begin{corollary}\label{cor:rho}
	Let  $(b_\epsilon,\sigma,r_\epsilon,f_\epsilon)$ and $\xi$  be a quadruple of functions and an initial value, respectively, satisfying Assumptions	
	\ref{assume:pert_defi}, \ref{assume:perturbed} and \ref{assume:condi1_2}. Then, we obtain Eq.\eqref{eqn:final_eqn}
	$$\lim_{T\rightarrow\infty}\frac{1}{T}\frac{\partial}{\partial\epsilon}\Big|_{\epsilon=0}\ln p_T^\epsilon
	=  -\frac{\partial }{\partial\epsilon}\Big|_{\epsilon=0}\lambda_\epsilon$$
	if the hypothesis of  either Theorem \ref{thm:rho_expo_condi} or \ref{thm:expo_rho} is satisfied. 
\end{corollary}

\subsection{Vega}
\label{sec:vega}
\subsubsection{The Lamperti transformation for univariate processes}
\label{sec:Lamperti_trans}

This section conducts a sensitivity analysis with 
a univariate underlying process.
Let $(b_\epsilon,\sigma_\epsilon,r_\epsilon,f_\epsilon)$ and $\xi$ be a quadruple of functions and an initial value, respectively, satisfying Assumptions \ref{assume:pert_defi} - \ref{assume:perturbed}. The functions 
$b_\epsilon,\sigma_\epsilon,r_\epsilon,f_\epsilon$
are univariate scalar functions, and $\xi$ is a scalar.
The initial value $\xi$ is not perturbed.
In this section, we assume that the function $\sigma_\epsilon(x)$ is  twice continuously differentiable in $(\epsilon,x)$ on $I\times \mathbb{R}^d.$   
The notations
\begin{equation}\label{eqn:obj}
X^\epsilon,\,\mathcal{P}^\epsilon,\,\mathcal{L}^\epsilon,\,M^\epsilon,\,\mathbb{Q}^\epsilon,\,(\lambda_\epsilon,\phi_\epsilon),\,\varphi_\epsilon,\,\nu_\epsilon
\end{equation}
are self-explanatory.
The perturbed process $X^\epsilon$ is given by \begin{equation} \label{eqn:X_ep_Lamp}
dX_{t}^{\epsilon}=b_\epsilon(X_{t}^{\epsilon})\,dt+\sigma_\epsilon (X_{t}^{\epsilon})\,dB_{t}\,,\,X_0^\epsilon=\xi \,.
\end{equation}
The evaluation of vega is covered by this   perturbation form.

The main idea of this section is to utilize the 
Lamperti transformation.
Fix any real number $c$ and define 
\begin{equation}\label{eqn:Lamperti}
\begin{aligned}
u_\epsilon(x):=\int_{c}^{x}\frac{1}{\sigma_\epsilon(y)}\,dy\,,\;U_t^\epsilon:=u_\epsilon(X_t^\epsilon)\,,\;
\zeta_\epsilon:=\int_{c}^{\xi}\frac{1}{\sigma_\epsilon(y)}\,dy\,,\;\zeta:=\zeta_0\,.
\end{aligned}
\end{equation}
The process $U^\epsilon=(U_t^\epsilon)_{t\geq0}$ satisfies
\begin{equation}\label{eqn:Lamperti_transformed_SDE}
dU_t^\epsilon=\delta_\epsilon(U_t^\epsilon)\,dt
+dB_t\;,\;U_0^\epsilon=\zeta_\epsilon\;,
\end{equation}
where $\delta_\epsilon:=(\frac{ b_\epsilon}{\sigma_\epsilon}-\frac{1}{2}\sigma_\epsilon')\circ v_\epsilon.$ 
Here, the notation $\circ$ represents a composition of functions, and 
the function $v_\epsilon$ is the inverse function of $u_\epsilon.$ It is noteworthy that $v_\epsilon$ is continuously  differentiable in $\epsilon$ on $I$ since $\sigma_\epsilon(x)$ is twice continuously differentiable in two variables $(\epsilon,x)$ on $I\times\mathbb{R}^d.$
This mapping
$X^\epsilon\mapsto U^\epsilon$
is called the Lamperti transformation.

The Lamperti transformation is useful 
because it converts
the perturbed diffusion term into the constant $1,$ and transfers the perturbation  to the drift term and the initial value.
As shown in Eq.\eqref{eqn:Lamperti_transformed_SDE}, the diffusion term of the process $U^\epsilon$ is not perturbed.
Thus, we can utilize the results in Sections  \ref{sec:delta} and \ref{sec:rho}. 
Define $R_\epsilon:=r_\epsilon\circ v_\epsilon$ and  $F_\epsilon:=f_\epsilon\circ v_\epsilon.$ The perturbed expectation $p_{T}^{\epsilon}$ is expressed as  
\begin{equation}\label{eqn:p_T_Lamp}
\begin{aligned}
p_{T}^{\epsilon}
=&\mathbb{E}_{X_0^\epsilon
	=\xi}^{\mathbb{P}}[e^{-\int_{0}^{T}r_\epsilon(X_{s}^{\epsilon})ds}\,f_\epsilon(X_{T}^{\epsilon})]
=\mathbb{E}_{U_0^\epsilon=\zeta_\epsilon}^{\mathbb{P}}[e^{-\int_{0}^{T}R_\epsilon(U_{s}^{\epsilon})ds}\,F_\epsilon(U_{T}^{\epsilon})]\,.
\end{aligned}
\end{equation}

Through the Lamperti transformation defined above, we obtain the quadruple of functions
$(\delta_\epsilon,1,R_\epsilon,F_\epsilon)$ and the initial value $\zeta_\epsilon.$ 
Here,  $1$ represents a constant function that is identically equal to one.
The following proposition concerns  Assumptions \ref{assume:pert_defi} - \ref{assume:perturbed}.
See Appendix \ref{app:Lamp_equiv} for the proof.

\begin{proposition}\label{prop:Lamp_equiv} 
	Consider a quadruple of functions $(b_\epsilon,\sigma_\epsilon,r_\epsilon,f_\epsilon
	)$ and a real number $\xi.$ Assume that the function $\sigma_\epsilon(x)$ is twice  continuously  differentiable in $(\epsilon,x)$ on $I\times\mathbb{R}^d.$
	The quadruple $(b_\epsilon,\sigma_\epsilon,r_\epsilon,f_\epsilon
	)$ and $\xi$ satisfy Assumptions \ref{assume:pert_defi} - \ref{assume:perturbed} if and only if the 
	quadruple $(\delta_\epsilon,1,R_\epsilon,F_\epsilon)$ and $\zeta_\epsilon,$ defined by the Lamperti transformation, satisfy Assumptions \ref{assume:pert_defi} - \ref{assume:perturbed}. In this case, the recurrent eigenvalue and the recurrent eigen-measure are invariant under the Lamperti transformation.	
\end{proposition}

Assumptions \ref{assume:pert_defi} - \ref{assume:perturbed} and 
Proposition \ref{prop:Lamp_equiv} say  that the 
quadruple $(\delta_\epsilon,1,R_\epsilon,F_\epsilon)$ and $\zeta_\epsilon$ also satisfy Assumptions \ref{assume:pert_defi} - \ref{assume:perturbed}.
The notations 
$$U^\epsilon,\,\mathcal{P}^{\textnormal{Lamp}\,\epsilon},\,\mathcal{L}^{\textnormal{Lamp}\,\epsilon},\,M^{\textnormal{Lamp}\,\epsilon},\,\mathbb{Q}^\epsilon,\,(\lambda_\epsilon,\Phi_\epsilon),\,\varphi_\epsilon^{\textnormal{Lamp}},\,\nu_\epsilon^{\textnormal{Lamp}}$$
are straightforward. These objects will be used in Appendix \ref{app:Lamp_equiv}.
It is noteworthy that the recurrent eigenfunction satisfies $\Phi_\epsilon=\phi_\epsilon\circ v_\epsilon.$ 
Using the Hansen--Scheinkman decomposition, the expectation $p_T^\epsilon$ in Eq.\eqref{eqn:p_T_Lamp} is written as  
\begin{equation} \label{eqn:HS_decomp_Lamp}
\begin{aligned}
p_{T}^{\epsilon}
&=\Phi_{\epsilon}(\zeta_\epsilon)\,e^{-\lambda_\epsilon T} \mathbb{E}_{U_0^\epsilon=\zeta_\epsilon}^{\mathbb{Q_{\epsilon}}}[(F_\epsilon/\Phi_{\epsilon})(U_{T}^{\epsilon})]\,. 
\end{aligned}
\end{equation}

The main purpose of Section \ref{sec:Lamperti_trans} is to introduce Theorem \ref{thm:Lamp}.
This theorem can be easily proven by using Theorem \ref{thm:total_chain} and Corollary \ref{cor:rho}, so we omit the proof.
In Theorem \ref{thm:Lamp}, we emphasize that
the initial value $\zeta$ is not perturbed in condition (ii), whereas in condition (iii)  the initial value $\zeta_\eta$ is perturbed.
\begin{theorem}\label{thm:Lamp}
	Let $(b_\epsilon,\sigma_\epsilon,r_\epsilon,f_\epsilon
	)$ and $\xi$ be a quadruple of functions and a real number, respectively, satisfying Assumptions \ref{assume:pert_defi} - \ref{assume:perturbed}. Assume that the function $\sigma_\epsilon(x)$ is twice continuously differentiable in $(\epsilon,x)$ on $I\times\mathbb{R}^d.$ Define a quadruple $(\delta_\epsilon,1,R_\epsilon,F_\epsilon)$ and $\zeta_\epsilon$ by the Lamperti transformation (thus,  satisfying Assumptions \ref{assume:pert_defi} - \ref{assume:perturbed} by Proposition \ref{prop:Lamp_equiv}).
	Suppose that the following conditions hold.
	\begin{itemize}
		\item[(i)] The quadruple $(\delta_\epsilon,1,R_\epsilon,F_\epsilon)$ and $\zeta_\epsilon$ 
		satisfy Assumption  \ref{assume:condi1_2}. 
		\item[(ii)] The quadruple  $(\delta_\epsilon,1,R_\epsilon,F_\epsilon)$  and $\zeta$ 
		satisfy  the hypothesis of either Theorem \ref{thm:rho_expo_condi} or \ref{thm:expo_rho}. (Thus, the partial derivative $\frac{\partial}{\partial\epsilon}\mathbb{E}_{U_0^\epsilon=\zeta}^{\mathbb{Q}^\epsilon}
		[(f_\eta/\phi_\eta)(U_T^\epsilon)]$ exists and is continuous in $(\eta,\epsilon)$ on $I^2$.)	
		\item[(iii)] The partial derivative	$\frac{\partial}{\partial\epsilon}\mathbb{E}_{U_0^\epsilon=\zeta_\eta}^{\mathbb{Q}^\epsilon}
		[(f_\eta/\phi_\eta)(U_T^\epsilon)]$ is continuous in $(\eta,\epsilon)$ on $I^2.$				
	\end{itemize}
	Then, 
	$p_T^\epsilon$ is differentiable at $\epsilon=0$ and the long-term behavior of the partial derivative is 
	$$\lim_{T\rightarrow\infty}\frac{1}{T}\frac{\partial}{\partial\epsilon}\Big|_{\epsilon=0}\ln p_T^\epsilon =  - \frac{\partial }{\partial\epsilon}\Big|_{\epsilon=0}\lambda_\epsilon\,.$$
\end{theorem}

In the Lamperti transformation, one can choose any real number $c$ in Eq.\eqref{eqn:Lamperti} as a reference point. As a special case, if we put $c=\xi,$ then 
$\zeta_\epsilon=0$ for all $\epsilon\in I.$  Since the initial value $U_0^\epsilon=\zeta_\epsilon=0$ is not perturbed, condition (iii) in Theorem \ref{thm:Lamp} is automatically guaranteed by (ii). 

\begin{corollary}   If $c=\xi$ in the Lamperti transformation in Eq.\eqref{eqn:Lamperti}, then condition (iii) in Theorem \ref{thm:Lamp} can be omitted.
\end{corollary}

\subsubsection{The Fournie et al. method with bounded-derivative coefficients}
\label{sec:vega_fournie}

This section presents how the method developed by \cite{fournie1999applications} can be applied to analyze the long-term vega value. 
Let $(b,\sigma_\epsilon,r_\epsilon,f_\epsilon)$ and $\xi$ be a quadruple of functions and an initial value satisfying Assumptions \ref{assume:pert_defi}, \ref{assume:perturbed} and \ref{assume:condi1_2} with the linear perturbation form $$\sigma_\epsilon:=\sigma+\epsilon\overline{\sigma}\,.$$
In this section, assume the hypothesis in \cite{fournie1999applications} holds, that is, the functions $b,$ $\sigma$ and $\overline{\sigma}$ are continuously differentiable with bounded derivatives.  
The perturbed process $X^\epsilon$ satisfies
\begin{equation}
\label{eqn:vega_Fournie}
dX_{t}^{\epsilon}=b(X_{t}^{\epsilon})\,dt+(\sigma +\epsilon\overline{\sigma})(X_{t}^{\epsilon})\,dW_{t}\,.
\end{equation}

We find a sufficient condition for
(iii) in Theorem \ref{thm:total_chain}. Condition (iii) in Theorem \ref{thm:total_chain} states that the partial derivative	$\frac{\partial}{\partial\epsilon}\mathbb{E}_{\xi}^{\mathbb{Q}^\epsilon}[(f_\eta/\phi_\eta)(X_T^\epsilon)]$ exists and 
is continuous in $(\eta,\epsilon)$ on $I^2,$ and moreover,
$$\lim_{T\to\infty}\frac{1}{T}\frac{\partial}{\partial \epsilon}\;\Big|_{\epsilon =0}\mathbb{E}_\xi
^{\mathbb{Q_{\epsilon}}}
[(f/\phi)(X_T^\epsilon)]=0\,.$$
Observe that
the $\epsilon$-perturbation of $\mathbb{Q}^\epsilon$ induces the drift perturbation and that the $\epsilon$-perturbation of $X^\epsilon$ induces the diffusion perturbation in Eq.\eqref{eqn:vega_Fournie}.
From Eq.\eqref{eqn:Q-dynamics}, the $\mathbb{Q}^\epsilon$-dynamics of $X^\epsilon$ is
\begin{equation*}
\begin{aligned}
dX_{t}^\epsilon
=(b+(\sigma+\epsilon\overline{\sigma})\varphi_\epsilon)(X_{t}^\epsilon)\, dt+(\sigma+\epsilon\overline{\sigma})(X_{t}^\epsilon)\, dW_{t}^\epsilon \,.
\end{aligned}
\end{equation*}
Motivated by this expression, we define a process $\hat{X}^{\rho,\nu}$ with two parameters $\rho$ and $\nu$ by
\begin{equation}\label{eqn:two_parameters}
d\hat{X}_{t}^{\rho,\nu}=(b+(\sigma+\rho\overline{\sigma})\varphi_\rho)(\hat{X}_{t}^{\rho,\nu})\, dt+(\sigma+\nu\overline{\sigma})(\hat{X}_{t}^{\rho,\nu})\, dW_{t} 
\end{equation}
where $W$ is the $\mathbb{Q}$-Brownian motion. The 
$\mathbb{Q}^\epsilon$-distribution of $X_T^\epsilon$ is equal to the $\mathbb{Q}$-distribution of $\hat{X}_T^{\epsilon,\epsilon},$ thus
$\mathbb{E}_\xi
^{\mathbb{Q_{\epsilon}}}
[(f_\eta/\phi_\eta)(X_T^\epsilon)]=\mathbb{E}_\xi
^{\mathbb{Q}}
[(f_\eta/\phi_\eta)(\hat{X}_T^{\epsilon,\epsilon})].$ Applying the chain rule,
it follows that for a given $\eta\in I$ 
\begin{align}\label{eqn:Fournie_chain}
\frac{\partial}{\partial \epsilon}\mathbb{E}_\xi
^{\mathbb{Q_{\epsilon}}}
[(f_\eta/\phi_\eta)(X_T^\epsilon)]
=\frac{\partial}{\partial \rho}\Big|_{\rho =\epsilon}\mathbb{E}_\xi
^{\mathbb{Q}}
[(f_\eta/\phi_\eta)(\hat{X}_T^{\rho,\epsilon})]
+ \frac{\partial}{\partial \nu}\Big|_{\nu=\epsilon}\mathbb{E}_\xi
^{\mathbb{Q}}
[(f_\eta/\phi_\eta)(\hat{X}_T^{\epsilon,\nu})]   
\end{align}
under the assumption that the two partial derivatives
\begin{equation}
\label{eq:vega_deriva}
\frac{\partial}{\partial \rho}\mathbb{E}_\xi
^{\mathbb{Q}}
[(f_\eta/\phi_\eta)(\hat{X}_T^{\rho,\nu})]
\quad\textnormal{and}\quad\frac{\partial}{\partial \nu}\mathbb{E}_\xi
^{\mathbb{Q}}
[(f_\eta/\phi_\eta)(\hat{X}_T^{\rho,\nu})]
\end{equation} 
exist and 
are continuous
in $(\rho,\nu)$
on $I^2.$ Thus, we obtain the following proposition, which can be directly proven 
using  Eq.\eqref{eqn:Fournie_chain}
and the chain rule.

\begin{proposition}
	Let $(b,\sigma_\epsilon,r_\epsilon,f_\epsilon)$ and $\xi$ be a quadruple of functions and an initial value satisfying Assumptions \ref{assume:pert_defi}, \ref{assume:perturbed}, \ref{assume:condi1_2}, and let $\sigma_\epsilon$ have the linear perturbation form $\sigma_\epsilon=\sigma+\epsilon\overline{\sigma}.$
	Condition (iii) in Theorem \ref{thm:total_chain} holds if the two partial derivatives
	in Eq.\eqref{eq:vega_deriva} exist and are continuous  in $(\eta,\rho,\nu)$ on $I^3$ and if
	\begin{align}
	&\lim_{T\to\infty}\frac{1}{T} \frac{\partial}{\partial \rho}\Big|_{\rho=0}\mathbb{E}_\xi
	^{\mathbb{Q}}
	[(f/\phi)(\hat{X}_T^{\rho,0})]=0\label{eqn:1st_vega}\,,\\
	&\lim_{T\to\infty}\frac{1}{T} \frac{\partial}{\partial \nu}\Big|_{\nu=0}\mathbb{E}_\xi
	^{\mathbb{Q}}
	[(f/\phi)(\hat{X}_T^{0,\nu})]=0\,.\label{eqn:2nd_vega}
	\end{align}
\end{proposition}

We now discuss our approach to the three assumptions of this proposition. In the first assumption, the continuous differentiability on $I^3$ of the two partial derivatives is not easy to check in general; however, if the $\mathbb{Q}$-density function of $\hat{X}_T^{\rho,\nu}$  is known, then it can easily be checked case by case,
as we will see in later examples. 
In the second assumption, Eq.\eqref{eqn:1st_vega}, only the drift is perturbed because $\nu=0$ in Eq.\eqref{eqn:two_parameters}. Therefore,
Eq.\eqref{eqn:1st_vega} can be checked by using the method presented in Section \ref{sec:rho}. We do not discuss the first two assumptions further here.

For the rest of this section, we shift our attention to the third assumption, Eq.\eqref{eqn:2nd_vega}, in which
only the diffusion term is perturbed because $\rho=0$ in Eq.\eqref{eqn:two_parameters}.  
Our purpose is to find a sufficient condition such that Eq.\eqref{eqn:2nd_vega} holds.
For convenience, we define $\hat{X}^\nu:=\hat{X}^{0,\nu}$ and 
$\hat{X}_{T}:=\hat{X}_{T}^{0}.$
The $\mathbb{Q}$-dynamics of $\hat{X}^\nu$ is
$$d\hat{X}_{t}^{\nu}=(b+\sigma\varphi)(\hat{X}_{t}^{\nu})\,dt+(\sigma+\nu\overline{\sigma})(\hat{X}_{t}^{\nu})\, dW_{t}\,,\;t\geq 0\,.$$
Suppose that $b+\sigma\varphi$ and $f/\phi$ are continuously differentiable with bounded derivatives.
Define a variation process $Z$ by
\begin{equation*}
\begin{aligned}
dZ_t=(b+\sigma\varphi)'(\hat{X}_t)Z_t\,dt+\overline{\sigma}(\hat{X}_t)\,dB_t+\sum_{i=1}^{d}\sigma_i'(\hat{X}_t)Z_t\,dB_{i,t}\,,\;Z_0=0_d\,.
\end{aligned}
\end{equation*}
Here, $\sigma_i$ is  the $i$-th column vector of $\sigma$, and $0_d$ is the $d$-dimensional zero column vector. 
Proposition 3.3 in \cite{fournie1999applications} says that $\mathbb{E}_\xi^{\mathbb{Q}}[(f/\phi)(\hat{X}_T^{\nu})]$ is differentiable in $\nu$ on $I$ and that
\begin{equation}\label{eqn:vaga_Z}
\frac{\partial}{\partial \nu}\Big|_{\nu=0}\mathbb{E}_\xi	^{\mathbb{Q}}
[(f/\phi)(\hat{X}_T^{\nu})]=\mathbb{E}_\xi^{\mathbb{Q}}[\nabla(f/\phi)(\hat{X}_{T})\, Z_T]\,.
\end{equation}
From these observations, we obtain the following theorem.
\begin{theorem}\label{thm:vega_Fournie_condi}
	Let $(b,\sigma_\epsilon,r_\epsilon,f_\epsilon)$ and $\xi$ be a quadruple of functions and an initial value satisfying Assumptions \ref{assume:pert_defi}, \ref{assume:perturbed} and \ref{assume:condi1_2},
	and let $\sigma_\epsilon$ have the linear perturbation form $\sigma_\epsilon=\sigma+\epsilon\overline{\sigma}.$
	Assume that the functions $b,$ $\sigma,$ $\overline{\sigma},$ $b+\sigma\varphi$ and $f/\phi$ are continuously differentiable with bounded derivatives. 
	Then, $\mathbb{E}_\xi^{\mathbb{Q}}[(f/\phi)(\hat{X}_T^{\nu})]$ is differentiable in  $\nu$ on $I.$ Moreover, if 
	$\frac{1}{T}\mathbb{E}_\xi^{\mathbb{Q}}[|Z_T|]\rightarrow0$ as $T\rightarrow\infty,$ then  Eq.\eqref{eqn:2nd_vega}	holds.
\end{theorem}
\noindent The proof is straightforward because 
$$\Big|\mathbb{E}_\xi^{\mathbb{Q}}[\nabla(f/\phi)(\hat{X}_{T})\, Z_T]\Big|\leq M\, \mathbb{E}_\xi^{\mathbb{Q}}[|Z_T|]$$
where the derivative of $f/\phi$ is bounded by a constant $M>0.$

\begin{remark}		
	In our analysis, there are two primary differences between the sensitivities of delta and rho/vega. 
	First, for delta, we explore  
	the zeroth-order growth rate 
	$$\lim_{T\rightarrow\infty}\nabla_\xi\ln p_T=\frac{\nabla_\xi\,\phi}{\phi(\xi)} \;, $$
	which is given in Theorem \ref{thm:delta}.	
	For rho/vega, we determine the first-order growth rate
	$$ \lim_{T\rightarrow\infty}\frac{1}{T}\frac{\partial}{\partial\epsilon}\Big|_{\epsilon=0}\ln p_T^\epsilon=-\frac{\partial }{\partial\epsilon}\Big|_{\epsilon=0}\lambda_\epsilon
	$$
	which is presented in Eq.\eqref{eqn:final_eqn}.
	Second, the long-term delta is determined in terms of the eigenfunction $\frac{\nabla_\xi\,\phi}{\phi(\xi)} .$	
	However, the long-term rho/vega is determined in terms of 
	the eigenvalue $-\left.\frac{\partial }{\partial\epsilon}\right|_{\epsilon=0}\lambda_\epsilon.$	
\end{remark}

\section{Examples of option prices}
\label{sec:ex_option_prices}

\subsection{The CIR model}	
\label{sec:CIR}

We conduct
a sensitivity analysis of option prices whose underlying process is the Cox-Ingersoll-Ross (CIR) short-rate model. Let $\mathbb{P}$ be a risk-neutral measure.
Define a quadruple of functions $(b,\sigma,r,f)$ and an initial value $\xi$ as follows.
Let
$$b(x)=\theta- ax\,,\;\sigma(x)=\sigma\sqrt{|x|}\,,\;r(x)=x$$
and let $f:\mathbb{R}\to\mathbb{R}$ be a 
non-negative, non-zero,  continuous function whose growth rate
is less than or equal to the polynomial growth rate. 
Fix a positive initial value $\xi>0.$ 
Here, the parameters $a$ and $\sigma$ are positive constants and 
$2\theta>\sigma^{2}$ so that both the original short-rate process and the perturbed process
stay strictly positive for small perturbations of the parameters.
The short interest rate is a solution of the SDE 
\begin{equation}\label{eqn:SDE_CIR}
dX_{t}=(\theta-aX_{t})\,dt + \sigma \sqrt{|X_{t}|} \,dB_{t}\,,\,X_0=\xi\,.
\end{equation} 
This SDE has a unique strong solution, as proven by Yamada-Watanabe (Proposition 2.13 on page 291 in \cite{shreve1991brownian}).
The option price whose payoff is $f(X_T)$ at maturity $T$ is given by Eq.\eqref{eqn:p_T},
\begin{equation}\label{eqn:CIR_p_T}
p_T=\mathbb{E}_\xi^{\mathbb{P}}[e^{-\int_0^T r(X_s)\,ds} f(X_T)]\,. 
\end{equation}
The sensitivities with respect to the parameters $\theta,$ $a,$ $\sigma,$   $\xi$ are   of interest to us. 

The parameters $\theta,$ $a,$ $\sigma,$ $\xi$ can also be regarded as perturbation parameters. For example, consider the perturbation of $\theta.$ The perturbed drift in Eq.\eqref{eqn:SDE_CIR} is 
$b_\epsilon(x)=(\theta+\epsilon)-ax$, while
the other functions $\sigma,$ $r,$ $f$ and the initial value $\xi$ are not perturbed.
Let $p_T^\epsilon$ be the expectation corresponding to this $\epsilon$-perturbation defined by Eq.\eqref{eqn:p_T_eps}.  
It is clear that
\begin{equation}\label{eqn:dtheta}
\frac{d}{d\theta}\ln p_T=\frac{d}{d\epsilon}\Big|_{\epsilon=0}\ln p_T^\epsilon\,,
\end{equation}
and thus, we may regard $\theta$ itself as a perturbation parameter.
The other parameters can also be understood to be
perturbation parameters through this approach.

The quadruple of functions
$(b,\sigma,r,f)$ and the initial value $\xi$ satisfy
Assumptions \ref{assume:pert_defi} - \ref{assume:perturbed}. The recurrent eigenpair is
$(\lambda,\phi(x)):=(\theta \kappa,e^{-\kappa x})$
where $\kappa:=\frac{\sqrt{a^{2}+2\sigma^{2}}-a}{\sigma^{2}}.$
Using this eigenpair $(\lambda,\phi),$ the sensitivities of the long-term option prices with respect to parameters $\theta,$ $a,$ $\sigma,$ $\xi$ are   
\begin{equation}\label{eqn:sen_CIR}
\begin{aligned}
&\lim_{T\rightarrow\infty}\frac{\partial}{\partial \xi}\ln p_T=\frac{\phi'(\xi)}{\phi(\xi)}
=-\frac{\sqrt{a^{2}+2\sigma^{2}}-a}{\sigma^{2}}\,,\\
&\lim_{T\rightarrow\infty}\frac{1}{T}\frac{\partial}{\partial\theta}\ln p_T=-\frac{\partial\lambda}{\partial\theta}= -\frac{\sqrt{a^{2}+2\sigma^{2}}-a}{\sigma^{2}}\,,\\
&\lim_{T\rightarrow\infty}\frac{1}{T}\frac{\partial}{\partial a}\ln p_T=-\frac{\partial\lambda}{\partial a}
=\frac{\theta(\sqrt{a^{2}+2\sigma^{2}}-a)}{\sigma^{2}\sqrt{a^{2}+2\sigma^{2}}}\,,\\
&\lim_{T\rightarrow\infty}\frac{1}{T}\frac{\partial}{\partial\sigma}\ln p_T=-\frac{\partial\lambda}{\partial\sigma}
=\frac{\theta(\sqrt{a^{2}+2\sigma^{2}}-a)^2}{\sigma^{3}\sqrt{a^{2}+2\sigma^{2}}}\,.
\end{aligned}
\end{equation}
For more details about the sensitivities of the CIR model, see  Appendix \ref{app:CIR}.

\subsection{Quadratic models}
We investigate the sensitivities of option prices whose underlying process is a 
short interest rate given by a quadratic-term structure model.
This section is based on Section 6.2 in \cite{qin2016positive}.
Let $\mathbb{P}$ be a risk-neutral measure.
Define a quadruple of functions $(b,\sigma,r,f)$ and an initial value $\xi$ as follows. Let 
$$b(x)=b+Bx\,,\;\sigma(x)=\sigma\,,$$
where $b=(b_i)_{1\leq i\leq d}$ is a $d$-dimensional column vector, $B=(B_{ij})_{1\leq i,j\leq d}$ is a $d\times d$ matrix, and $\sigma=(\sigma_{ij})_{1\leq i,j\leq d}$ is a non-singular $d\times d$ matrix.
It follows that  $a:=\sigma\sigma^{\top}$ is strictly positive definite. 
The short interest rate function $r$ is given by 
$$r(x)=\beta+\langle\alpha,x\rangle+\langle \Gamma x,x\rangle$$
where the constant $\beta,$ vector $\alpha$ and symmetric positive definite $\Gamma$ are such that the short interest rate $r(x)$ is non-negative for all $x\in\mathbb{R}^d.$ 
Let $f:\mathbb{R}^d\to\mathbb{R}$ be a non-negative, non-zero, bounded function with bounded support. 
Fix an initial value $\xi\in\mathbb{R}^d.$ 
The underlying process $X$ is a $d$-dimensional Ornstein-Uhlenbeck process satisfying the SDE 
$$dX_t=(b+BX_t)\,dt+\sigma\,dB_t\,,\;X_0=\xi\,.$$

We analyze the sensitivities of option prices given in 
Eq.\eqref{eqn:p_T}
with respect to the perturbations of the parameters $b,$ $B,$ $\sigma$ and $\xi$ in the underlying process $X.$
As discussed in Eq.\eqref{eqn:dtheta}, these parameters 
can be regarded as perturbation parameters.
The quadruple of functions
$(b,\sigma,r,f)$ and the initial value $\xi$ satisfy
Assumptions \ref{assume:pert_defi} - \ref{assume:perturbed}. The recurrent eigenpair is
\begin{equation}\label{eqn:eigenpair_QTSM}
(\lambda,\phi(x))=(\beta-\frac{1}{2}u^\top au+tr(aV)+u^\top b\,,\,e^{-\langle u,x\rangle-\langle Vx,x\rangle}),
\end{equation}
where   $V$ is the  {\em stabilizing solution}, defined as a unique solution of 
$$2VaV-B^\top V-VB-\Gamma=0$$ 
such that all eigenvalues of $B-2aV$ have negative real parts, 
and $u:=(2Va-B^\top)^{-1}(2Vb+\alpha).$
By using the eigenpair in Eq.\eqref{eqn:eigenpair_QTSM},
we obtain the long-term sensitivities   
\begin{equation*}
\begin{aligned}
\lim_{T\rightarrow\infty} \nabla_\xi\ln p_T =-u-2V\xi\,,\;	
\lim_{T\rightarrow\infty}\frac{1}{T} \frac{\partial }{\partial b_i}\ln p_T=-\frac{\partial \lambda}{\partial b_i}\,,\;
\lim_{T\rightarrow\infty} \frac{1}{T} \frac{\partial }{\partial B_{ij}}\ln p_T=-\frac{\partial \lambda}{\partial B_{ij}}
\end{aligned}
\end{equation*}
for $1\leq i,j\leq d.$ 
If $f$ is continuously differentiable with compact support, then we have
$$\lim_{T\rightarrow\infty}\frac{1}{T}\frac{\partial }{\partial \sigma_i}\ln p_T=-\frac{\,\partial \lambda\,}{\partial \sigma_i}\;.$$
See Appendix \ref{app:QTSM} for more details.

\section{Examples of expected utilities}
\label{sec:ex_utility}

\subsection{The Heston model}
This section conducts a sensitivity analysis of
the expected utility from holding an asset. 
Under the physical measure $\mathbb{L},$ suppose that
the asset $S=(S_t)_{t\geq0}$
follows the Heston model,
\begin{equation*}
\begin{aligned}
&dS_t=\mu S_t\, dt+ \sqrt{v_t}S_t\,dZ_t^S\;,\\
&dv_t=(\gamma-\beta v_t)\,dt+\delta\sqrt{|v_t|}\,dZ_t^v\;,
\end{aligned}
\end{equation*}
where $Z^S$ and $Z^v$ are correlated $\mathbb{L}$-Brownian motions with $\langle Z_t^S,Z_t^v\rangle_t=\rho t$  for  $-1\leq \rho\leq 1.$  
Let the parameters be $\mu,\gamma,\beta,\delta>0$ and $2\gamma>\delta^2.$
We consider a power utility function of the form
$u(c)={c}^{\alpha}$ for $0<\alpha< 1.$
The long-term sensitivities of the expected utility  
$$p_T:=\mathbb{E}^\mathbb{L}[u(S_T)]=\mathbb{E}^\mathbb{L}[S_T^\alpha]=\mathbb{E}^\mathbb{L}[e^{\alpha\int_0^T\sqrt{v_s}\,dZ_s-\frac{\alpha}{2}\int_0^Tv_s\,ds}]\,e^{\alpha\mu T}S_0^\alpha$$
are of interest to us.

The Heston model and the above expectation $p_T$ do not satisfy the underlying framework (Assumption \ref{assume:SDE} and Eq.\eqref{eqn:p_T}) of this paper.
We manipulate the setting to fit the underlying framework as follows.
Define a measure $\mathbb{P}$ on $\mathcal{F}_T$ by 
$$\frac{d\mathbb{P}}{d\mathbb{L}}\Big|_{\mathcal{F}_T}=e^{\alpha\int_0^T\sqrt{v_s}\,dZ_s-\frac{\alpha^2}{2}\int_0^Tv_s\,ds}$$
so that the expectation $p_T$ is expressed as
\begin{equation*}
\begin{aligned}
p_T=\mathbb{E}^\mathbb{P}[e^{-\frac{1}{2}\alpha(1-\alpha)\int_0^Tv_s\,ds}]\,e^{\alpha\mu T}S_0^\alpha\;.
\end{aligned}
\end{equation*}
By the Girsanov theorem, the $\mathbb{P}$-dynamics of $v$ is
\begin{equation*}
\begin{aligned}
dv_t
=(\gamma-(\beta-\alpha\rho\delta) v_t)\,dt+\delta\sqrt{v_t}\,dB_t
\end{aligned}
\end{equation*}
with a $\mathbb{P}$-Brownian motion $B.$
Define a process $X$ by $X_t=\frac{1}{2}\alpha(1-\alpha)v_t.$ It follows that 
$X$ is a CIR model satisfying 
\begin{equation*}
\begin{aligned}
dX_t
&=\Big(\frac{1}{2}\alpha(1-\alpha)\gamma-(\beta-\alpha\rho\delta)X_t\Big)dt+\frac{\delta\sqrt{2\alpha(1-\alpha)}}{2}\sqrt{|X_t|}\,dB_t\,,\;X_0=\frac{1}{2}\alpha(1-\alpha)v_0\,.
\end{aligned}
\end{equation*}
We define $\xi:=\frac{1}{2}\alpha(1-\alpha)v_0$ and 
$q_T:=\mathbb{E}_\xi^\mathbb{P}[e^{-\int_0^TX_s\,ds}].$ Then  $p_T=q_T\,e^{\alpha\mu T}S_0^\alpha.$
In conclusion, the quadruple of functions of $x$
$$\Big(\frac{1}{2}\alpha(1-\alpha)\gamma-(\beta-\alpha\rho\delta)x,\frac{\delta\sqrt{2\alpha(1-\alpha)}}{2}\sqrt{|x|},x,1\Big)$$
and the initial value $\xi$ satisfy the underlying framework of this paper. Here, $1$ is a constant function that is identically equal to one.

The sensitivities of $q_T$ for large $T$ 
are already analyzed in Section \ref{sec:CIR}.  
With new parameters $\theta=\frac{1}{2}\alpha(1-\alpha)\gamma,$ $a=\beta-\rho\alpha\delta$ and $\sigma=\frac{\delta}{2}\sqrt{2\alpha(1-\alpha)},$ 
the process $X$ is expressed as 
$$dX_t=(\theta-aX_t)\,dt+\sigma\sqrt{|X_t|}\,dB_t\,,\;X_0=\xi\,,$$
which is equal to   Eq.\eqref{eqn:SDE_CIR}. 
Employing the results of Section \ref{sec:CIR} and the chain rule,
we obtain the following sensitivities.
For the drift term sensitivities of $S$ and $v,$ we have
\begin{equation*}
\begin{aligned}
\lim_{T\rightarrow\infty}\frac{1}{T}\frac{\partial}{\partial\mu}\ln p_T
&=\alpha\,,\\
\lim_{T\rightarrow\infty}\frac{1}{T}\frac{\partial}{\partial\gamma}\ln p_T
&=\lim_{T\rightarrow\infty}\frac{1}{T}\frac{\partial}{\partial\gamma}\ln q_T
=\lim_{T\rightarrow\infty}\frac{1}{T}\frac{\partial}{\partial\theta}\ln q_T\frac{\partial\theta}{\partial\gamma}\\
&=\frac{1}{2}\alpha(1-\alpha)\lim_{T\rightarrow\infty}\frac{1}{T}\frac{\partial}{\partial\theta}\ln q_T
=-\frac{1}{2}\alpha(1-\alpha)\frac{\sqrt{a^{2}+2\sigma^{2}}-a}{\sigma^{2}}\\
&=-\frac{1}{2}\alpha(1-\alpha)\frac{\sqrt{(\beta-\rho\alpha\delta)^{2}+\delta^2\alpha(1-\alpha)}-\beta+\rho\alpha\delta}{\delta^2}\,,\\ 
\lim_{T\rightarrow\infty}\frac{1}{T}\frac{\partial}{\partial\beta}\ln p_T
&=\lim_{T\rightarrow\infty}\frac{1}{T}\frac{\partial}{\partial\beta}\ln q_T
=\lim_{T\rightarrow\infty}\frac{1}{T}\frac{\partial}{\partial a}\ln q_T\frac{\partial a}{\partial\beta}
=\lim_{T\rightarrow\infty}\frac{1}{T}\frac{\partial}{\partial a}\ln q_T\\
&=\frac{\theta(\sqrt{a^{2}+2\sigma^{2}}-a)}{\sigma^{2}\sqrt{a^{2}+2\sigma^{2}}}
=\frac{\sqrt{(\beta-\rho\alpha\delta)^{2}+\delta^2\alpha(1-\alpha)}-\beta+\rho\alpha\delta}{\delta^2\sqrt{(\beta-\rho\alpha\delta)^{2}+\delta^2\alpha(1-\alpha)}}\,.
\end{aligned}
\end{equation*}
For the volatility term sensitivities of $S$ and $v,$  it can be shown that  
\begin{equation*}
\begin{aligned}
\lim_{T\rightarrow\infty}\frac{1}{T}\frac{\partial}{\partial\delta}\ln p_T
&=-\rho\alpha\frac{\sqrt{(\beta-\rho\alpha\delta)^{2}+\delta^2\alpha(1-\alpha)}-\beta+\rho\alpha\delta}{\delta^2\sqrt{(\beta-\rho\alpha\delta)^{2}+\delta^2\alpha(1-\alpha)}}\\
&+\frac{(\sqrt{(\beta-\rho\alpha\delta)^{2}+\delta^2\alpha(1-\alpha)}-\beta+\rho\alpha\delta)^2}{\delta^3\sqrt{(\beta-\rho\alpha\delta)^{2}+\delta^2\alpha(1-\alpha)}}\,,\\
\lim_{T\rightarrow\infty}\frac{1}{T}\frac{\partial}{\partial\rho}\ln p_T
&=-\frac{\alpha\sqrt{(\beta-\rho\alpha\delta)^{2}+\delta^2\alpha(1-\alpha)}-\alpha\beta+\rho\alpha^2\delta}{\delta\sqrt{(\beta-\rho\alpha\delta)^{2}+\delta^2\alpha(1-\alpha)}}\;.
\end{aligned}
\end{equation*}
For the delta values, we have
\begin{equation*}
\begin{aligned}
&\lim_{T\rightarrow\infty}\frac{\partial}{\partial S_0}\ln p_T
=\frac{\alpha}{S_0}\,,\\
&\lim_{T\rightarrow\infty}\frac{\partial}{\partial v_0}\ln p_T
=-\frac{1}{2}\alpha(1-\alpha)\frac{\sqrt{(\beta-\rho\alpha\delta)^{2}+\delta^2\alpha(1-\alpha)}-\beta+\rho\alpha\delta}{\delta^2}\;.
\end{aligned}
\end{equation*}

\subsection{The $3/2$ LETF model}
\label{sec:3/2_model}

A sensitivity analysis of the expected utility from holding an exchange-traded fund (ETF) is now performed.
A leveraged ETF (LETF) $L=(L_t)_{t\geq0}$ is a fund 
designed to amplify the return by  
using a reference process.
Typical reference processes are index, stock, currency and commodity.
Assume that the reference process $X$ stays positive and the short interest rate is a constant $r>0.$
The  LETF is 
a constant proportion portfolio invested in the reference process and the money market account. The proportion is called the leverage ratio of the LETF and is denoted by $\beta.$

An LETF with leverage ratio 
$\beta\ge1$ (respectively, $\beta\leq-1$) is said to be long-leveraged (respectively, short-leveraged) and is characterized as follows.
Fix an initial investment $L_0$ in the LETF 
for a given 
initial value $X_0$ of the reference process.
At time $t\ge0,$ the cash amount of $\beta L_t$  is
invested in the reference at price $X_t,$ and the amount $(\beta-1)L_t$ is financed at the risk-free rate $r.$ 
In the actual financial market, common leverage ratios are $\beta=1,2,3$ (long) and $\beta=-1,-2,-3$ (short).
Refer to \cite{leung2015implied} for more details about LETFs.

Based on this characterization,
the LETF $L=(L_t)_{t\ge0}$ satisfies
\begin{equation*}
\begin{aligned}
\frac{dL_t}{L_t}
&=\beta\Big(\frac{dX_t}{X_t}\Big)-((\beta-1)r)\,dt\\
&=\Big(\beta\Big(\frac{\mu(X_t)}{X_t}\Big)-(\beta-1)r\Big)\,dt+\beta\Big(\frac{\sigma(X_t)}{X_t}\Big)\,dB_t\,,
\end{aligned}
\end{equation*}
which can be written as 
\begin{equation}\label{eqn:LETF_undelying}	
\frac{L_t}{L_0}=e^{(1-\beta)rt-\frac{1}{2}\beta(\beta-1)\int_0^t  \sigma^2(X_s)/X_s^2\,ds}\Big(\frac{X_t}{X_0}\Big)^\beta\,.
\end{equation}
Assume that the utility is a power function of the form
$u(c)={c}^{\alpha}$ for $0<\alpha< 1,$
and the reference process $X$ is given by the $3/2$ model 
$$dX_t=(\theta-aX_t)X_t\,dt+\sigma |X_t|^{3/2}\,dB_t\;,\,X_0=\xi$$
with positive constants 
$\theta,$ $a,$ $\sigma,$ $\xi$ under the physical measure $\mathbb{P}.$
This process stays positive and is mean-reverting. For a  practical example, one can consider the LETF on commodity prices or volatility indices.
The expected utility from holding the LETF $L$ is 
\begin{equation}\label{eqn:LETF_3/2}
p_T:=\mathbb{E}^\mathbb{P}[u(L_T)]=\mathbb{E}^\mathbb{P}[e^{-\frac{1}{2}\alpha\beta(\beta-1)\sigma^2\int_0^tX_u\,du}X_t^{\alpha\beta}]\,e^{\alpha(1-\beta)rt}\xi^{-\alpha\beta}L_0^\alpha \,.
\end{equation}
The sensitivity of $p_T$ is of interest to us. 
For convenience, we define
\begin{equation}\label{eqn:q_T_3_2}
q_T:=\mathbb{E}^\mathbb{P}[e^{-\frac{1}{2}\alpha\beta(\beta-1)\sigma^2\int_0^tX_u\,du}X_t^{\alpha\beta}] 
\end{equation}  
so that  $p_T=q_Te^{\alpha(1-\beta)rt}\xi^{-\alpha\beta}L_0^\alpha.$ 

The above expectation  $q_T$ fits into the underlying framework of this paper.
Observe that the  quadruple of functions is
\begin{equation}\label{eqn:quad_3_2}
(b(x),\sigma(x),r(x),f(x)):=((\theta-ax)x,\sigma |x|^{3/2},\frac{1}{2}\alpha\beta(1-\beta)\sigma^2x,x^{\alpha\beta}) 
\end{equation} 
and the initial value is $\xi>0.$ 
These satisfy Assumptions \ref{assume:pert_defi} - \ref{assume:perturbed}, and the recurrent eigenpair is  
$(\lambda,\phi(x)):=(\theta\ell\,,\,x^{-\ell})$
where
\begin{equation}
\label{eqn:3/2_ell}
\ell:=\sqrt{\Big(\frac{1}{2}+\frac{a}{\sigma^2}\Big)^2+\alpha\beta(\beta-1)}-\Big(\frac{1}{2}+\frac{a}{\sigma^2}\Big)\,.
\end{equation} 
It can be shown that if $|\beta|\leq3,$ which is a realistic condition, then the function $f(x)=x^{\alpha\beta}$ satisfies Assumption \ref{assume:ergodic}.

With this eigenpair, we obtain the long-term sensitivities of the expectation $p_T.$
The sensitivity of $\xi$ is 
\begin{equation}\label{eqn:xi_3_2}
\lim_{T\rightarrow\infty}\frac{\partial}{\partial \xi}\ln p_T=-\alpha\beta \xi^{-\alpha\beta-1}\lim_{T\rightarrow\infty}\frac{\partial}{\partial \xi}\ln q_T
= -\alpha\beta \xi^{-\alpha\beta-1}\frac{\phi'(\xi)}{\phi(\xi)}
=\alpha\beta\ell \xi^{-\alpha\beta-2}\,. 
\end{equation}
For the sensitivity of $\theta,$ 
\begin{equation}\label{eqn:theta_3_2}
\begin{aligned}
\lim_{T\rightarrow\infty}\frac{1}{T} \frac{\,\partial }{\partial \theta}\ln p_T=-\sqrt{\Big(\frac{1}{2}+\frac{a}{\sigma^2}\Big)^2+\alpha\beta(\beta-1)}-\Big(\frac{1}{2}+\frac{a}{\sigma^2}\Big)\,.
\end{aligned}
\end{equation}
For the sensitivities of $a$ and $\sigma,$ when $\frac{a}{\sigma^2}+1-\alpha\beta>0,$ we have
\begin{equation}\label{eqn:a_3_2}
\begin{aligned}
\lim_{T\rightarrow\infty}\frac{1}{T} \frac{\,\partial }{\partial a}\ln p_T=\frac{\theta(\sqrt{(\frac{1}{2}+\frac{a}{\sigma^2})^2+\alpha\beta(\beta-1)}-(\frac{\sigma}{2}+a))}{\sigma^2\sqrt{(\frac{1}{2}+\frac{a}{\sigma^2})^2+\alpha\beta(\beta-1)}} 
\end{aligned}
\end{equation}
and 
\begin{equation}\label{eqn:sigma_3_2}
\begin{aligned}
\lim_{T\rightarrow\infty}\frac{1}{T} \frac{\,\partial }{\partial \sigma}\ln p_T=\frac{{2a\theta}(\sqrt{(\frac{1}{2}+\frac{a}{\sigma^2})^2+\alpha\beta(\beta-1)}-(\frac{1}{2}+\frac{a}{\sigma^2}))}{{\sigma^3}\sqrt{(\frac{1}{2}+\frac{a}{\sigma^2})^2+\alpha\beta(\beta-1)}} \;.
\end{aligned}
\end{equation}
See Appendix \ref{app:3/2_model} for more details.

\section{Conclusion}

This paper conducts a sensitivity analysis of long-term cash flows. The price of the cash flow at time zero is given 
by the expectation form 
$p_T=\mathbb{E}_\xi^{\mathbb{P}}[e^{-\int_0^T r(X_s)\,ds} f(X_T)].$
We explore the extent to which the price of the cash flow is affected by small perturbations of the underlying Markov diffusion process $X.$
Essentially, two types of perturbations are presented.
First, the sensitivity with respect to the initial value $\xi=X_0$ is investigated. Using the Hansen--Scheinkman decomposition, 
the expectation $p_T$ can be expressed as 
\begin{equation*} 
\begin{aligned}
p_T=\phi(\xi)\,e^{-\lambda T}
\mathbb{E}_\xi^{\mathbb{Q}}[(f/\phi)(X_T)], 
\end{aligned}
\end{equation*} 
with the recurrent eigenpair $(\lambda,\phi)$ and the recurrent eigen-measure $\mathbb{Q}.$ 
Under appropriate conditions, we obtain the long-term sensitivity $$\lim_{T\rightarrow\infty}\nabla_\xi\ln p_T=\frac{\nabla_\xi\,\phi}{\phi(\xi)}\,.$$

Second, the sensitivities
with respect to the drift and volatility terms are studied.
From the Hansen--Scheinkman decomposition,
the perturbed expectation $p_T^\epsilon=\mathbb{E}_\xi^{\mathbb{P}}[e^{-\int_0^T r(X_s^\epsilon)\,ds} f(X_T^\epsilon)]$ induced by the perturbed process $X^\epsilon$ is 
expressed as 
\begin{equation*} 
\begin{aligned}
p_{T}^{\epsilon}
&=e^{-\lambda_\epsilon T}\phi_{\epsilon}(\xi)\,\mathbb{E}_{\xi}^{\mathbb{Q_{\epsilon}}}[(f_\epsilon/\phi_{\epsilon})(X_{T}^{\epsilon})],
\end{aligned}
\end{equation*}
with the recurrent eigenpair $(\lambda_\epsilon,\phi_\epsilon)$ and the recurrent eigen-measure $\mathbb{Q}^{\epsilon}.$ 
We prove that  the long-term sensitivity of $p_T^\epsilon$ with respect to the perturbation parameter $\epsilon$  can be expressed in a simple form as 
$$\lim_{T\rightarrow\infty}\frac{1}{T}\frac{\partial}{\partial \epsilon}\Big|_{\epsilon=0}\ln p_T^\epsilon=- \frac{\partial }{\partial\epsilon}\Big|_{\epsilon=0}\lambda_\epsilon\,.$$


$$ $$

\noindent {\small {\bf Acknowledgement}. 
The author  sincerely appreciates 	the valuable suggestions received from Jonathan Goodman, Stephan Sturm and Srinivasa Varadhan.
The author is grateful to the Editor, the Associate Editor and two anonymous referees for
their helpful comments and insights that have greatly improved the quality of the paper.
This work was supported by the National Research Foundation of Korea (NRF) Grant funded by the Korean
Government (MSIP) (No.2017R1A5A1015626).}

\appendix\normalsize

\section{Proof of Theorem \ref{thm:rho_expo_condi}}
\label{app:pf_rho_expo_condi}

Propositions \ref{prop:rho} and \ref{prop:expo_p_th_power_inequ} 
are  essential steps for proving Theorem \ref{thm:rho_expo_condi}.
Proposition \ref{prop:rho} is a generalization of Proposition 3.1  in \cite{fournie1999applications}. 
We modify their proof.
Recall the functions  $\overline{k}_\epsilon$ and $\overline{k}$  defined in Eq.\eqref{eqn:bar_k}.

\begin{proposition} \label{prop:rho}
	Let $(b_\epsilon,\sigma,r_\epsilon,f_\epsilon)$ and $\xi$ be a quadruple of functions and an initial value, respectively, satisfying Assumptions \ref{assume:pert_defi} - \ref{assume:perturbed}. 
	Assume that  both $\phi_\epsilon(x)$ and
	$\nabla\phi_\epsilon(x)$ (thus, $k_\epsilon(x)$) are continuously differentiable in $\epsilon$ on $I$ for each $x$ and that there exist functions $g,\psi:\mathbb{R}^d\rightarrow\mathbb{R}$ satisfying Eq.\eqref{eqn:g_bound} and Eq.\eqref{eqn:psi_bound} for $(\epsilon,x)$ in $ I\times\mathbb{R}^d.$ 	
	Suppose that for each $T>0$ there exist positive constants $\epsilon_0,$ $\epsilon_1,$ $p,$ $q$ with $p\geq2$ and  $1/p+1/q=1$ such that 
	\begin{align}
	&\mathbb{E}_\xi^{\mathbb{Q}}[e^{\epsilon_0\int_0^Tg^2(X_s)\,ds}]<\infty\label{eqn:expo_condi}\,,\\
	&\mathbb{E}_\xi^{\mathbb{Q}}\Big[\int_0^Tg^{p+\epsilon_1}(X_s)\,ds\Big]<\infty\label{eqn:g_condi}\,,\\
	&\mathbb{E}_\xi^{\mathbb{Q}}[\psi^q(X_T)]<\infty\label{eqn:psi_condi}\,.
	\end{align}	
	Then 
	the partial derivative	$\frac{\partial}{\partial\epsilon}\mathbb{E}_{\xi}^{\mathbb{Q}^\epsilon}[(f_\eta/\phi_\eta)(X_T^\epsilon)]$ exists and 
	\begin{equation}
	\begin{aligned}\label{eqn:deriva_rho}
	\frac{\partial}{\partial\epsilon}\mathbb{E}_{\xi}^{\mathbb{Q}^\epsilon}[(f_\eta/\phi_\eta)(X_T^\epsilon)]=\mathbb{E}_\xi^{\mathbb{Q}^\epsilon}
	\Big[(f_\eta/\phi_\eta)(X_{T}^\epsilon)\int_{0}^{T}\overline{k}_\epsilon(X_{s}^\epsilon)\, dW_{s}^\epsilon\Big]\,.
	\end{aligned}
	\end{equation}
	Moreover, this partial derivative is continuous in $(\eta,\epsilon)$ on $I^2.$  	
\end{proposition}

\noindent The  proof is organized as follows.
\begin{itemize} 
	\item[(I)] First, prove Eq.\eqref{eqn:deriva_rho} for $\epsilon=0,$ that is,  
	\begin{equation}\label{eqn:rho_Fournie}
	\frac{\partial}{\partial \epsilon}\;\Big|_{\epsilon =0}\mathbb{E}_\xi^{\mathbb{Q_{\epsilon}}}
	[(f_\eta/\phi_\eta)(X_{T}^{\epsilon})] =\mathbb{E}_\xi^{\mathbb{Q}}
	\Big[(f_\eta/\phi_\eta)(X_{T})\int_{0}^{T}\overline{k}(X_{s})\, dW_{s}\Big]\,.
	\end{equation} 
	We conduct the following sub-steps to show the above equality.
	\begin{itemize}[noitemsep,nolistsep]
		\item[(i)] Show that
		$$ \frac{\partial}{\partial \epsilon}\;\Big|_{\epsilon =0}\mathbb{E}_\xi^{\mathbb{Q}^\epsilon}
		[(f_\eta/\phi_\eta)(X_{T}^{\epsilon})]
		=\lim_{\epsilon\to0}\,\mathbb{E}_\xi^{\mathbb{Q}}\Big[(f_\eta/\phi_\eta)(X_{T})\int_0^TZ^\epsilon_s\,\ell_\epsilon(X_s)\,dW_s\Big]$$ 	
		for the first-order approximation $\ell_\epsilon$ of $k_\epsilon$ in
		$\epsilon$-perturbation, and for an exponential martingale  $Z^\epsilon.$ Both $\ell_\epsilon$ and $Z^\epsilon$ are defined later. 
		Then it is enough to show that
		\begin{equation}\label{eqn:ss}
		\lim_{\epsilon\to0}\,\mathbb{E}_\xi^{\mathbb{Q}}\Big[(f_\eta/\phi_\eta)(X_{T})\int_0^T(Z^\epsilon_s\ell_\epsilon(X_s) -\overline{k}(X_{s}))\,dW_s\Big]=0
		\end{equation}
		which gives the desired result stated in Eq.\eqref{eqn:rho_Fournie}. 
		\item[(ii)] 
		To show \eqref{eqn:ss}, it suffices to show that
		$\int_0^T(Z^\epsilon_s\ell_\epsilon(X_s) -\overline{k}(X_s))\,dW_s\to 0$ in $L^p$ as $\epsilon\to 0.$   Observing the equality
		\begin{equation} 
		\int_0^T(Z^\epsilon_s\ell_\epsilon(X_s) -\overline{k}(X_s))\,dW_s
		=\int_0^T(Z^\epsilon_s-1)\ell_\epsilon(X_s)\,dW_s+\int_0^T(\ell_\epsilon-\overline{k})(X_s)\,dW_s\,, 
		\end{equation}
		we prove that the two terms on the right-hand side converge to zero in $L^p$  as $\epsilon\to0.$	
	\end{itemize}
	\item[(II)] Using Eq.\eqref{eqn:rho_Fournie}, verify Eq.\eqref{eqn:deriva_rho} for arbitrary $\epsilon\in I.$
	\item[(III)] Prove that the partial derivative in Eq.\eqref{eqn:deriva_rho} is continuous in $(\eta,\epsilon)$ on $I^2.$
\end{itemize}

\begin{proof}	The proof of Proposition \ref{prop:rho} will be given in several steps. 
	
	\noindent {\em Step} (I) - (i).	
	From Eq.\eqref{eqn:BM_Girsa} and Eq.\eqref{eqn:Q-dynamics}, a process
	$(W_{t}^\epsilon)_{t\ge0}:=(B_{t}-\int_{0}^{t}\varphi_\epsilon(X_{s}^\epsilon)\,ds)_{t\ge0}$
	is a $\mathbb{Q}^\epsilon$-Brownian motion,
	and the $\mathbb{Q}^\epsilon$-dynamics of $X^\epsilon$ is
	\begin{equation} 
	\begin{aligned}
	dX_{t}^\epsilon
	&=(b_\epsilon+\sigma\varphi_\epsilon)(X_{t}^\epsilon)\, dt+\sigma(X_{t})\, dW_{t}^\epsilon
	=(\sigma k_\epsilon)(X_t^\epsilon)\,dt+\sigma (X_t^\epsilon)\,dW_t^\epsilon\;.
	\end{aligned}
	\end{equation}
	Because $k_\epsilon$ is continuously differentiable in $\epsilon$ on $I,$ by the Taylor expansion, we write $k_\epsilon=k+\epsilon\ell_\epsilon$ for some $d\times 1$ vector function $\ell_\epsilon.$ The $\mathbb{Q}^\epsilon$-dynamics of $X^\epsilon$ is expressed by
	$$dX_t^\epsilon=(\sigma k+\epsilon\sigma\ell_\epsilon )(X_t^\epsilon)\,dt+\sigma(X_t^\epsilon)\,dW_t^\epsilon\;.$$
	
	By Assumption \eqref{eqn:expo_condi}, the process  	
	\begin{equation}\label{eqn:Z}
	(Z^\epsilon_t)_{0\leq t\leq  T}
	:=(e^{\epsilon\int_0^t\ell_\epsilon(X_s)\,dW_s
		-\frac{\epsilon^2}{2}\int_0^t|\ell_\epsilon|^2(X_s)\,ds })_{0\leq t\leq  T}  
	\end{equation}
	is a martingale for small $\epsilon$ because the Novikov condition is satisfied. Here we used the mean-value theorem 
	$$|\ell_\epsilon(x)|=\left|\frac{k_\epsilon(x)-k(x)}{\epsilon}\right|=\left| \frac{\partial}{\partial\epsilon}\Big|_{\epsilon=\epsilon^*}k_\epsilon(x) \right|\leq g(x)$$
	for some $\epsilon^*\in I.$
	By the Girsanov theorem, we have	
	$\mathbb{E}_\xi^{\mathbb{Q}^\epsilon}
	[(f_\eta/\phi_\eta)({X}_{T}^{\epsilon})]=\mathbb{E}_\xi^{\mathbb{Q}}[(f_\eta/\phi_\eta)(X_{T})\,Z^\epsilon_T],$ and thus
	\begin{equation}\label{eqn:aa}
	\begin{aligned}
	& \frac{\partial}{\partial \epsilon}\;\Big|_{\epsilon =0}\mathbb{E}_\xi^{\mathbb{Q}^\epsilon}
	[(f_\eta/\phi_\eta)(X_{T}^{\epsilon})]
	= \frac{\partial}{\partial \epsilon}\;\Big|_{\epsilon =0}\mathbb{E}_\xi^{\mathbb{Q}}[(f_\eta/\phi_\eta)(X_{T})\,Z^\epsilon_T]\\
	=&\lim_{\epsilon\to0}\,\mathbb{E}_\xi^{\mathbb{Q}}	\Big[(f_\eta/\phi_\eta)(X_{T})\,\frac{Z^\epsilon_T-1}{\epsilon}\Big]
	=\lim_{\epsilon\to0}\,\mathbb{E}_\xi^{\mathbb{Q}}	\Big[(f_\eta/\phi_\eta)(X_{T})\int_0^TZ^\epsilon_s\,\ell_\epsilon(X_s)\,dW_s\Big]\,.
	\end{aligned}
	\end{equation}
	For the last equality, we used $\frac{Z^\epsilon_T-1}{\epsilon}=\int_0^TZ^\epsilon_s\,\ell_\epsilon(X_s)\,dW_s,$ which is easily obtained by the Ito formula.  		
	From  Eq.\eqref{eqn:aa}, it suffices to prove  that	
	$$\lim_{\epsilon\to0}\,\mathbb{E}_\xi^{\mathbb{Q}}\Big[(f_\eta/\phi_\eta)(X_{T})\int_0^T(Z^\epsilon_s\ell_\epsilon(X_s) -\overline{k}(X_{s}))\,dW_s\Big]=0\,,$$
	which gives  Eq.\eqref{eqn:rho_Fournie}.\newline

	\noindent {\em Step} (I) - (ii). 
	From the condition $\mathbb{E}_\xi^{\mathbb{Q}}[(f_\eta/\phi_\eta)^q(X_T)]\le \mathbb{E}_\xi^{\mathbb{Q}}[\psi^q(X_T)]<\infty,$  by the Holder inequality, it is enough to show that
	$\int_0^T(Z^\epsilon_s\ell_\epsilon(X_s) -\overline{k}(X_s))\,dW_s$ converges to $0$ in $L^p$ as $\epsilon\to 0.$	
	Using the equality
	\begin{equation}\label{eqn:split}
	\int_0^T(Z^\epsilon_s\ell_\epsilon(X_s) -\overline{k}(X_s))\,dW_s
	=\int_0^T(Z^\epsilon_s-1)\ell_\epsilon(X_s)\,dW_s+\int_0^T(\ell_\epsilon-\overline{k})(X_s)\,dW_s\,, 
	\end{equation}
	we show that each term on the right-hand side converges to zero in $L^p.$	
	For the second term on the right-hand side, we use the Lebesgue dominated convergence theorem.  
	Because $|\ell_\epsilon-\overline{k}|^p\leq c\, (|\ell_\epsilon|^p+|\overline{k}|^p)\leq 2cg^p$ for some positive constant $c,$ and 
	$\mathbb{E}_\xi^{\mathbb{Q}}[\int_0^Tg^p(X_s)\,ds]$ is finite, we have 
	\begin{equation*}
	\begin{aligned}
	\mathbb{E}_\xi^{\mathbb{Q}}\Big[\Big|\int_0^T(\ell_\epsilon-\overline{k})(X_s)\,dW_s\Big|^p\Big]
	&\leq c_q\,\mathbb{E}_\xi^{\mathbb{Q}}\Big[\Big(\int_0^T|\ell_\epsilon-\overline{k}|^2(X_s)\,ds\Big)^{\frac{p}{2}}\Big]\\&\leq c_q\, T^{\frac{p}{2}-1}\, \mathbb{E}_\xi^{\mathbb{Q}}\Big[\int_0^T|\ell_\epsilon-\overline{k}|^p(X_{t})\,dt\Big]\to0 
	\end{aligned}
	\end{equation*}
	as $\epsilon\rightarrow0$ for some positive constant $c_q$ which is independent of $\epsilon.$ The Burkholder-Davis-Gundy inequality and the Jensen inequality were used.

	For the first term on the right-hand side of Eq.\eqref{eqn:split}, we   prove that
	$$\lim_{\epsilon\rightarrow0}\,\mathbb{E}_\xi^{\mathbb{Q}}\Big[\Big|\int_0^T(Z^\epsilon_s-1)\,\ell_\epsilon(X_s)\,dW_s\Big|^p\Big]=0\;.$$
	Let $r>0$ be such that $1/r+1/(1+\epsilon_1/p)=1.$
	Applying the Burkholder-Davis-Gundy inequality, the Jensen inequality and the Holder inequality,     we have 
	\begin{equation*}
	\begin{aligned}
	&\mathbb{E}_\xi^{\mathbb{Q}}\Big[\Big|\int_0^T(Z^\epsilon_s-1)\,\ell_\epsilon(X_s)\,dW_s\Big|^p\Big]\\
	\leq&\,c_q\,\mathbb{E}_\xi^{\mathbb{Q}}\Big[\Big(\int_0^T(Z^\epsilon_s-1)^2\,|\ell_\epsilon|^2(X_s)\,ds\Big)^\frac{p}{2}\Big]\\
	\leq&\,c_q\,T^{\frac{p}{2}-1}\,\mathbb{E}_\xi^{\mathbb{Q}}\Big[\int_0^T|Z^\epsilon_s-1|^p\,|\ell_\epsilon|^p(X_s)\,ds \Big] \\
	\leq\,&\,c_q\,T^{\frac{q}{2}-1}\Big(\mathbb{E}_\xi^{\mathbb{Q}}\Big[\int_0^T|Z^\epsilon_s-1|^{pr}\,ds\Big]\Big)^{\frac{1}{r}}\,\Big(\mathbb{E}_\xi^{\mathbb{Q}}\Big[\int_0^T|\ell_\epsilon|^{p+\epsilon_1}(X_s)\,ds\Big]\Big)^{\frac{p}{p+\epsilon_1}}\\
	\leq\,&\,c_q\,T^{\frac{q}{2}-1}\Big(\mathbb{E}_\xi^{\mathbb{Q}}\Big[\int_0^T|Z^\epsilon_s-1|^{pr}\,ds\Big]\Big)^{\frac{1}{r}}\,\Big(\mathbb{E}_\xi^{\mathbb{Q}}\Big[\int_0^Tg^{p+\epsilon_1}(X_s)\,ds\Big]\Big)^{\frac{p}{p+\epsilon_1}}\,.
	\end{aligned}
	\end{equation*}
	In the last inequality, since  $\mathbb{E}_\xi^{\mathbb{Q}}[\int_0^Tg^{p+\epsilon_1}(X_s)\,ds]$ is finite from  Assumption \eqref{eqn:g_condi}, it suffices to prove that   $\mathbb{E}_\xi^{\mathbb{Q}}[\int_0^T|Z^\epsilon_s-1|^{pr}\,ds]\to 0$ as $\epsilon\to0.$ 
	Choose a positive even integer $m$ such that $m>pr.$

	We will show that 
	$\mathbb{E}_\xi^{\mathbb{Q}}[\int_0^T(Z^\epsilon_s-1)^{m}\,ds]$ converges to zero as $\epsilon\to0.$ 
	Observe that
	\begin{equation}\label{eqn:decompose_Z}
	\begin{aligned}
	(Z^\epsilon_t-1)^{m}=\sum_{j=0}^{m}{m \choose j} (-1)^{j}Z^{\epsilon\, j}_t\,. 
	\end{aligned}
	\end{equation} 
	Because 
	\begin{equation}\label{eqn:binomial_expan}
	\mathbb{E}_\xi^{\mathbb{Q}}\Big[\int_0^T(Z^\epsilon_s-1)^{m}\,ds\Big]\!=\!\sum_{j=0}^{m}{m \choose j} (-1)^{j}\int_0^T \mathbb{E}_\xi^{\mathbb{Q}}[Z^{\epsilon\,j}_t]\,dt\rightarrow T\sum_{j=0}^{m}{m\choose j} (-1)^{j}=0\;,
	\end{equation}
	it is enough to show that $\int_0^T\mathbb{E}_\xi^{\mathbb{Q}}[Z^{\epsilon\,j}_t]\,dt$ converges to $T$ 
	as $\epsilon\rightarrow0$	for $j=0,1,\cdots,m.$  
	To achieve this,  the Lebesgue dominated convergence theorem is used. We prove that the expectation $\mathbb{E}_\xi^{\mathbb{Q}}[Z^{\epsilon\,j}_t]$ is uniformly bounded  by a constant for small $\epsilon$ and $0\leq t\leq T,$ and that $\mathbb{E}_\xi^{\mathbb{Q}}[Z^{\epsilon\,j}_t]$ converges to $1$ as $\epsilon\to 0$ for each fixed $t.$	Observe that
	\begin{equation*}
	\begin{aligned}
	&\;\quad\mathbb{E}_\xi^{\mathbb{Q}}[Z^{\epsilon\,j}_t]\\
	&=\mathbb{E}_\xi^{\mathbb{Q}}[e^{j\epsilon\int_0^t\ell_\epsilon(X_s)\,dW_s
		-\frac{1}{2}j\epsilon^2\int_0^t|\ell_\epsilon|^2(X_s)\,ds  }]\\
	&=\mathbb{E}_\xi^{\mathbb{Q}} [e^{j\epsilon\int_0^t\ell_\epsilon(X_s)\,dW_s
		-j^2\epsilon^2\int_0^t|\ell_\epsilon|^2(X_s)\,ds}	   e^{ j(j-1/2)\epsilon^2\int_0^t|\ell_\epsilon|^2(X_s)\,ds }] \\
	&\leq (\mathbb{E}_\xi^{\mathbb{Q}}[e^{2j\epsilon\int_0^t\ell_\epsilon(X_s)\,dW_s
		-2j^2\epsilon^2\int_0^t|\ell_\epsilon|^2(X_s)\,ds}] )^{\frac{1}{2}}  \, (\mathbb{E}_\xi^{\mathbb{Q}}[e^{j(2j-1)\epsilon^2\int_0^t|\ell_\epsilon|^2(X_s)\,ds}])^{\frac{1}{2}}\\
	&= (\mathbb{E}_\xi^{\mathbb{Q}}[e^{j(2j-1)\epsilon^2\int_0^t|\ell_\epsilon|^2(X_s)\,ds}] )^{\frac{1}{2}} \quad(\because\textnormal{the former term is a martingale for small } \epsilon) \\
	&\leq (\mathbb{E}_\xi^{\mathbb{Q}}[e^{j(2j-1)\epsilon^2\int_0^tg^2(X_s)\,ds}] )^{\frac{1}{2}} \,.
	\end{aligned}
	\end{equation*}
	By choosing smaller $I$ if necessary, we may assume that $j(2j-1)\epsilon^2\le \epsilon_0$ for all $\epsilon\in I$ and $j=0,1,\dots,m.$ 
	For  $0\le t\le T$ and   $\epsilon\in I,$ 
	we have
	\begin{equation}\label{eqn:UB_Z}
	\mathbb{E}_\xi^{\mathbb{Q}}[Z^{\epsilon\,j}_t]
	\leq(\mathbb{E}_\xi^{\mathbb{Q}}[e^{\epsilon_0\int_0^Tg^2(X_s)\,ds}])^{\frac{1}{2}}\,.
	\end{equation}
	Thus, for $\epsilon\in I$ and $0\leq t\leq T,$ the expectation
	$\mathbb{E}_\xi^{\mathbb{Q}}[Z^{\epsilon\,j}_t]$ is uniformly bounded by the constant $(\mathbb{E}_\xi^{\mathbb{Q}}[e^{\epsilon_0\int_0^Tg^2(X_s)\,ds}])^{\frac{1}{2}}$ which is a finite number by Assumption \eqref{eqn:expo_condi}.

	We now show that $\mathbb{E}_\xi^{\mathbb{Q}}[Z^{\epsilon\,j}_t]$ converges to $1$ as $\epsilon\to 0$ for fixed $t\in[0,T].$
	Apply the Lebesgue dominated convergence theorem to
	$e^{j(2j-1)\epsilon^2\int_0^tg^2(X_s)\,ds}$
	as $\epsilon\to0.$
	Because this is dominated  by
	$e^{\epsilon_0\int_0^Tg^2(X_s)\,ds}$
	whose expectation  is finite,
	we know that $\mathbb{E}_\xi^{\mathbb{Q}}[e^{j(2j-1)\epsilon^2\int_0^tg^2(X_s)\,ds}]$ converges to $1$ as $\epsilon\to0.$ It follows that
	\begin{equation}\label{eqn:limit_Z}
	\begin{aligned}
	1=\mathbb{E}_\xi^{\mathbb{Q}}[\liminf_{\epsilon\rightarrow0}Z^{\epsilon\,j}_t]
	&\leq \liminf_{\epsilon\rightarrow0}\mathbb{E}_\xi^{\mathbb{Q}}[Z^{\epsilon\,j}_t]\\
	&\leq
	\limsup_{\epsilon\rightarrow0}\mathbb{E}_\xi^{\mathbb{Q}}[Z^{\epsilon\,j}_t] \leq\lim_{\epsilon\rightarrow0}\mathbb{E}_\xi^{\mathbb{Q}}[e^{j(2j-1)\epsilon^2\int_0^tg^2(X_s)\,ds}]=1\,.
	\end{aligned}
	\end{equation}
	This completes the proof.\newline
	
	\noindent {\em Step} (II).  We show Eq.\eqref{eqn:deriva_rho} for arbitrary $\epsilon\in I.$
	Fix $\epsilon\in I$ and choose an    open interval $J$ at $0$ small enough so that $\epsilon+J$ is still in $I.$  
	To utilize Eq.\eqref{eqn:rho_Fournie}, 
	we introduce another parameter $h$
	in the interval $J.$  	  
	Consider the quadruple of functions   $(b_{\epsilon+h},\sigma,r_{\epsilon+h},f_{\epsilon+h})$ and $\xi$ with perturbation parameter $h.$
	This quadruple and initial value satisfy the hypothesis of  Proposition \ref{prop:rho} because  $\epsilon+J$ is a subset of $I.$  
	Thus, from the result of Step (I), we know that 
	$\mathbb{E}_\xi^{\mathbb{Q}^{\epsilon+h}}
	[(f_\eta/\phi_\eta)(X_{T}^{\epsilon+h})]$ is   differentiable at $h=0,$ and   
	$$ \frac{\partial}{\partial \epsilon}\;\Big|_{h =0}\mathbb{E}_\xi^{\mathbb{Q}^{\epsilon+h}}
	[(f_\eta/\phi_\eta)(X_{T}^{\epsilon+h})] =\mathbb{E}_\xi^{\mathbb{Q}^\epsilon}
	\Big[(f_\eta/\phi_\eta)(X_{T}^\epsilon)\int_{0}^{T}\overline{k}_\epsilon(X_{s}^\epsilon)\, dW_{s}^\epsilon\Big]$$
	for 
	$\overline{k}_\epsilon(x)= \frac{\partial}{\partial\epsilon} k_\epsilon(x)
	=\left.\frac{\partial}{\partial h}\right|_{h=0} k_{\epsilon+h}(x).$ This gives Eq.\eqref{eqn:deriva_rho}.\newline
	
	\noindent {\em Step}  (III). Prove that the partial derivative 
	$$	 \frac{\partial}{\partial\epsilon}\mathbb{E}_{\xi}^{\mathbb{Q}^\epsilon}
	[(f_\eta/\phi_\eta)(X_T^\epsilon)]=\mathbb{E}_\xi^{\mathbb{Q}^\epsilon}
	\Big[(f_\eta/\phi_\eta)(X_{T}^\epsilon)\int_{0}^{T}\overline{k}_\epsilon(X_{s}^\epsilon)\, dW_{s}^\epsilon\Big]$$
	is continuous in $(\eta,\epsilon)$ on $I^2.$
	Without loss of generality, we prove the continuity at the origin $(\eta,\epsilon)=(0,0).$ Observe that
	\begin{equation}
	\begin{aligned}
	&\,\mathbb{E}_\xi^{\mathbb{Q}^\epsilon}
	\Big[(f_\eta/\phi_\eta)(X_{T}^\epsilon)\int_{0}^{T}\overline{k}_\epsilon(X_{s}^\epsilon)\, dW_{s}^\epsilon\Big]\\
	=&\,\mathbb{E}_\xi^{\mathbb{Q}}
	\Big[(f_\eta/\phi_\eta)(X_{T})\Big(\int_{0}^{T}\overline{k}_\epsilon(X_{s})\, dW_{s}+\epsilon\int_0^T(\overline{k}_\epsilon\ell_\epsilon)(X_s)\,ds\Big) Z_T^\epsilon\Big]\,.
	\end{aligned}
	\end{equation}
	For convenience, define 
	\begin{equation}
	\begin{aligned}
	&H_T^\epsilon:=\Big(\int_{0}^{T}\overline{k}_\epsilon(X_{s})\, dW_{s}+\epsilon\int_0^T(\overline{k}_\epsilon\ell_\epsilon)(X_s)\,ds\Big)  Z_T^\epsilon\\
	&H_T:=H_T^0\,.
	\end{aligned}
	\end{equation}
	We want to show that as $(\eta,\epsilon)\to(0,0),$
	$$\mathbb{E}_\xi^{\mathbb{Q}}
	[(f_\eta/\phi_\eta)(X_{T})  H_T^\epsilon ]
	\to\mathbb{E}_\xi^{\mathbb{Q}}
	[(f/\phi)(X_{T})  H_T]\,.$$
	Since $\mathbb{E}_\xi^{\mathbb{Q}}[\psi^q(X_T)]$  is finite,  from the Lebesgue dominated convergence theorem, we know that $(f_\eta/\phi_\eta)(X_{T})$ converges to $(f/\phi)(X_{T})$ in $L^q$ as $\eta\to0.$ Thus, it suffices to show that $H_T^\epsilon$ converges to $H_T$ in $L^p$ as $\epsilon\to0.$		
	This convergence can be obtained by showing that
	\begin{align}
	&Z_T^\epsilon \int_{0}^{T}\overline{k}_\epsilon(X_{s})\, dW_{s}  \to \int_{0}^{T}\overline{k}(X_{s})\, dW_{s}\label{eqn:one}\,,\\
	&\epsilon Z_T^\epsilon \int_0^T(\overline{k}_\epsilon\ell_\epsilon)(X_s)\,ds \to 0\label{eqn:onea}
	\end{align}	
	in $L^p$ as $\epsilon\to0.$

	We prove Eq.\eqref{eqn:one}. Choose a sufficiently  large  positive even integer $m$ such that $\frac{1}{p+\epsilon_1}+\frac{1}{m}<\frac{1}{p}.$ 
	From the generalized Holder inequality, it suffices to show that as $\epsilon\to0$ 
	\begin{equation}
	\begin{aligned}
	\int_{0}^{T}\overline{k}_\epsilon(X_{s})\, dW_{s} \to \int_{0}^{T}\overline{k}(X_{s})\, dW_{s} \quad\textnormal{ in } L^{p+\epsilon_1} 
	\end{aligned}
	\end{equation}
	and
	\begin{equation}
	\begin{aligned}
	Z_T^\epsilon\to 1 \quad\textnormal{ in } L^{m} \,.
	\end{aligned}
	\end{equation}
	The second $L^{m}$-convergence is obtained from Eq.\eqref{eqn:decompose_Z} and the fact that
	$\lim_{\epsilon\to0}\mathbb{E}_\xi^{\mathbb{Q}}[Z^{\epsilon\,j}_T]=1$ 
	which was
	shown in Eq.\eqref{eqn:limit_Z}	for $j=0,1,\cdots,m$.
	For the first $L^{p+\epsilon_1}$-convergence, observe that  
	$$\mathbb{E}_\xi^\mathbb{Q}\Big[\Big|\int_0^T(\overline{k}_\epsilon-\overline{k})(X_{s})\, dW_{s}\Big|^{p+\epsilon_1}\Big]\leq c_{p+\epsilon_1}\,\mathbb{E}_\xi^\mathbb{Q} \Big[\int_0^T|\overline{k}_\epsilon-\overline{k}|^{p+\epsilon_1}(X_{s})\,ds\Big] $$
	where $c_{p+\epsilon_1}$ is the constant from the David-Burkholder-Gundy inequality.  
	From Assumption \eqref{eqn:g_condi}, the Lebesgue dominated convergence theorem says
	$$\mathbb{E}_\xi^\mathbb{Q} \Big[\int_0^T|\overline{k}_\epsilon-\overline{k}|^{p+\epsilon_1}(X_{s})\,ds\Big] \to 0\;\;\textnormal{ as } \epsilon\to 0\,.$$

	Now we prove Eq.\eqref{eqn:onea}. 
	It is enough to show that $Z_T^\epsilon \int_0^T(\overline{k}_\epsilon\ell_\epsilon)(X_s)\,ds $ is uniformly  bounded in $\epsilon$ on $I$ in $L^p.$ 
	This is achieved from  
	\begin{equation}
	\begin{aligned}
	\mathbb{E}_\xi^\mathbb{Q}\Big[\Big|Z_T^\epsilon\int_0^T(\overline{k}_\epsilon\ell_\epsilon)(X_s)\,ds  \Big|^p\Big]
	&\leq \mathbb{E}_\xi^\mathbb{Q}\Big[Z_T^{\epsilon\,p}\Big(\int_0^Tg^2(X_s)\,ds\Big)^p  \Big]\\
	&\leq \Big(\mathbb{E}_\xi^\mathbb{Q}[Z_T^{\epsilon\,2p}]\Big)^{1/2}\,  \Big(\mathbb{E}_\xi^\mathbb{Q}\Big[\Big(\int_0^Tg^2(X_s)\,ds\Big)^{2p}\Big]\Big)^{1/2}  \,.
	\end{aligned}
	\end{equation}
	The first expectation $\mathbb{E}_\xi^\mathbb{Q}[Z_T^{\epsilon\,2p}]$ is uniformly bounded in $\epsilon$ on $I$ by constant $(\mathbb{E}_\xi^{\mathbb{Q}}[e^{\epsilon_0\int_0^Tg^2(X_s)\,ds}])^{\frac{1}{2}}$ by  using the same method as
	in Eq.\eqref{eqn:UB_Z}. 
	The second expectation $\mathbb{E}_\xi^\mathbb{Q}[(\int_0^Tg^2(X_s)\,ds)^{2p}]$ is finite since  	$\mathbb{E}_\xi^{\mathbb{Q}}[e^{\epsilon_0\int_0^Tg^2(X_s)\,ds}]$ is finite
	by Assumption \eqref{eqn:expo_condi}.  
\end{proof}

The following proposition says that the exponential growth rate of the expectation of $e^{Y_T}$ in time $T$ guarantees the $T^p$-order growth rate of the  expectation of $Y_T^p$ in time $T$ for any positive constant $p.$

\begin{proposition}\label{prop:expo_p_th_power_inequ}
	Let $(Y_t)_{t\geq0}$ be a non-negative stochastic process and $p$ be a positive constant. Suppose that  there are positive constants $a$ and $c$ such that for all $T>0,$
	$$\mathbb{E}[e^{Y_T}]\leq c\,e^{aT}\,.$$  Then there exists a  positive constant $d$ such that   $$\mathbb{E}[Y_T^p]\leq d\,T^p$$
	for all sufficiently large $T>0.$   
\end{proposition}

\begin{proof}
	It suffices to show that
	$\limsup_{T\rightarrow\infty}\frac{\mathbb{E}[Y_T^p]}{T^p}$ is finite. Suppose that the $\limsup$ is divergent to infinity. There exists a sequence $\{T_n\}_{n=1}^\infty$ such that  $T_n\to\infty$ and  
	$b_n:= \frac{\mathbb{E}[Y_{T_n}^p]}{T_n^p}\to\infty$
	as $n\rightarrow\infty.$
	Let $\hat{p}$ be a non-negative integer such that $\hat{p}<p\leq\hat{p}+1.$
	For any non-negative random variable $Y,$ we know that
	\begin{equation*}
	\begin{aligned}
	\mathbb{E}[e^{Y}]
	&=\sum_{j=0}^\infty\frac{\mathbb{E}[Y^j]}{j!}
	\geq\sum_{j=\hat{p}+1}^\infty\frac{\mathbb{E}[Y^j]}{j!}
	\geq\sum_{j=\hat{p}+1}^\infty\frac{(\mathbb{E}[Y^p])^{\frac{j}{p}}}{j!}\\
	&=\sum_{j=0}^\infty\frac{(\mathbb{E}[Y^p])^{\frac{j}{p}}}{j!}-\sum_{j=0}^{\hat{p}}\frac{(\mathbb{E}[Y^p])^{\frac{j}{p}}}{j!}=e^{(\mathbb{E}[Y^p])^{\frac{1}{p}}}-\sum_{j=0}^{\hat{p}}\frac{(\mathbb{E}[Y^p])^{\frac{j}{p}}}{j!} \;.	
	\end{aligned}
	\end{equation*}
	Here, we used the Taylor expansion and the Jensen inequality. Substituting $Y=Y_{T_n},$ because 
	$\mathbb{E}[Y_{T_n}^p]\to\infty$ as $n\to\infty$ and 
	the exponential growth rate is faster than the polynomial growth rate,   it follows that
	$$\mathbb{E}[e^{Y_{T_n}}]\geq e^{(\mathbb{E}[Y_{T_n}^p])^{\frac{1}{p}}}-\sum_{j=0}^{\hat{p}}\frac{(\mathbb{E}[Y_{T_n}^p])^{\frac{j}{p}}}{j!}\geq\frac{1}{\,2\,}e^{(\mathbb{E}[Y_{T_n}^p])^{\frac{1}{p}}}=\frac{1}{\,2\,}e^{b_n^{\,\frac{1}{p}}  T_n}$$
	for sufficiently large $n.$
	From the assumption, we obtain 
	$$c\,e^{aT_n}\geq \mathbb{E}[e^{Y_{T_n}}]\geq\frac{1}{\,2\,}e^{b_n^{\,\frac{1}{p}}  T_n}$$
	which is a contradiction because $\lim_{n\rightarrow\infty}b_n=\infty.$  
\end{proof}

\begin{proof}
	We now prove Theorem \ref{thm:rho_expo_condi}.   
	By Proposition \ref{prop:rho}, the existence and 
	the continuity of the partial derivative are directly obtained. From Eq.\eqref{eqn:rho_Fournie}, it suffices to show that
	$$\lim_{T\to\infty}\frac{1}{T} \,\mathbb{E}_\xi^{\mathbb{Q}}
	\Big[(f/\phi)(X_{T})\int_{0}^{T}\overline{k}(X_s)\, dW_s\Big]=0\,.$$	
	Since the growth rate of $\mathbb{E}_\xi^{\mathbb{Q}}[e^{\epsilon_0\int_0^Tg^2(X_s)\,ds}]$ is less than or equal to the exponential rate,   Proposition \ref{prop:expo_p_th_power_inequ} says that  the growth rate of 
	$\mathbb{E}_\xi^{\mathbb{Q}}[(\int_{0}^{T}g^2(X_s)\, ds)^{\frac{p}{2}}]$ is less than or equal to the $T^{\frac{p}{2}}$-order growth rate. Thus, there is a constant $d_p$ depending on $p$ but not on $T$ such that
	\begin{equation}
	\begin{aligned}
	\Big(\mathbb{E}_\xi^{\mathbb{Q}}\Big[\Big(\int_{0}^{T}|\overline{k}|^2(X_s)\,ds\Big)^{\frac{p}{2}}\Big]\Big)^{\frac{2}{p}}
	&\leq\Big(\mathbb{E}_\xi^{\mathbb{Q}}\Big[\Big(\int_{0}^{T}g^2(X_s)\, ds\Big)^{\frac{p}{2}}\Big]\Big)^{\frac{2}{p}}\leq d_pT 
	\end{aligned}
	\end{equation} 
	for sufficiently large $T.$
	Using the Holder inequality and the Burkholder-Davis-Gundy inequality, it follows that
	\begin{equation}\label{eqn:q_order_growth}
	\begin{aligned}
	&\frac{1}{T}\mathbb{E}_\xi^{\mathbb{Q}}
	\Big[\Big|(f/\phi)(X_{T})\int_{0}^{T}\overline{k}(X_s)\, dW_s\Big|\Big]\\
	\leq\;&\frac{1}{T}\Big(\mathbb{E}_\xi^{\mathbb{Q}} 
	[(f/\phi)^{q}(X_{T})] \Big)^{\frac{1}{q}}\,\Big(\mathbb{E}_\xi^{\mathbb{Q}} \Big[\Big|\int_{0}^{T}\overline{k}(X_s)\, dW_s\Big|^{p}\Big]\Big)^{\frac{1}{p}}\\
	\leq\;&\frac{c_q}{T}\Big(\mathbb{E}_\xi^{\mathbb{Q}} 
	[(f/\phi)^{q}(X_{T})] \Big)^{\frac{1}{q}}\,\Big(\mathbb{E}_\xi^{\mathbb{Q}}\Big[\Big(\int_{0}^{T}|\overline{k}|^2(X_s)\, ds\Big)^{\frac{p}{2}}\Big]\Big)^{\frac{1}{p}}\\
	\leq\;&\frac{c_q}{T}\Big(\mathbb{E}_\xi^{\mathbb{Q}} 
	[(f/\phi)^{q}(X_{T})] \Big)^{\frac{1}{q}}\,(d_pT)^{\frac{1}{2}}\\	
	=\;&\frac{c_qd_q^{\frac{1}{2}}}{T^{\frac{1}{2}}}\Big(\mathbb{E}_\xi^{\mathbb{Q}} 
	[(f/\phi)^{q}(X_{T})] \Big)^{\frac{1}{q}} \rightarrow0
	\end{aligned}
	\end{equation}
	as $T\rightarrow\infty$ because $\mathbb{E}_\xi^{\mathbb{Q}}[(f/\phi)^q(X_T)]$ is uniformly bounded in $T$ on $[0,\infty).$ 
	This completes the proof.
\end{proof}

\section{Proof of Theorem \ref{thm:expo_rho}}
\label{app:pf_rho_thm}

\begin{proof}
	Let $c_1$ be a positive constant such that 
	$\mathbb{E}_\xi^{\mathbb{Q}}[e^{\epsilon_0\,g^2(X_T)}]\leq c_1$ for all $T\geq 0.$
	Replacing $q$ by a sufficiently small positive number, we may assume that  
	$1<q\leq 2,$ and that a constant $p$ defined by $1/p+1/q=1$ (thus, $p\ge2$) is an even integer.

	For a fixed $T>0,$ we first show that 
	the conditions of Proposition \ref{prop:rho} are satisfied with these constants $p$ and $q.$ Since Eq.\eqref{eqn:psi_condi} is already assumed to hold, it remains to prove Eq.\eqref{eqn:expo_condi} and Eq.\eqref{eqn:g_condi}.
	The expectation
	$\mathbb{E}_\xi^{\mathbb{Q}}[e^{\frac{1}{T}\int_0^T\epsilon_0\,g^2(X_s)\,ds}]$
	is uniformly bounded in $T$ on $[0,\infty)$	because
	\begin{equation} \label{eqn:expo_rate} 
	\begin{aligned}
	\mathbb{E}_\xi^{\mathbb{Q}}[e^{\frac{1}{T}\int_0^T\epsilon_0\,g^2(X_s)\,ds}]
	\leq 	 \mathbb{E}_\xi^{\mathbb{Q}}\Big[\frac{1}{T}\int_0^T e^{\epsilon_0\,g^2(X_s)}ds\Big]=\frac{1}{T}\int_0^T\mathbb{E}_\xi^{\mathbb{Q}}[e^{\epsilon_0\,g^2(X_s)}]\,ds\leq c_1\,.
	\end{aligned}
	\end{equation}
	Replacing $\epsilon_0/T$ by $\epsilon_0,$  Eq.\eqref{eqn:expo_condi} is satisfied.
	For Eq.\eqref{eqn:g_condi}, observe that  for any $n\in\mathbb{N}$ such that $2n>p+1,$ the expectation $\mathbb{E}_\xi^{\mathbb{Q}}[\int_0^Tg^{2n}(X_s)\,ds]$ 
	is finite since
	\begin{equation}\label{eqn:liner_growth}
	\mathbb{E}_\xi^{\mathbb{Q}}\Big[\int_0^T(\epsilon_0g^{2})^n(X_s)\,ds\Big]\leq \mathbb{E}_\xi^{\mathbb{Q}}\Big[\int_0^Tn! e^{\epsilon_0\,g^2(X_s)}\,ds\Big]\leq Tn!\,c_1\,. 
	\end{equation} 
	Thus, the expectation	
	$\mathbb{E}_\xi^{\mathbb{Q}}[\int_0^Tg^{p+1}(X_s)\,ds]$ is also finite.	
	All conditions of Proposition  \ref{prop:rho} are satisfied. 
	
	Therefore,  the partial derivative   $\frac{\partial }{\partial\epsilon}\mathbb{E}_\xi^{\mathbb{Q_{\epsilon}}}[(f_\eta/\phi_\eta)(X_{T}^{\epsilon})]$  exists and is continuous in $(\eta,\epsilon)$ on $I^2.$ Moreover,     
	\begin{equation} \label{eqn:app_Fournie}
	\frac{\partial}{\partial \epsilon}\;\Big|_{\epsilon =0}\mathbb{E}_\xi^{\mathbb{Q_{\epsilon}}}
	[(f/\phi)(X_{T}^{\epsilon})] =\mathbb{E}_\xi^{\mathbb{Q}}
	\Big[(f/\phi)(X_{T})\int_{0}^{T}\overline{k}(X_{s})\, dW_{s}\Big] \,.
	\end{equation}	
	Now it remains to show that 
	$$\frac{1}{T} \frac{\partial}{\partial \epsilon}\;\Big|_{\epsilon =0}\mathbb{E}_\xi^{\mathbb{Q_{\epsilon}}}
	[(f/\phi)(X_{T}^{\epsilon})] =\frac{1}{T}\mathbb{E}_\xi^{\mathbb{Q}}
	\Big[(f/\phi)(X_{T})\int_{0}^{T}\overline{k}(X_{s})\, dW_{s}\Big] \rightarrow0$$   as $T\rightarrow\infty.$ From Eq.\eqref{eqn:q_order_growth}, it is enough to prove that the growth rate of
	$\mathbb{E}_\xi^{\mathbb{Q}}[(\int_{0}^{T}g^2(X_s)\, ds)^{\frac{p}{2}}]$ is less than or  equal to the order of   $T^{\frac{p}{2}}$ as $T\to\infty.$  Recall that $p$ is a positive even integer.
	Using  Eq.\eqref{eqn:expo_rate} and the inequality $x^{\frac{p}{2}}\leq (p/2)!\,e^{x}$ for $x>0,$ it follows that 
	$$\mathbb{E}_\xi^{\mathbb{Q}}\Big[\Big(\frac{1}{T}\int_{0}^{T}\epsilon_0g^2(X_{t})\,dt\Big)^{\frac{p}{2}}\Big]\leq  (p/2)!\, \mathbb{E}_\xi^{\mathbb{Q}}[e^{\frac{1}{T}\int_0^T\epsilon_0\,g^2(X_s)\,ds}] \leq   (p/2)!\,c_1\,,$$
	which gives the desired result. This completes the proof. 
\end{proof}

\section{Proof of Proposition \ref{prop:Lamp_equiv}}
\label{app:Lamp_equiv}

This section provides the proof of Proposition \ref{prop:Lamp_equiv}.
The pricing operator $\mathcal{P}^\epsilon$ in Eq.\eqref{eqn:obj} is
$$\mathcal{P}_T^\epsilon f(x)=\mathbb{E}_{X_0^\epsilon=x}^{\mathbb{P}}[e^{-\int_0^T r_\epsilon(X_s^\epsilon)\,ds} f(X_T^\epsilon)]\,.$$
Define an operator $\mathcal{P}_T^{\textnormal{Lamp}\,\epsilon}$ corresponding to the Lamperti transformation by
$$\mathcal{P}_T^{\textnormal{Lamp}\,\epsilon}F(\zeta)=
\mathbb{E}_{U_0^\epsilon=\zeta}^\mathbb{P}[e^{-\int_0^TR_\epsilon(U_s^\epsilon)\,ds}F(U_T^\epsilon)]\,.$$

\begin{lemma}\label{lem:Lamperti}  
	Let $\beta$ be a real number and $h$ be a positive function on $\mathbb{R}.$ The following statements are equivalent.
	\begin{itemize}
		\item The pair $(e^{-\beta T},h)$ is an eigenpair of $\mathcal{P}_T^\epsilon,$ that is,
		$$\mathcal{P}_T^\epsilon h(x)=e^{-\beta T}h(x)\,,\;x\in\mathbb{R}\,.$$ 
		\item The pair $(e^{-\beta T},h\circ v_\epsilon)$ is an eigenpair of $\mathcal{P}_T^{\textnormal{Lamp}\,\epsilon},$ that is, 	$$\mathcal{P}_T^{\textnormal{Lamp}\,\epsilon} (h\circ v_\epsilon)(\zeta)=e^{-\beta T}(h\circ v_\epsilon)(\zeta)\,,\;\zeta\in\mathbb{R}\,.$$ 
	\end{itemize} 
\end{lemma}
\begin{proof}
	The proof is straightforward from 
	\begin{equation}
	\begin{aligned}
	\mathcal{P}_T^\epsilon h(x)
	&=\mathbb{E}_{X_0^\epsilon=x}^{\mathbb{P}}[e^{-\int_0^T r_\epsilon(X_s^\epsilon)\,ds} h(X_T^\epsilon)]\\
	&=\mathbb{E}_{U_0^\epsilon=u_\epsilon(x)}^{\mathbb{P}}[e^{-\int_{0}^{T}R_\epsilon(U_{s}^{\epsilon})ds}\,(h\circ v_\epsilon)(U_{T}^{\epsilon})]=\mathcal{P}_T^{\textnormal{Lamp}\,\epsilon}(h\circ v_\epsilon)(\zeta) 
	\end{aligned}
	\end{equation}
	and $\zeta=u_\epsilon(x).$
\end{proof}

\begin{proof} 
	We now show Proposition \ref{prop:Lamp_equiv}.
	The `only-if' condition will be proven. The `if' condition can be shown in a similar way, so we omit it.
	Assume that the quadruple $(b_\epsilon,\sigma_\epsilon,r_\epsilon,f_\epsilon
	)$ and $\xi$ satisfy Assumptions \ref{assume:pert_defi} - \ref{assume:perturbed}.  Define 
	$X^\epsilon,\mathcal{P}^\epsilon , M^\epsilon,\mathbb{Q}^\epsilon,(\lambda_\epsilon,\phi_\epsilon),\varphi_\epsilon$ accordingly. It is easy to check  that the 
	quadruple $(\delta_\epsilon,1,R_\epsilon,F_\epsilon)$ satisfies Assumption \ref{assume:pert_defi}.
	We show that  the 
	quadruple $(\delta_\epsilon,1,R_\epsilon,F_\epsilon)$ and $\zeta_\epsilon$ satisfy Assumption \ref{assume:perturbed} (that is, Assumptions \ref{assume:SDE} - \ref{assume:f} and \ref{assume:HS} - \ref{assume:ergodic}).
	Assumption \ref{assume:SDE} is satisfied because $U^\epsilon$ defined in Eq.\eqref{eqn:Lamperti} is a  strong solution of the SDE \eqref{eqn:Lamperti_transformed_SDE}, and because
	the strong solution  $X^\epsilon$ of SDE
	\eqref{eqn:X_ep_Lamp} (thus, the strong solution $U^\epsilon$ of SDE \eqref{eqn:Lamperti_transformed_SDE}) is unique and  non-explosive.
	Assumptions \ref{assume:r} and \ref{assume:f} are trivial. 
	For Assumption \ref{assume:HS}, we observe that 
	the pair $(e^{-\lambda_\epsilon T},\phi_\epsilon\circ v_\epsilon)$ is an eigenpair of $\mathcal{P}_T^{\textnormal{Lamp}\,\epsilon}$
	by Lemma \ref{lem:Lamperti}. 
	From Eq.\eqref{eqn:M}, it follows  that the corresponding martingale process is 
	\begin{equation}
	\begin{aligned}
	M_{t}^{\textnormal{Lamp}\,\epsilon}:=&\,e^{\lambda_\epsilon t-\int_{0}^{t} R_\epsilon(U_s^\epsilon)\,ds}\,\frac{(\phi_\epsilon\circ v_\epsilon)(U_t^\epsilon)}{(\phi_\epsilon\circ v_\epsilon)(\zeta_\epsilon)}  
	=\,e^{\lambda_\epsilon t-\int_{0}^{t} r_\epsilon(X_s^\epsilon)\,ds}\,\frac{\phi_\epsilon(X_t^\epsilon)}{\phi_\epsilon(\xi)} =M_t^\epsilon\,,\;0\leq t\leq T\,.
	\end{aligned}
	\end{equation}
	The recurrent eigen-measure $\mathbb{Q}^{\textnormal{Lamp}\,\epsilon}$ 
	defined by this martingale $	M^{\textnormal{Lamp}\,\epsilon}$ satisfies
	$$\frac{d\mathbb{Q}^{\textnormal{Lamp}}_{\epsilon}}{d\mathbb{P}^{\quad\;}}\Big|_{\mathcal{F}_T}=M_T^{\textnormal{Lamp}\,\epsilon}=M_T^\epsilon=\frac{d\mathbb{Q}^\epsilon}{d\mathbb{P}\,}\Big|_{\mathcal{F}_T}\;.$$
	Thus, $\mathbb{Q}^{\textnormal{Lamp}\,\epsilon}=\mathbb{Q}^\epsilon.$ It follows that  Assumption \ref{assume:HS} is satisfied because the recurrence of $X^\epsilon$ implies the recurrence of $U^\epsilon.$ Assumption \ref{assume:diff_phi} is clearly satisfied since 
	the recurrent eigenfunction   $\phi_\epsilon\circ v_\epsilon$  
	is twice continuously differentiable.  Assumptions \ref{assume:inv} and  \ref{assume:ergodic} are directly obtained from $\mathbb{Q}^{\textnormal{Lamp}\,\epsilon}=\mathbb{Q}^\epsilon.$ By Lemma \ref{lem:Lamperti} and the above argument, the two corresponding recurrent eigenvalues coincide.
\end{proof}

\section{The CIR model}
\label{app:CIR}

\subsection{Hansen--Scheinkman decomposition}
\label{app:HS_CIR}

First, we show that $(b,\sigma,r,f)$ and $\xi$ satisfy Assumptions \ref{assume:pert_defi} - \ref{assume:perturbed} (that is, Assumptions \ref{assume:SDE} - \ref{assume:f} and \ref{assume:HS} - \ref{assume:ergodic}). 
We only prove Assumptions \ref{assume:HS} - \ref{assume:ergodic}; the others are trivial.
It can be shown that
a pair 
$(\lambda,\phi(x)):=(\theta\kappa,e^{-\kappa x})$
is the recurrent eigenpair, where
$\kappa:=\frac{\sqrt{a^{2}+2\sigma^{2}}-a}{\sigma^{2}}$ (Section 6.1.1 in \cite{qin2016positive}).
This proves Assumptions \ref{assume:HS} - \ref{assume:diff_phi}.
Consider the recurrent eigen-measure $\mathbb{Q}.$
The corresponding Girsanov kernel is
$\varphi(X_t)=-\sigma \kappa\sqrt{X_t}$, and
the $\mathbb{Q}$-dynamics of $X$ is
\begin{equation}\label{eqn:CIR_under_Q}
dX_t= (
\theta-\sqrt{a^{2}+2\sigma^{2}}\,X_t)dt + \sigma \sqrt{X_t}\, dW_{t}\;,\,X_0=\xi\,. 
\end{equation}
Here, $W$ is a $\mathbb{Q}$-Brownian motion.
This process is a re-parameterized CIR model.
It is well known that the CIR model has an invariant distribution $\nu$, which implies Assumption \ref{assume:inv}.
For convenience, we define $b:=\sqrt{a^2+2\sigma^2}.$

To show Assumption \ref{assume:ergodic}, consider the $\mathbb{Q}$-density function $\ell(x;t)$ of $X_t$
$$\ell(x;t):=h_t\,e^{-u-v}\Big(\frac{v}{u}\Big)^{q/2}I_q(2\sqrt{uv}),$$
where $I_q$ is the modified Bessel function of the first type of order $q$ and 
$$h_t=\frac{2b}{\sigma^2(1-e^{-bt})}\,,\;q=\frac{2\theta}{\sigma^2}-1\,,\;
u=h_t\xi e^{-bt}\,,\;v=h_tx\,.$$
After slightly rewriting, we find
\begin{equation}\label{eqn:CIR_density_Re}
\ell(x;t)=k_t\,h_t\,e^{-h_tx}x^{q/2}I_q(2h_te^{-bt/2}\sqrt{\xi x})\,.
\end{equation}
Here, $k_t=e^{-h_t\xi e^{-bt}}(\xi e^{-bt})^{-q/2}$ and
\begin{equation}\label{eqn:I_bdd}
I_q(z)=\frac{(z/2)^q}{\pi^{1/2}\,\Gamma(q+1/2)}\int_0^\pi (e^{z\cos u}\sin^{2q}u)\,du\leq
\frac{\pi^{1/2}(z/2)^qe^z}{\Gamma(q+1/2)}
\;.
\end{equation}
For large $t>0,$ we have
$\ell(x;t)\leq B\,e^{-h_tx} x^{q}e^{2h_t\sqrt{\xi r}}$
for a positive constant $B.$
To show Assumption \ref{assume:ergodic} for function $f$, whose growth rate is less than or equal to the polynomial growth rate, it suffices to prove
Eq.\eqref{eqn:CIR_inv} below for a constant $c>\kappa$ because the growth rate of $(f/\phi)(x)=f(x)e^{\kappa x}$ is less than the growth rate of $e^{cx}$ as $x\to\infty.$
Choose a constant $c$ such that $\kappa=\frac{b-a}{\sigma^{2}}<c<\frac{2b}{\sigma^2}.$
Because 
$\frac{2b}{\sigma^2}<h_t,$ we know that $e^{cx}\ell(x;t)$ is dominated by 
$Be^{(c-\frac{2b}{\sigma^2})x} x^{q}e^{2h_1\sqrt{\xi x}},$
whose integration over $(0,\infty)$ is finite.
By the Lebesgue dominated convergence theorem, it follows that  
\begin{equation}\label{eqn:CIR_inv}
\mathbb{E}_\xi^{\mathbb{Q}}
[e^{cX_t}]=\int_0^\infty e^{cx}\ell(x;t)\,ds\to\int_0^\infty e^{cx}\ell(x;\infty)\,dx=\int e^{cx}\,d\nu(x),
\end{equation} 
where $\ell(x;\infty)=\lim_{t\rightarrow\infty} \ell(x;t),$ which is equal to the invariant density function of $X$ under $\mathbb{Q}.$
For more details regarding the density of the CIR model, see page 19 in \cite{benth2005pde}.

In summary, 
we showed that  the quadruple of functions $(b,\sigma,r,f)$ and the initial value $\xi,$ which were defined in Section \ref{sec:CIR}, satisfy Assumptions \ref{assume:pert_defi} and \ref{assume:perturbed}.
From now, in the context of the CIR model, the 
notations 
$X,$ $\mathcal{P},$ $\mathcal{L},$ $M,$ $\mathbb{Q},$ $(\lambda,\phi),$  $\varphi,$ $\nu$
are self-explanatory.

\subsection{Sensitivity of $\xi$}
\label{app:CIR_delta}

In this section, we show the long-term sensitivity of $\xi$ 
in Eq.\eqref{eqn:sen_CIR}.
By Theorem \ref{thm:delta}, it suffices to show that
$\mathbb{E}_{\xi}^{\mathbb{Q}}[(f/\phi)(X_T)]$ is continuously differentiable in $\xi,$ and $\frac{\partial}{\partial \xi}\mathbb{E}_{\xi}^{\mathbb{Q}}[(f/\phi)(X_T)]\rightarrow0$ as $T\rightarrow\infty.$  The continuous differentiability and the convergence to zero can be easily justified by observing the $\mathbb{Q}$-density function $\ell(x;T)$  of $X_T.$
Indeed,
\begin{equation}\label{eqn:pf_xi_CIR}
\begin{aligned}
\lim_{T\rightarrow\infty}\frac{\partial}{\partial \xi}\mathbb{E}_{\xi}^{\mathbb{Q}}[(f/\phi)(X_{T})]
&=\lim_{T\rightarrow\infty}\frac{\partial}{\partial \xi}\int_0^\infty (f/\phi)(x) \,\ell(x;T)\,dx\\
&=\lim_{T\rightarrow\infty}\int_0^\infty (f/\phi)(x)\,\frac{\partial \ell(x;T)}{\partial \xi}\,dx\\
&=\int_0^\infty (f/\phi)(x)\,\lim_{T\rightarrow\infty}\frac{\partial \ell(x;T)}{\partial\xi}\,dx=0\,.
\end{aligned}
\end{equation}
The interchange of the differentiation and the integration in the second equality can be checked by the standard argument.
For the last equality, we used the following lemma.

\begin{lemma}
	Let $\ell(x;t)$ be the $\mathbb{Q}$-density function of $X_t$ given in Eq.\eqref{eqn:CIR_density_Re}. Then,
	$$\lim_{t\rightarrow\infty}\frac{\partial \ell(x;t)}{\partial \xi}=0\,.$$  
\end{lemma}

\begin{proof}
	From the  $\mathbb{Q}$-density function 
	$$\ell(x;t)=e^{-h_t\xi e^{-bt}}(\xi e^{-bt})^{-q/2}\,h_t\,e^{-h_tx}x^{q/2}I_q(2h_te^{-bt/2}\sqrt{\xi x})\,,$$
	we have 
	\begin{equation}\label{eqn:deri_CIR}
	\begin{aligned}
	\frac{\partial \ell(x;t)}{\partial\xi}
	=\Big(-h_te^{-bt}-\frac{q}{2\xi}+\frac{1}{2}\sqrt{\frac{x}{\xi}}\,h_te^{-bt/2}\frac{I_{q-1}(z)+I_{q+1}(z)}{I_q(z)}\Big)\ell(x;t)
	\end{aligned}
	\end{equation}
	where $z=2h_te^{-bt/2}\sqrt{\xi x}.$  
	Here, we used the equality $I_q'(\cdot)=\frac{1}{2}(I_{q-1}(\cdot)+I_{q+1}(\cdot)).$	
	Observe that $z\rightarrow0$ as $t\rightarrow\infty.$
	The modified Bessel function $I_q$ of order $q$ satisfies 
	$\lim_{z\rightarrow0}\frac{I_q(z)}{\;\frac{(z/2)^q}{\Gamma(q+1)}\;}=1,$ thus 	
	$$\lim_{t\rightarrow\infty}h_te^{-bt/2}\frac{I_{q-1}(z)+I_{q+1}(z)}{I_q(z)}
	=\lim_{t\rightarrow\infty}h_te^{-bt/2}\frac{\frac{(h_te^{-bt/2}\sqrt{\xi x})^{q-1}}{\Gamma(q)}+\frac{(h_te^{-bt/2}\sqrt{\xi x})^{q+1}}{\Gamma(q+2)}}{\frac{(h_te^{-bt/2}\sqrt{\xi x})^q}{\Gamma(q+1)}}=\frac{q}{\sqrt{\xi x}}\,.$$	
	In conclusion,  we have  	$\lim_{t\rightarrow\infty}\frac{\partial \ell(x;t)}{\partial \xi}=0$ from Eq.\eqref{eqn:deri_CIR}.
\end{proof}

\subsection{Sensitivity of $\theta$}
\label{sec:sen_theta_CIR}
In this section, we analyze the long-term sensitivity of $\theta$ in the CIR model.
As discussed in Eq.\eqref{eqn:dtheta} in Section \ref{sec:CIR}, 
the parameter $\theta$ can be regarded as a perturbation parameter.
The aim is to show the long-term sensitivity of $\theta$ 
in Eq.\eqref{eqn:sen_CIR} 
using Corollary \ref{cor:rho}.
Only the hypothesis of Theorem \ref{thm:rho_expo_condi} will be checked because the other conditions are easy to prove.
Recall the definitions of the functions $k$ and $g$ in Section \ref{sec:rho} 
\begin{equation}\label{eqn:k_CIR}
k(x)=\frac{\theta}{\sigma \sqrt{|x|}}-\frac{\sqrt{a^{2}+2\sigma^{2}}}{\sigma}\,\sqrt{|x|}\,,\;g(x)=\frac{1}{\sigma\sqrt{|x|}}\,. 
\end{equation} 
Condition (i) in Theorem \ref{thm:rho_expo_condi} can be proven by Proposition \ref{prop:CIR_expo} below.
For (ii) and (iii), we set $p=q=2.$
Let $\epsilon_1$ be a positive number such that $\epsilon_1<\frac{2\theta}{\sigma^2}-1.$
By the same method as in Eq.\eqref{eqn:CIR_inv}, it follows that  
\begin{equation}\label{eqn:CIR_theta_ii}
\begin{aligned}
\mathbb{E}_\xi^\mathbb{Q}[g^{2+\epsilon_1}(X_t)]
&=\mathbb{E}_\xi^{\mathbb{Q}}\Big[\Big(\frac{1}{\sigma\sqrt{X_t}}\Big)^{2+\epsilon_1}\Big]
=\int_0^\infty \Big(\frac{1}{\sigma\sqrt{x}}\Big)^{2+\epsilon_1}\ell(x;t)\,dx
\end{aligned}
\end{equation}
is uniformly bounded in $t$ on $[0,\infty).$
Thus, $\mathbb{E}_\xi^\mathbb{Q}[\int_0^Tg^{2+\epsilon_1}(X_t)\,dt]$ is finite for each $T$, which means that (ii) is satisfied. For (iii),
we put $\psi=f/\phi$ since $f$ and $\phi$  are independent of $\theta.$  
Using the method in Eq.\eqref{eqn:CIR_inv},
we find that
\begin{equation}\label{eqn:f_phi_conv}
\mathbb{E}_\xi^{\mathbb{Q}}[(f/\phi)^2(X_T)]\to\int(f/\phi)^2(x)\,d\nu(x)
\end{equation}
as $T\to\infty$   since the exponential growth rate of $(f/\phi)^2(x)$ is ${2(b-a)x}/{\sigma^{2}}$ as $x\to\infty.$ 

\begin{proposition}	\label{prop:CIR_expo}
	For any $\epsilon_0$ with $0<\epsilon_0\leq \frac{1}{2}(\frac{\sigma}{2}-\frac{\theta}{\sigma})^2,$ there exist constants $a$ and $c$ such that for all $T>0$ 
	$$\mathbb{E}_\xi^{\mathbb{Q}}[e^{\epsilon_0\int_0^T(1/X_s)\,ds}]
	\leq c \,e^{aT}\,.$$ 
\end{proposition}
\begin{proof}
	We modify the proof found in  Appendix C in \cite{ahn1999parametric}.
	Their proof evaluates the above expectation under the condition $\epsilon_0<0$, whereas our proof evaluates the expectation under the condition  $\epsilon_0>0.$ 
	Our proof is given in several steps. 
	\begin{itemize}
		\item[(i)] 	Let $Y:=1/X.$ We find a positive function $V(y,t)$ on $\mathbb{R}^+\times [0,T] $
		such that $$V(Y_t,t)e^{\epsilon_0\int_0^tY_s\,ds}\,,\;0\leq t\leq T$$
		is a local martingale and $V(y,T)$ is a constant, independent of $y$ and  $T.$
		\item[(ii)]  Show that the function satisfies	$$V(y,0)\leq c_1  e^{-\gamma b T}$$ 
		for  positive constants $c_1,$ $\gamma,$ $b.$ In other words, the decay rate of the function $V(y,0)$ is less than or equal to an exponential rate in $T.$   
		\item[(iii)] 	Because $V(Y_t,t)e^{\epsilon_0\int_0^tY_s\,ds}$
		is a positive local martingale for $0\leq t\leq T,$ it is a supermartingale. Thus, we have	
		\begin{equation*}
		\begin{aligned}
		c_1 e^{-a T}\geq V(Y_0,0)
		&\geq\mathbb{E}_\xi^\mathbb{Q}[V(Y_T,T)e^{\epsilon_0\int_0^TY_s\,ds}]\geq\textnormal{(constant)}\,\mathbb{E}_\xi^\mathbb{Q}[e^{\epsilon_0\int_0^TY_s\,ds}]\,,
		\end{aligned}
		\end{equation*}
		which is the desired result.\newline
	\end{itemize}
	
	\noindent {\em Step} (i).
	From Eq.\eqref{eqn:CIR_under_Q}, the $\mathbb{Q}$-dynamics of $X$ is
	$dX_{t}=(
	\theta-bX_{t})dt + \sigma \sqrt{X_{t}}\, dW_{t}$
	where $b=\sqrt{a^{2}+2\sigma^{2}}.$
	Define $Y:=1/X.$ The Ito formula yields 	$$dY_t=((\sigma^2-\theta)Y_t+b)Y_t\,dt-\sigma Y_t^{3/2}\,dW_t\,.$$	
	We find a positive function $V(y,t)$ on $\mathbb{R}^+\times [0,T] $
	such that $$V(Y_t,t)e^{\epsilon_0\int_0^tY_s\,ds},\;0\leq t\leq T$$
	is a local martingale and $V(y,T)$ is a constant, independent of $y$ and  $T.$
	Such a function $V(y,t)$ satisfies
	\begin{equation}\label{eqn:FK}
	V_t+\frac{1}{2}\sigma^2x^3V_{xx}+((\sigma^2-\theta)x+b)xV_x+\epsilon_0xV=0\,.
	\end{equation}
	We make the Ansatz $V(y,t)=f(x)x^\gamma$ where $x=a(t)/y.$ 
	\begin{equation*}
	\begin{aligned}
	&V_y=-\frac{1}{a(t)}f'(x)x^{\gamma+2}-\frac{\gamma}{a(t)}f(x)x^{\gamma+1}\,,\\
	&V_{yy}=\frac{1}{a^2(t)}f''(x)x^{\gamma+4}+\frac{2(\gamma+1)}{a^2(t)}f'(x)x^{\gamma+3}+\frac{\gamma(\gamma+1)}{a^2(t)}f(x)x^{\gamma+2}\,,\\
	&V_t=\frac{a'(t)}{a(t)}f'(x)x^{\gamma+1}+\frac{a'(t)}{a(t)}\gamma f(x)x^{\gamma}\,.
	\end{aligned}
	\end{equation*}
	Then Eq.\eqref{eqn:FK} gives
	\begin{equation*}
	\begin{aligned}
	&\quad\frac{1}{2}\sigma^2a(t)x^{\gamma+1}f''(x)
	+(\frac{a'(t)}{a(t)}x^{\gamma+1}-bx^{\gamma+1}-(\sigma^2-\theta)a(t)x^{\gamma}
	+\sigma^2(\gamma+1)a(t)x^\gamma)f'(x)\\
	&+\Big(\frac{a'(t)}{a(t)}\gamma x^{\gamma}-b\gamma x^{\gamma}
	+\frac{1}{2}\sigma^2\gamma(\gamma+1)a(t)x^{\gamma-1}-(\sigma^2-\theta)\gamma a(t)x^{\gamma-1}+\epsilon_0a(t)x^{\gamma-1}\Big)f(x)=0\;.
	\end{aligned}
	\end{equation*}
	Assume that $a(t)$ and $\gamma$ satisfy
	\begin{equation}\label{eqn:two}
	\left\{\quad
	\begin{aligned}
	&\frac{a'(t)}{a(t)}-b=a(t) \\
	&\frac{1}{2}\sigma^2\gamma(\gamma+1)-(\sigma^2-\theta)\gamma+\epsilon_0=0 
	\end{aligned}\right.
	\end{equation}
	so that the above equation becomes
	$\frac{1}{2}\sigma^2xf''(x)+(x+\sigma^2\gamma+\theta)f'(x)+\gamma f(x)=0.$
	We define a new variable $z$ such that $x=-\frac{1}{2}\sigma^2z,$ and define a function $g$ as $g(z):=f(x).$ Then 	$zg''(z)+(\kappa-z)g'(z)-\gamma g(z)=0$
	where $\kappa:=2(\gamma+\frac{\theta}{\sigma^2}).$ It is well known that the standard confluent hypergeometric function
	$f(x)=g(z)=M(\gamma,\kappa;z)$ is a solution of this equation. 
	
	We now find an explicit expression for $V(y,t).$
	Eq.\eqref{eqn:two} yields
	$a(t)=\frac{b}{e^{b(T-t)}-1}$ for $0\leq t\leq T$ and $$\gamma=\frac{1}{2}-\frac{\theta}{\sigma^2}+\sqrt{\Big(\frac{1}{2}-\frac{\theta}{\sigma^2}\Big)^2-\frac{2\epsilon_0}{\sigma^2}}\,.$$
	The number $\gamma$ is a real number because of the assumption  that $0<\epsilon_0\leq \frac{1}{2}(\frac{\sigma}{2}-\frac{\theta}{\sigma})^2.$
	We know that  $\kappa=2(\gamma+\frac{\theta}{\sigma^2})>0,$ and  that  by using the Feller condition in the CIR process, $\gamma<0.$  	
	The solution $V(y,t)$ is  
	\begin{equation}\label{eqn:CIR_V}
	\begin{aligned}
	V(y,t)=f(x)x^\gamma=g(z)\Big(-\frac{1}{2}\sigma^2z\Big)^\gamma
	&=\Big(\frac{1}{2}\sigma^2\Big)^\gamma 
	M(\gamma,\kappa;z)(-z)^\gamma \\
	&=\Big(\frac{1}{2}\sigma^2\Big)^\gamma M(\kappa-\gamma,\kappa;-z)(-z)^\gamma e^z
	\end{aligned}
	\end{equation}
	where 	$z=-\frac{2x}{\sigma^2}=-\frac{2a(t)}{\sigma^2y}=-\frac{2b}{\sigma^2(e^{b(T-t)}-1)y}.$	
	Here, we used the equality  $M(\gamma,\kappa;z)=M(\kappa-\gamma,\kappa;-z)e^z.$

	We now show that
	$V(y,T)$ is a constant independent of $y$ and $T.$ 
	By direct calculation,
	\begin{equation*}
	\begin{aligned}
	\lim_{t\rightarrow T}V(y,t)
	&=\Big(\frac{1}{2}\sigma^2\Big)^\gamma\,\lim_{z\rightarrow -\infty}
	M(\kappa-\gamma,\kappa;-z)(-z)^\gamma e^z\\
	&=\Big(\frac{1}{2}\sigma^2\Big)^\gamma\,\lim_{u\rightarrow \infty}
	M(\kappa-\gamma,\kappa;u)u^\gamma e^{-u} \\
	&=\Big(\frac{1}{2}\sigma^2\Big)^\gamma\,\frac{\Gamma(\kappa)}{\Gamma(\kappa-\gamma)\Gamma(\gamma)}\,\lim_{u\rightarrow \infty}u^\gamma e^{-u} 
	\int_0^1 e^{us}s^{\kappa-\gamma-1}(1-s)^{\gamma-1}\,ds \\
	&=\Big(\frac{1}{2}\sigma^2\Big)^\gamma\,\frac{\Gamma(\kappa)}{\Gamma(\kappa-\gamma)\Gamma(\gamma)}\,\lim_{u\rightarrow \infty}u^\gamma 
	\int_0^1 e^{-us}(1-s)^{\kappa-\gamma-1}s^{\gamma-1}\,ds \\
	&=\Big(\frac{1}{2}\sigma^2\Big)^\gamma\,\frac{\Gamma(\kappa)}{\Gamma(\kappa-\gamma)\Gamma(\gamma)}\,\lim_{u\rightarrow \infty} 
	\int_0^u e^{-t}(1-t/u)^{\kappa-\gamma-1}t^{\gamma-1}\,dt \\
	&=\Big(\frac{1}{2}\sigma^2\Big)^\gamma\,\frac{\Gamma(\kappa)}{\Gamma(\kappa-\gamma)\Gamma(\gamma)}\,\int_0^\infty e^{-t}t^{\gamma-1}\,dt \\
	&=\Big(\frac{1}{2}\sigma^2\Big)^\gamma\,\frac{\Gamma(\kappa)}{\Gamma(\kappa-\gamma)} 
	\end{aligned}
	\end{equation*}
	where $\Gamma$ is the gamma function. \newline

	\noindent {\em Step} (ii). We now show that the function  $V(y,0)$ satisfies
	$V(y,0)\leq c_1 e^{-\gamma b T}$ 	
	for some positive constant $c_1$ independent of $T.$  
	From Eq.\eqref{eqn:CIR_V}, we know 
	$$V(y,0)=c_2(T;y)\Big(\frac{\sigma^2(1-e^{-bT})y}{2b}\Big)^{-\gamma} e^{-\gamma bT}$$
	where
	$$c_2(T;y):=
	\Big(\frac{1}{2}\sigma^2\Big)^\gamma M\Big(\kappa-\gamma,\kappa;\frac{2b}{\sigma^2(e^{bT}-1)y}\Big)\,e^{-\frac{2b}{\sigma^2(e^{bT}-1)y}}\,.$$
	Observe that  $c_2(T;x)$ is uniformly bounded for large $T$ because $\lim_{u\rightarrow 0}M(\kappa-\gamma,\kappa,u)=1.$ This gives the desired result. \newline

	\noindent {\em Step} (iii).
	Because $V(Y_t,t)e^{\epsilon_0\int_0^tY_s\,ds}$
	is a positive local martingale for $0\leq t\leq T,$ it is a supermartingale. Thus, we have
	\begin{equation*}
	\begin{aligned}
	&\Big(\frac{1}{2}\sigma^2\Big)^\gamma\,\frac{\Gamma(\kappa)}{\Gamma(\kappa-\gamma)}
	\mathbb{E}_\xi^\mathbb{Q}[e^{\epsilon_0\int_0^TY_s\,ds}]=\mathbb{E}_\xi^\mathbb{Q}[V(Y_T,T)e^{\epsilon_0\int_0^TY_s\,ds}]\leq  V(Y_0,0)\leq c_1 e^{-\gamma b T}\,.
	\end{aligned}
	\end{equation*}
	This completes the proof.
\end{proof}

\subsection{Sensitivity of $a$}
\label{sec:sen_CIR_a}

This section analyzes the long-term sensitivity
of $a$ in the drift coefficient of the CIR model.
The parameter $a$ can be regarded as a perturbation parameter. 
The purpose is to show the long-term sensitivity of $a$
in Eq.\eqref{eqn:sen_CIR}
by applying Corollary \ref{cor:rho}.
Only the hypothesis of Theorem \ref{thm:rho_expo_condi} will be checked because the other conditions are easy to prove.
Recall the definitions of $k$ and $g$ in Section \ref{sec:rho}. From the function $k$ in Eq.\eqref{eqn:k_CIR}, we define
$g(x):=\frac{1}{\sigma}\sqrt{|x|}$
so that for all $a$
$$\left|\frac{\partial k(x)}{\partial a}\right|
= \left|\frac{a\sqrt{|x|}}{\sigma\sqrt{a^2+2\sigma^2}}\right|\leq\frac{1}{\sigma}\sqrt{|x|} =g(x)\,.$$  
For condition (i) in Theorem \ref{thm:rho_expo_condi}, it is sufficient to show that there exist constants $c$ and $d$ such that
\begin{equation} \label{eqn:expo_CIR}
\begin{aligned}
\mathbb{E}_\xi^{\mathbb{Q}}[e^{\int_0^TX_s\,ds}]
\leq c \,e^{dT}
\end{aligned}
\end{equation}
for all $T.$ This is proven by Lemma 3.1 on page 6 in \cite{wong2006changes}.
For (ii) and (iii), we set $p=q=2$ and let $\epsilon_1=2.$ Then, it can easily be shown that the expectation
\begin{equation}\label{eqn:CIR_a_ii} 
\begin{aligned}
\mathbb{E}_\xi^\mathbb{Q}[g^{2+\epsilon_1}(X_t)]=\frac{1}{\sigma^4}\mathbb{E}_\xi^{\mathbb{Q}}[X_t^2]
\end{aligned} 
\end{equation}
is uniformly bounded in $t$ on $[0,\infty)$ 
because $X$ is a CIR process. 
Thus, $\mathbb{E}_\xi^\mathbb{Q}[\int_0^Tg^{2+\epsilon_1}(X_t)\,dt]$ is finite for each $T$, which implies (ii).
For (iii), we define
$\psi(x)=f(x)e^{cx}$
for a constant $c$ such that ${(b-a)}/{\sigma^2}<c<{b}/{\sigma^2},$ and then  Eq.\eqref{eqn:psi_bound} follows.
By using the method in Eq.\eqref{eqn:CIR_inv}, we obtain condition (iii) since  the exponential growth rate of $\psi^2(x)$ is $2cx$    as $x\to\infty$.

\subsection{Sensitivity of $\sigma$}
\label{app:sens_sigma_CIR}

This section conducts a sensitivity analysis with respect to the variable $\sigma$ in the CIR model.
The parameter $\sigma$ in the diffusion term can be regarded as a perturbation parameter.
Define a quadruple $(\delta,1,R,F)$ and $\zeta$ by the Lamperti transformation given in  Section \ref{sec:Lamperti_trans}. Let $u(x):=\int_0^x\frac{1}{\sigma\sqrt{|y|}}\,dy=\frac{2}{\sigma}\sqrt{x}$ for $x>0,$ then 
\begin{equation}
\begin{aligned}
&\delta(u)= \Big(\frac{2\theta}{\sigma^2}-\frac{1}{2}\Big)\frac{1}{u}-\frac{au}{2}\,,\;R(u)=\sigma^2u^2/4\,,\;F(u)=f(\sigma^2u^2/4)\,,\;\zeta=\frac{2}{\sigma}\sqrt{\xi}\,.
\end{aligned}
\end{equation} 
Because
Section \ref{app:HS_CIR} shows that the quadruple $(b,\sigma,r,f)$ and the initial value $\xi$ satisfy Assumptions \ref{assume:pert_defi} - \ref{assume:perturbed},
the 
quadruple $(\delta,1,R,F)$ and $\zeta$ also satisfy Assumptions \ref{assume:pert_defi} - \ref{assume:perturbed}
by Proposition \ref{prop:Lamp_equiv}. 
The notations
$U,$ $\mathbb{Q},$ $(\lambda,\Phi)$ 
are now self-explanatory. The recurrent eigenfunction and the payoff function are $\Phi(u)=\phi(\sigma^2u^2/4)$ and $F(u)=f(\sigma^2u^2/4),$ respectively.

The goal of this section is to
show the long-term sensitivity of $\sigma$  
in Eq.\eqref{eqn:sen_CIR} by using Theorem \ref{thm:Lamp}.
Conditions (i) and (ii) in Theorem \ref{thm:Lamp} will be discussed below and 
condition (iii) will be  proven in Proposition \ref{prop:sen_sigma_sigma}.
To check (i), observe that  
$\lambda={\theta(\sqrt{a^{2}+2\sigma^{2}}-a)}/{\sigma^{2}}$ and  
$\Phi(\zeta)=\phi(\sigma^2\zeta^2/4)$ are continuously differentiable in variable $\sigma,$ 
which with Proposition \ref{prop:sens_sigma_CIR} below gives (i). 
To check (ii)  in Theorem \ref{thm:Lamp}, we apply Theorem \ref{thm:rho_expo_condi}.
Recall the definitions of $k$ and $g$ in Section \ref{sec:rho} and that $b=\sqrt{a^2+2\sigma^2}.$
We define
$k(u)=(\frac{2\theta}{\sigma^2}-\frac{1}{2})\frac{1}{u}-\frac{bu}{2}$ and
$g(u)=C(u+\frac{1}{u})$
for sufficiently large $C>0$ such that 
$|\frac{\partial}{\partial \sigma}k(u)|
\leq C(\frac{1}{u}+u)=g(u).$
Observe that 
$g^2(U_t)\leq C_1(\frac{1}{X_t}+{X_t})$
for sufficiently large $C_1>0.$
To prove the exponential condition (i) in Theorem \ref{thm:rho_expo_condi}, 
it suffices to show that there exist 
positive constants $a,$ $c$ and $\epsilon_0$ such that 
$\mathbb{E}_\xi^{\mathbb{Q}}[e^{\epsilon_0\int_0^T(X_s+\frac{1}{X_s})\,ds}]
\leq c \,e^{aT}$ for all $T>0.$ 
This can be shown by combining Proposition \ref{prop:CIR_expo} and Eq.\eqref{eqn:expo_CIR}.
Condition (ii) of Theorem \ref{thm:rho_expo_condi}
can be confirmed with $p=2$ and $0<\epsilon_1<\min\{\frac{2\theta}{\sigma^2}-1,2\}$
by combining the methods in  Eq.\eqref{eqn:CIR_theta_ii} and
Eq.\eqref{eqn:CIR_a_ii}.
To check (iii) of Theorem \ref{thm:rho_expo_condi}, choose a real number $c$ such that
$(b-a)/4<c<b/4,$ and   define $\psi(u):=e^{cu^2}.$ For sufficiently large $u,$ 
$$F(u)/\Phi(u)=f(\sigma^2u^2/4)e^{\kappa\sigma^2u^2/4}
=f(\sigma^2u^2/4)e^{(b-a)u^2/4}\leq  \psi(u)$$
since $f$ is of polynomial growth rate.
Using the  method in Eq.\eqref{eqn:CIR_inv}, 
it is easy to show that the expectation $\mathbb{E}^\mathbb{Q}[\psi^2(U_T)]=\mathbb{E}^\mathbb{Q}[e^{2cU_T^2}]=\mathbb{E}^\mathbb{Q}[e^{\frac{8c}{\sigma^2}X_T}]$ is uniformly bounded in $T$ on $[0,\infty)$
because ${8c}/{\sigma^2}<{2b}/{\sigma^2}.$
This proves (iii) of Theorem \ref{thm:rho_expo_condi} with $q=2.$

Proposition \ref{prop:sens_sigma_CIR} below is useful for checking (i) in Theorem \ref{thm:Lamp}.
The parameter $\sigma$ is a variable in $F/\Phi
,$ $\zeta,$ and the dynamics of $U.$
We temporarily employ a new parameter $s$ to distinguish the parameter $\sigma$ in $F/\Phi$ and $\zeta$ from the parameter $\sigma$ in the dynamics of $U.$  Define
\begin{equation}\label{eqn:ftns_s}
\begin{aligned}
&\eta(s)=\frac{\sqrt{a^2+2s^2}-a}{s^2}\,,\;
\pi_s(r)=e^{-\eta(s)r}\,,\;
\Pi_s(u)=\pi_s(s^2u^2/4)\,,\\
&G_s(u)=f(s^2u^2/4)\,,\;
q(s)=\frac{2}{s}\sqrt{\xi} \,,
\end{aligned}
\end{equation}	
so that
$\mathbb{E}_{\zeta}^{\mathbb{Q}}
[(F/\Phi)(U_T)]=\mathbb{E}_{q(\sigma)}^\mathbb{Q}[(G_\sigma/\Pi_\sigma)(U_T)].$

\begin{proposition}\label{prop:sens_sigma_CIR}
	Fix a positive real number $\sigma_0.$ 
	The partial derivative	$\frac{\partial}{\partial s}\mathbb{E}_{q(s)}^\mathbb{Q}[(G_s/\Pi_s)(U_T)]$
	exists and is continuous in $(s,\sigma)$ on a   neighborhood of $(\sigma_0,\sigma_0).$ Moreover,  we have  
	$$\lim_{T\to\infty}\frac{1}{T}  \frac{\partial}{\partial s}\Big|_{s=\sigma} \mathbb{E}_{q(s)}^\mathbb{Q}[(G_s/\Pi_s)(U_T)]=0$$
	for any positive number $\sigma$ in a neighborhood of $\sigma_0.$
\end{proposition}

\begin{proof}
	The proof will be given in several steps. 
	\begin{itemize}
		\item[(i)] Define a process $Z=(Z_t)_{t\geq0}$ by $Z_t=Z_t(s)=s^2U_t^2/4$ so that  
		$$\mathbb{E}_{q(s)}^\mathbb{Q}[(G_s/\Pi_s)(U_T)]
		=\mathbb{E}_{\xi}^\mathbb{Q}[(f/\pi_s)(Z_T)]\,.$$
		The right-hand side is more manageable.
		\item[(ii)] Show that	the partial derivative
		$\frac{\partial }{\partial s}\mathbb{E}_{\xi}^\mathbb{Q}[(f/\pi_s)(Z_T)]$
		exists and
		$$\frac{\partial}{\partial s}\mathbb{E}_{\xi}^\mathbb{Q}[(f/\pi_s)(Z_T)] =\int_0^\infty f(z)\,\frac{\partial}{\partial s} \frac{\ell(z;T,s)}{\pi_s(z)}\,dz$$
		for $(s,\sigma)$ near $(\sigma_0,\sigma_0),$
		where $\ell(z;t,s)$ is the density function of $Z_t.$ 
		Then, deduce that  this partial derivative is continuous in $(s,\sigma)$ on a neighborhood of $(\sigma_0,\sigma_0).$  
		\item[(iii)] Finally, we show that
		$$\int_0^\infty f(z)\,\frac{\partial}{\partial s}\Big|_{s=\sigma} \frac{\ell(z;T,s)}{\pi_s(z)}\,dz$$
		converges to a finite constant as $T\rightarrow\infty,$ which gives the desired result. \newline
	\end{itemize}

	\noindent {\em Step} (i).
	Define a process $Z=(Z_t)_{t\geq0}$ by $Z_t=Z_t(s)=s^2U_t^2/4$ so that  $(G_s/\Pi_s)(U_T)=(f/\pi_s)(s^2U^2_T/4)=(f/\pi_s)(Z_T)$ and $Z_0=\xi.$ Then,  
	$ \mathbb{E}_{q(s)}^\mathbb{Q}[(G_s/\Pi_s)(U_T)]
	=\mathbb{E}_{\xi}^\mathbb{Q}[(f/\pi_s)(Z_T)].$
	The Ito formula gives 
	$$dZ_t=\Big(\frac{\theta s^2}{\sigma^2}-bZ_t\Big)\,dt+s\sqrt{Z_t}\,dW_t\;,\,Z_0=\xi\,.$$
	It is noteworthy that both the parameters $\sigma$ and $s$ are components in the dynamics of $Z$, but we are only interested in the sensitivity of $s.$
	One of the notable properties of this process $Z$ is that the initial value is not perturbed.\newline

	\noindent {\em Step} (ii).
	The process $Z$ is a CIR process  and 
	the density function of $Z_t$ (from Eq.\eqref{eqn:CIR_density_Re}) is 
	\begin{equation}\label{eqb:CIR_density_s}
	\ell(z;t,s)=e^{-h_t\xi e^{-bt}}(\xi e^{-bt})^{-q/2}\,h_t\,e^{-h_tz}z^{q/2}I_q(2h_te^{-bt/2}\sqrt{\xi z})\,,
	\end{equation}
	where 
	$h_t=\frac{2b}{s^2(1-e^{-bt})},$ $q=\frac{2\theta}{\sigma^2}-1,$ and	
	$I_q$ is the modified Bessel function of the first type of order $q.$
	For $(s,\sigma)$ near $(\sigma_0,\sigma_0),$
	we will prove that  
	\begin{equation}\label{eqn:interchange_int_diff_cir_sigma}
	\begin{aligned}
	\frac{\partial}{\partial s}\mathbb{E}_{\xi}^\mathbb{Q}[(f/\pi_s)(Z_T)]
	&=\frac{\partial}{\partial s}\int_0^\infty (f/\pi_s)(z)\,\ell(z;T,s)\,dz
	=\int_0^\infty f(z)\,\frac{\partial}{\partial s} \frac{\ell(z;T,s)}{\pi_s(z)}\,dz\,.
	\end{aligned}
	\end{equation} 
	To prove the interchangeability of the differentiation and the integration in the second equality, it  suffices to show that 
	for $(s,\sigma)$ near $(\sigma_0,\sigma_0)$ and for all $z>0,$
	$$\left|f(z)\,\frac{\partial}{\partial s} \frac{\ell(z;t,s)}{\pi_s(z)}\right|\leq Ce^{-\frac{1}{\sigma_0^2}\sqrt{a^2+2\sigma_0^2}\,z}=: G(z)  $$
	for a positive constant $C$ because the function $G(z)$ is integrable over $(0,\infty).$	
	Let us estimate how fact the function 
	$|f(z)\,\frac{\partial}{\partial s}({\ell(z;t,s)}/{\pi_s(z)})|$ grows as $z\to\infty$ 
	by observing  the growth rates of $f(z),$ $\frac{1}{\pi_s(z)},$  $\frac{\partial}{\partial s}\frac{1}{\pi_s(z)},$ $\ell(z;t,s),$ and each term of
	\begin{equation}\label{eqn:deriva_ell} 
	\begin{aligned}
	\frac{\partial}{\partial s}\,\ell(z;t,s)&=\frac{\,2\,}{s}h_t\xi e^{-bt}\ell(z;t,s)-\frac{\,2\,}{s}\ell(z;t,s)+\frac{\,2\,}{s}zh_t\ell(z;t,s)\\
	&-\frac{\,2\,}{s}e^{-h_t\xi e^{-bt}}\xi^{(-q+1)/2}e^{(q-1)bt/2}\,h_t^2\,e^{-h_tz}z^{(q+1)/2}(I_{q-1}+I_{q+1})\,.\\
	\end{aligned}
	\end{equation}
	Given $\sigma>0$ and large $t>0,$ for $s$ near $\sigma_0,$ each term of $\frac{\partial}{\partial s} \ell(z;t,s)$ is dominated by one of 
	\begin{equation}\label{eqn:partial_g}
	\ell(z;t,s)\,,\;z\ell(z;t,s)\,,\; z^qe^{-h_tz+2h_te^{-bt/2}\sqrt{\xi z}}\,,\;z^{q+1}e^{-h_tz+2h_te^{-bt/2}\sqrt{\xi z}}
	\end{equation}	 up to constant multiples.
	We used the upper bound of $I_q$ given in Eq.\eqref{eqn:I_bdd}.	
	The growth rate of each term is essentially dominated by $e^{-\frac{2b}{s^2}z}$ because  $\frac{2b}{s^2}<h_t.$ Thus, the growth rate of $|\frac{\partial}{\partial s}(\ell(z;t,s)/\pi_s (z))|$ is less than or equal to that of  $e^{(\eta(s)-\frac{2b}{s^2})z}.$  Because the growth rate of $f(z)$ is less than or equal to the polynomial growth rate, it follows that the growth rate of  $|f(z)\frac{\partial}{\partial s}(\ell(z;t,s)/\pi_s (z))|$ is less than or 
	equal to that of $e^{(\eta(s)-\frac{2b}{s^2})z}.$ 
	For $(s,\sigma)$ near $(\sigma_0,\sigma_0),$ 
	the exponent of $e^{(\eta(s)-\frac{2b}{s^2})z}$  satisfies 
	\begin{equation}
	\label{eqn:eta_exp_estimate}
	\eta(s)-\frac{2b}{s^2}=\frac{\sqrt{a^2+2s^2}-a}{s^2}-\frac{2\sqrt{a^2+2\sigma^2}}{s^2}<-\frac{\sqrt{a^2+2\sigma_0^2}}{\sigma_0^2}\,,
	\end{equation} 
	which is the desired inequality. Since Eq.\eqref{eqn:interchange_int_diff_cir_sigma} holds,
it is directly derived that the partial derivative 
	$\frac{\partial}{\partial s}\mathbb{E}_{q(s)}^\mathbb{Q}[(G_s/\Pi_s)(U_T)]=\frac{\partial}{\partial s}\mathbb{E}_{\xi}^\mathbb{Q}[(f/\pi_s)(Z_T)]$ exists and is continuous in $(s,\sigma)$ on a   neighborhood of $(\sigma_0,\sigma_0).$ 	\newline

	\noindent {\em Step} (iii).	
	Finally, we demonstrate that
	$$\int_0^\infty f(z)\,\frac{\partial}{\partial s}\Big|_{s=\sigma} \frac{\ell(z;T,s)}{\pi_s(z)}\,dz$$
	converges to a finite constant as $T\rightarrow\infty.$ 
	This can be proven by the Lebesgue  dominated convergence theorem and by observing how fact 
	the function  $|\frac{\partial}{\partial s}(\ell(z;t,s)/\pi_s (z))|$ grows  as $T\to\infty$ in the same manner as above.
	This completes the proof.
\end{proof}

We now prove (iii) in Theorem \ref{thm:Lamp}.
For simplicity, we omit the variable $s$ in the following notations, so  
$$\pi=\pi_s\,,\;\Pi=\Pi_s\,,\;G=G_s\,,\;q=q(s)$$
for functions defined in Eq.\eqref{eqn:ftns_s}.
\begin{proposition}\label{prop:sen_sigma_sigma}
	Fix a positive real number $\sigma_0.$ 
	The partial derivative
	$\frac{\partial}{\partial \sigma}\mathbb{E}_{q}^\mathbb{Q}[(G/\Pi)(U_T)]$
	is continuous in $(s,\sigma)$ on a   neighborhood of $(\sigma_0,\sigma_0).$	 
\end{proposition}

\begin{proof}
	We only sketch the main idea because the proof is similar to that of Proposition \ref{prop:sens_sigma_CIR}.   	Define a process $Z=(Z_t)_{t\geq0}$ by $Z_t=s^2U_t^2/4$ so that   
	$ \mathbb{E}_{q}^\mathbb{Q}[(G/\Pi)(U_T)]
	=\mathbb{E}_{\xi}^\mathbb{Q}[(f/\pi)(Z_T)].$
	Consider  
	the density function $\ell=\ell(z;t)$ of $Z_t$ 
	given in   Eq.\eqref{eqb:CIR_density_s}
	$$\ell(z;t)=e^{-h_t\xi e^{-bt}}(\xi e^{-bt})^{-q/2}\,h_t\,e^{-h_tz}z^{q/2}I_q(2h_te^{-bt/2}\sqrt{\xi z})\,,$$
	where 
	$h_t=\frac{2b}{s^2(1-e^{-bt})}$ and $q=\frac{2\theta}{\sigma^2}-1.$
	For $(s,\sigma)$ near $(\sigma_0,\sigma_0),$
	we will prove that  
	\begin{equation} \label{eqn:CIR_sigma}
	\begin{aligned}
	\frac{\partial}{\partial \sigma}\mathbb{E}_{\xi}^\mathbb{Q}[(f/\pi)(Z_T)]
	&=\frac{\partial}{\partial\sigma }\int_0^\infty (f/\pi)(z)\,\ell(z;T)\,dz
	=\int_0^\infty (f/\pi)(z)\,\frac{\partial}{\partial \sigma}\ell(z;T)\,dz\,.
	\end{aligned}
	\end{equation}

	To prove the interchangeability of the differentiation and the integration in the above equality, it  suffices to show that 
	for $(s,\sigma)$ near $(\sigma_0,\sigma_0)$ and for all $z>0,$
	$$\left|(f/\pi)(z)\frac{\partial}{\partial \sigma}\ell(z;t)\right|\leq Ce^{-\frac{1}{\sigma_0^2}\sqrt{a^2+2\sigma_0^2}\,z}=: G(z)  $$
	for a positive constant $C$ because the function $G(z)$ is integrable over $(0,\infty).$
	Consider the growth rate of $(f/\pi)(z)\frac{\partial}{\partial \sigma}\ell(z;t).$
	Given $\sigma>0$ and large $t>0,$ for $s$ near $\sigma_0,$ each term of $\frac{\partial}{\partial \sigma}\ell(z;t)$ is dominated by one of
	\begin{equation}\label{eqn:deriva_ell_sigma}
	\ell(z;t),\,z\ell(z;t),\,\ln(z)\ell(z;t),\,	 z^{q/2}e^{-h_tz+2h_te^{-bt/2}\sqrt{\xi z}},\,z^{q+1}e^{-h_tz+2h_te^{-bt/2}\sqrt{\xi z}} 
	\end{equation}  
	up to constant multiples.
	In the calculation of $\frac{\partial}{\partial \sigma}\ell(z;t),$ we used 
	the upper bound of $I_q$ given in Eq.\eqref{eqn:I_bdd}
	and the equality 
	$$\frac{\partial}{\partial q}I_q(z)=I_q(z)\ln(z/2)+\frac{\Gamma'(q+1/2)}{\Gamma(q+1/2)}I_q(z)+\int_0^\pi (e^{z\cos u}\sin^{2q}u\ln(\sin^2 u))\,du\,.$$
	Let $x=\sin^2u$  for $u\in[0,\pi].$ Then for $x\in[0,1]$
	it is easy to check that the range of $x^{q}\ln x$ is $[-1/(qe),0].$ Thus,
	$$-\frac{\pi}{q} e^{z-1} 
	\leq \int_0^\pi (e^{z\cos u}\sin^{2q}u\ln(\sin^2 u))\,du\leq 0\,.$$
	The growth rate of each term in Eq.\eqref{eqn:deriva_ell_sigma} is essentially dominated by  $e^{-\frac{2b}{s^2}z}$ up to polynomial multiples. Thus, the growth rate of $|(f/\pi)(z)\frac{\partial}{\partial \sigma}\ell(z;t)|$ is less than or equal to that of $e^{(\eta(s)-\frac{2b}{s^2})z}$ up to polynomial multiples since  the growth rate of $f(z)$ is less than or equal to the polynomial growth rate.
	From the  argument in Eq.\eqref{eqn:eta_exp_estimate}, we obtain the desired result.
	Since Eq.\eqref{eqn:CIR_sigma} holds, it is directly derived that     the partial derivative $\frac{\partial}{\partial \sigma}\mathbb{E}_{q}^\mathbb{Q}[(G/\Pi)(U_T)]=\frac{\partial}{\partial \sigma}\mathbb{E}_{\xi}^\mathbb{Q}[(f/\pi)(Z_T)]$ 
		is continuous in $(s,\sigma)$ on a   neighborhood of $(\sigma_0,\sigma_0).$

\end{proof}

\section{The quadratic-term structure model}
\label{app:QTSM}

\subsection{Hansen--Scheinkman decomposition}
\label{app:HS_QTSM}

First, observe that $(b,\sigma,r,f)$ and $\xi$ satisfy Assumptions \ref{assume:pert_defi} - \ref{assume:perturbed} (that is, Assumptions \ref{assume:SDE} - \ref{assume:f} and \ref{assume:HS} - \ref{assume:ergodic}).
Assumptions \ref{assume:SDE} and \ref{assume:HS} - \ref{assume:inv} can be confirmed from Section 6.2 in \cite{qin2016positive}, and the other conditions are trivial. The notations $X,$ $\mathcal{P},$ $\mathcal{L},$ $M,$ $\mathbb{Q},$ $(\lambda,\phi),$  $\varphi,$ $\nu$
are self-explanatory.
The recurrent eigenpair is 
given in 
Eq.\eqref{eqn:eigenpair_QTSM}.
The $\mathbb{Q}$-dynamics of $X$ is
\begin{equation}\label{eqn:SDE_QTSM_under_Q}
dX_t=(b-au+(B-2aV)X_t)\,dt+\sigma\,dW_t
\end{equation}
where $W$ is a $\mathbb{Q}$-Brownian motion.

\subsection{Sensitivity of $\xi$}
\label{app:sen_xi_QTSM}
We want to find the long-term sensitivity of the expectation $p_T$ with respect to the initial value $\xi.$ The aim is to show 
$\lim_{T\rightarrow\infty} \nabla_\xi\ln p_T=\frac{\nabla_\xi\,\phi(\xi)}{\phi(\xi)} =-u-2V\xi$
by applying Proposition \ref{prop:delta}. 
The first variation process $Y$ is given by
$dY_t=(B-2aV)Y_t\,dt$ with $Y_0=I_d,$
where $I_d$ is the $d\times d$ identity matrix.
It follows that
$\mathbb{E}_\xi^\mathbb{Q}[|\!|Y_T|\!|^2]=|\!|Y_T|\!|^2=|\!|e^{(B-2aV)T}|\!|^2.$
Because all eigenvalues of $B-2aV$ have negative real parts, it follows that $\mathbb{E}_\xi^\mathbb{Q}[|\!|Y_T|\!|^2]$ is uniformly bounded in $T$ on $[0,\infty).$ This gives the desired result.

\subsection{Sensitivity of $b$}
\label{app:sen_b_QTSM}

We perform a sensitivity analysis of the expectation $p_T$ with respect to the drift coefficients $b=(b_1,b_2,\cdots,b_d)^\top.$ 
Fix $i=1,2,\cdots,d.$ The parameter $b_i$ can be regarded as a perturbation parameter.
The goal is to show that by applying Corollary \ref{cor:rho},
$\lim_{T\rightarrow\infty}\frac{1}{T}\frac{\partial}{\partial b_i}\ln p_T=-\frac{\partial\lambda}{\partial b_i}.$
Assumption \ref{assume:condi1_2} is easy to confirm from Eq.\eqref{eqn:eigenpair_QTSM} and the fact that $f$ is a bounded function with bounded support. 
We now apply Theorem \ref{thm:rho_expo_condi}.
Recall the definitions of $k$ and $g$ in Section \ref{sec:rho}. Define
\begin{equation}\label{eqn:k_QTSM}
k(x)=\sigma^{-1}b-\sigma^{\top}u+(\sigma^{-1}B-2\sigma^{\top}V)x
\end{equation} 
and let $g(x)=C$ be a constant function for sufficiently large $C>0$ such that
$|\frac{\partial}{\partial b_i}k(x)|
\leq |(\sigma^{-1})_i|<C=g(x)$
for $i=1,2,\cdots,d$ where $(\sigma^{-1})_i$ is the $i$-th column of $\sigma^{-1}.$
Because $g$ is a constant function,   (i) and (ii) of Theorem \ref{thm:rho_expo_condi} are trivially satisfied with $p=q=2.$ We now consider (iii) of Theorem \ref{thm:rho_expo_condi}.
As a function of two variables $(x,b_i)$, we write the function $\phi(x)$ as $\phi(x,b_i).$ Since $f$ has  bounded support, we choose a compact set $K$ such that supp$(f)\subseteq K.$
For a bounded open neighborhood $I\subseteq\mathbb{R}$ of $0,$ define
$\overline{I}_{b_i}:=\{b_i+r\in\mathbb{R}:r\in \overline{I}\}.$
Since $\phi$ is a positive and continuous function in two variables $(x,b_i)$,
the reciprocal $1/\phi$ has a positive maximum on compact set $K\times \overline{I}_{b_i}.$ 
We define
\begin{equation}
\label{eqn:M_phi}
M:=\max_{(x,z)\in K\times \overline{I}_{b_i} }\frac{1}{\phi(x,z)}\,,\;\;\psi(x):=Mf(x)\,.
\end{equation} 
Then Eq.\eqref{eqn:psi_bound}
is satisfied.
With this function $\psi,$ it is easy to check (iii) 
because $f$ is a bounded function and supp$(f)\subseteq K.$

\subsection{Sensitivity of $B$}
\label{app:sen_B_QTSM}

We investigate the long-term sensitivity of the expectation $p_T$ with respect to the matrix $B=(B_{ij})_{1\leq i,j\leq d}.$ 
The parameter $B_{ij}$ can be regarded as a perturbation parameter.
The goal is to show that 
$\lim_{T\rightarrow\infty}\frac{1}{T}\frac{\partial}{\partial B_{ij}}\ln p_T=-\frac{\partial\lambda}{\partial B_{ij}}$
by applying Corollary \ref{cor:rho}.
For condition (i) in Theorem \ref{thm:total_chain}, it is sufficient to check that $V$ (and hence, $u$) is continuously differentiable in $B_{ij}.$
Here, the continuous differentiability of a matrix and a vector means that all components are continuously differentiable.
The continuous differentiability of $V$ is from  
Eq.(2.5) on page 240 in \cite{sun2002condition}
and Theorem 3.1 in \cite{sun1998perturbation}.
Condition (ii) in Theorem \ref{thm:total_chain} is easy to check because $f$ is a bounded function with bounded support.

We now apply Theorem \ref{thm:expo_rho}.
Recall the definitions of $k$ and $g$ in Section \ref{sec:rho}. From Eq.\eqref{eqn:k_QTSM}, because $V$ and $u$ are continuously differentiable in $B_{ij},$ there exist sufficiently large constants $c_1$ and $c_2$ such that 
$$\left|\frac{\partial}{\partial B_{ij}}k(x)\right|\leq c_1+c_2|x|=:g(x).$$
To check condition (i) in Theorem \ref{thm:expo_rho}, it suffices to show that there exists a positive $\epsilon_0$ such that
$\mathbb{E}_\xi^{\mathbb{Q}}[e^{\epsilon_0|X_T|^2}]$ is uniformly bounded in $T$ on $[0,\infty).$
Consider the density function of $X_T,$ which is a multivariate normal random variable. We have 
$$\mathbb{E}_\xi^{\mathbb{Q}}[e^{\epsilon_0|X_T|^2}]
=\frac{1}{\sqrt{(2\pi)^d\det\Sigma_T}}\int_{\mathbb{R}^d} e^{\epsilon_0 |z|^2-\frac{1}{2}(z-\mu_T)^\top\Sigma_T^{-1}(z-\mu_T)}\,dz$$ 
where $\mu_T$ and $\Sigma_T$ are the mean vector and the covariance matrix of $X_T,$ respectively. Observe the exponent $\epsilon_0 |z|^2-\frac{1}{2}(z-\mu_T)^\top\Sigma_T^{-1}(z-\mu_T)$ of the integrand.
Under the recurrent eigen-measure $\mathbb{Q},$ because Assumption \ref{assume:inv} is satisfied,  the distribution of $X_T$ converges to an invariant distribution which is a non-degenerate multivariate normal distribution. Let $\Sigma_\infty$ be the covariance matrix of the invariant distribution. Choose $\epsilon_0$ less than the smallest eigenvalue of $\Sigma_\infty^{-1},$ then the above integral converges to a constant as $T\rightarrow\infty,$ which implies condition (i). Condition (ii) in Theorem \ref{thm:expo_rho}  can also be checked by the  method in Eq.\eqref{eqn:M_phi}.

\subsection{Sensitivity of $\sigma$}
\label{app:sen_sigma_QTSM}

This section investigates the sensitivity of the expectation $p_T$ with respect to the volatility matrix $\sigma=(\sigma_{ij})_{1\leq i,j\leq d}.$
Assume that $f$ is continuously differentiable with compact support.
It can be shown that 
$\lim_{T\rightarrow\infty}\frac{1}{T}\frac{\partial }{\partial \sigma_i}\ln p_T=-\frac{\,\partial \lambda\,}{\partial \sigma_i}$
by using Theorem \ref{thm:total_chain} and \ref{thm:vega_Fournie_condi}. We check only the hypothesis of Theorems \ref{thm:vega_Fournie_condi} because the other conditions are easy to prove.
The corresponding variation process $Z=(Z_t)_{t\geq0}$ is given by
$$dZ_t=(B-2aV)Z_t\,dt+\sigma\,dW_t\,.$$
It follows that  
$\mathbb{E}_\xi^\mathbb{Q}[|Z_T|]$ is 
convergent as $T\rightarrow\infty$ because
the process $Z$ is an Ornstein-Uhlenbeck (OU) process and all eigenvalues of  $B-2aV$ have negative real parts.
This gives the desired result.

\section{The $3/2$ model}
\label{app:3/2_model}

The aim of this section is to prove the sensitivities
discussed in Section \ref{sec:3/2_model}.
For the sensitivity of $\xi,$  Eq.\eqref{eqn:xi_3_2} is obtained by showing 
$\lim_{T\rightarrow\infty}\frac{\partial}{\partial \xi}\ln q_T
= \frac{\phi'(\xi)}{\phi(\xi)}
=-\ell \xi^{-1}$
where $q_T$ is the expectation in Eq.\eqref{eqn:q_T_3_2}.
The reciprocal $Y:=1/X$ satisfies
$$dY_t=(a+\sigma^2(\ell+1)-\theta Y_t)\,dt-\sigma\sqrt{Y_t}\,dW_t$$ 
which is a CIR model, and therefore we can use the results of Section \ref{app:CIR_delta}. 
By Theorem \ref{thm:delta}, it is enough to show that the expectation
$\mathbb{E}_{\xi}^{\mathbb{Q}}[(f/\phi)(X_T)]=\mathbb{E}_{\xi}^{\mathbb{Q}}[Y_{T}^{-\alpha\beta-\ell}]$ is continuously differentiable in $\xi$ and that $\lim_{T\rightarrow\infty}\frac{\partial}{\partial \xi}\mathbb{E}_{\xi}^{\mathbb{Q}}[(f/\phi)(X_{T})]
=\lim_{T\rightarrow\infty}\frac{\partial}{\partial \xi}\mathbb{E}_{\xi}^{\mathbb{Q}}[Y_{T}^{-\alpha\beta-\ell}]=0.$
This can be proven by the method in Eq.\eqref{eqn:pf_xi_CIR}.

For the sensitivity of $\theta,$
Corollary \ref{cor:rho} with  
Theorem \ref{thm:rho_expo_condi} will be used   to show
$\lim_{T\rightarrow\infty}\frac{1}{T}\frac{\partial }{\partial \theta}\ln p_T=-\frac{\partial \lambda}{\partial\theta} =-\ell.$
We only show the conditions of Theorem \ref{thm:rho_expo_condi} because the other conditions are easily  checked.
From
$k(x)=\frac{\theta}{\sigma\sqrt{x}}-(\frac{\,a\,}{\sigma}+\sigma\ell)\sqrt{x},$
we define
$g(x):=\frac{1}{\sigma \sqrt{x}}.$
Condition (i) is evident since $1/X$ is a CIR process. Consider (ii) and (iii) with $q=1+\epsilon$ for a sufficiently small $\epsilon>0.$ 
Observe that for any $n\in\mathbb{N},$ the expectation
$\mathbb{E}_\xi^{\mathbb{Q}} [(1/X_T)^n]$   converges to a constant as $T\rightarrow\infty$ since $1/X$ is a CIR process. 
This proves (ii) in Theorem \ref{thm:rho_expo_condi}.
For (iii),   since $f$ and $\phi$ are independent of the parameter $\theta,$ we define $\psi(x)$ as $f(x)/\phi(x).$
For sufficiently small positive number $\epsilon,$ it is easy to show that the expectation   
$\mathbb{E}_\xi^\mathbb{Q}[\psi^{1+\epsilon}(X_T)]=\mathbb{E}_\xi^\mathbb{Q}[X_T^{(1+\epsilon)(\alpha\beta+\ell)}]$
converges as $T\rightarrow\infty$ by considering the density function of the CIR process $1/X.$

For the sensitivity of $a,$ the goal is to 
prove Eq.\eqref{eqn:a_3_2} by using Corollary \ref{cor:rho} and Theorem \ref{thm:rho_expo_condi}.
We only check condition (ii) in Theorem \ref{thm:total_chain} because the other conditions are easy to prove.
Define $Y:=1/X$ so that $Y$ is a CIR process. 
Let  $h(y;t)$ be the density function of $Y_t.$ 
We temporarily employ new parameter $b$ to distinguish parameter $a$ in  $f/\phi$ from parameter $a$ in the drift of $X.$  
Define $$\ell_b:=\sqrt{\Big(\frac{1}{2}+\frac{b}{\sigma^2}\Big)^2+\alpha\beta(\beta-1)}-\Big(\frac{1}{2}+\frac{b}{\sigma^2}\Big)\,,\;\pi_b(x):=x^{-\ell_b}$$
so that $\ell_a=\ell$ and $\pi_a(x)=x^{-\ell}=\phi(x)$ for the constant $\ell$ in Eq.\eqref{eqn:3/2_ell}. 
First, we show that the partial derivative $\frac{\partial}{\partial b}\mathbb{E}_\xi^\mathbb{Q}[(f/\pi_b)(X_T)]$ exists and that
$\frac{\partial}{\partial b}\mathbb{E}_\xi^\mathbb{Q}[(f/\pi_b)(X_T)]
=\mathbb{E}_\xi^\mathbb{Q}[\frac{\partial}{\partial b}(f/\pi_b)(X_T)].$
The proof is  obtained from   Theorem \ref{thm:payoff} by defining the dominating function $g$ as $g(x)=c_1x^{\alpha\beta+\ell+1}+ c_2x^{\alpha\beta+\ell-1}$  
for sufficiently large  constants $c_1$ and $c_2$ since
$$\left|\frac{\partial}{\partial b}(f/\pi_b)(x)\right|=\left|\frac{\partial\ell_b}{\partial b}x^{\alpha\beta+\ell_b}\ln x\right|\leq c_1x^{\alpha\beta+\ell+1}+ c_2x^{\alpha\beta+\ell-1}=g(x)$$
for all $b$ in a small open neighborhood of $a.$
The expectation $\mathbb{E}_\xi^\mathbb{Q}[g(X_T)]=c_1\mathbb{E}_\xi^{\mathbb{Q}}[Y_t^{-\alpha\beta-\ell-1}]+c_2\mathbb{E}_\xi^{\mathbb{Q}}[Y_t^{-\alpha\beta-\ell+1}]$ is finite   when $\frac{a}{\sigma^2}+1-\alpha\beta>0$ because the growth rate of the density $h(y,t)$ is dominated by $e^{-\frac{2\theta }{\sigma^2 }y}$ as $y\rightarrow\infty$ and is dominated by $y^{\frac{2a}{\sigma^2}+2\ell+1}$ as $y\rightarrow0^+.$ 
The joint continuity  of $\frac{\partial}{\partial b}\mathbb{E}_\xi^\mathbb{Q}[(f/\pi_b)(X_T)]$ in two variables $(b,a)$ can be obtained from the  joint continuity of $h(y;T)$ and from the equality 
\begin{equation}
\begin{aligned}
\frac{\partial}{\partial b}\mathbb{E}_\xi^\mathbb{Q}[(f/\pi_b)(X_T)]
&=\mathbb{E}_\xi^\mathbb{Q}\Big[\frac{\partial}{\partial b}(f/\pi_b)(X_T)\Big]
=\frac{\partial\ell_b}{\partial b}\mathbb{E}_\xi^\mathbb{Q}[X_T^{\alpha\beta+\ell_b}\ln X_T]\\
&=-\frac{\partial\ell_b}{\partial b}\mathbb{E}_\xi^\mathbb{Q}[Y_T^{-\alpha\beta-\ell_b}\ln Y_T]=-\frac{\partial\ell_b}{\partial b}\int_0^\infty h(y;T) y^{-\alpha\beta-\ell_b}\ln y \,dy\,.
\end{aligned}
\end{equation}
It is easy to check that $\mathbb{E}_\xi^\mathbb{Q}[g(X_T)]$ is convergent as $T\to\infty,$ which gives Eq.\eqref{eqn:condi_2_conv_0}.

For the sensitivity of $\sigma,$
consider a quadruple $(\delta,1,R,F)$ and an initial value $\zeta$  defined by the   Lamperti transformation in Section \ref{sec:Lamperti_trans}. Defining 
$u(x) =\frac{2}{\sigma\sqrt{x}}$ for $x>0,$ we get
\begin{equation}
\begin{aligned}
&\delta(u)=\Big(\frac{2a}{\sigma^2}+2\ell+\frac{3}{2}\Big)\frac{1}{u}-\frac{\theta u}{2} ,\,R(u)=2\alpha\beta(1-\beta)u^2,\,F(u)=(\sigma u/2)^{-2\alpha\beta},\,\zeta=\frac{2}{\sigma\sqrt{\xi}}\,.
\end{aligned}
\end{equation}
By Proposition \ref{prop:Lamp_equiv},
the 
quadruple $(\delta,1,R,F)$ and  $\zeta$   satisfy Assumptions \ref{assume:pert_defi} and \ref{assume:perturbed}
because 
the quadruple $(b,\sigma,r,f)$ and the initial value $\xi$ also satisfy them. 
One can show Eq.\eqref{eqn:sigma_3_2}
by using Theorem \ref{thm:Lamp}. 
Conditions   (i) and (iii) can be proven by the  methods in Propositions \ref{prop:sens_sigma_CIR} and \ref{prop:sen_sigma_sigma}, and condition
(ii)
can be shown by Theorem  \ref{thm:rho_expo_condi} 
when  $\frac{a}{\sigma^2}+1-\alpha\beta>0$ 
by applying the same method as used in the sensitivity analysis of $a.$

\section{Perturbation of payoff function}
\label{sec:payoff}
In this section, we are interested in the partial derivative  $\frac{\partial}{\partial \epsilon}\mathbb{E}[h_\epsilon(X)]$ of an expectation. 
The  interchangeability of 
differentiation and expectation is an important issue in this paper. 
The following theorem is a well known fact, and it is noteworthy because  this theorem is useful for checking  condition (ii) in Theorem \ref{thm:total_chain} when the initial value $\xi$ is not perturbed.

\begin{theorem} \label{thm:payoff}
	Let $X$ be a random variable and let $h_\epsilon(x)$ be a function of two variables $(\epsilon,x)$ on $I\times \mathbb{R}^d$ where $I$ is an open neighborhood of $0.$ Assume that $\mathbb{E}[h_\epsilon(X)]<\infty$ for each $\epsilon$ in $I$ and that $h_\epsilon(x)$ is 
	continuously differentiable  in $\epsilon$ on $I$ for each $x.$ 
	Suppose that there exists a positive function $g,$ which is called a {\em dominating function}, such that    $\mathbb{E}[g(X)]<\infty$ and
	$$\left|\frac{\partial}{\partial\epsilon}\,h_\epsilon(x)\right|\leq g(x)\;\;\textnormal{on}\;\;
	I\times \mathbb{R}^d\,.$$ 
	Then, the expectation $\mathbb{E}[h_\epsilon(X)]$ is continuously differentiable in $\epsilon$ on $I,$ and
	$$\frac{\partial}{\partial \epsilon}\mathbb{E}[h_\epsilon(X)]=
	\mathbb{E}\Big[\frac{\partial}{\partial \epsilon} h_\epsilon(X)\Big]\,.$$
\end{theorem}


\bibliographystyle{plainnat}

\bibliography{sensitivity}

\end{document}